\documentclass[reprint, aps, pra, twocolumn, amsmath, amssymb, nofootinbib]{revtex4-1}

\usepackage[utf8]{inputenc}
\usepackage[T2A,T1]{fontenc}
\newcommand{\Zhe}{\mbox{\usefont{T2A}{\rmdefault}{m}{n}\CYRZH}}

\usepackage{graphicx}
\usepackage{units}
\usepackage{color}
\usepackage[pdftex, colorlinks=true, linkcolor=myblue, citecolor=myblue,
urlcolor=myblue]{hyperref}
\usepackage{times}
\usepackage{bbold}

\DeclareMathAlphabet{\mathantt}{OT1}{antt}{li}{it}

\definecolor{myblue}{rgb}{0,0,1}

\usepackage{graphicx}
\usepackage{amsmath}
\usepackage{amssymb}

\begin{document}

\title{Magnetic response of metallic nanoparticles: Geometric and weakly relativistic effects}

\author{Mauricio G\'omez Viloria}
\affiliation{Universit\'e de Strasbourg, CNRS, Institut de Physique et Chimie des Mat\'eriaux de Strasbourg,
UMR 7504, F-67000 Strasbourg, France}

\author{Guillaume Weick}
\affiliation{Universit\'e de Strasbourg, CNRS, Institut de Physique et Chimie des Mat\'eriaux de Strasbourg,
UMR 7504, F-67000 Strasbourg, France}

\author{Dietmar Weinmann}
\affiliation{Universit\'e de Strasbourg, CNRS, Institut de Physique et Chimie des Mat\'eriaux de Strasbourg,
UMR 7504, F-67000 Strasbourg, France}

\author{Rodolfo A.\ Jalabert}
\affiliation{Universit\'e de Strasbourg, CNRS, Institut de Physique et Chimie des Mat\'eriaux de Strasbourg,
UMR 7504, F-67000 Strasbourg, France}


\begin{abstract}
While the large paramagnetic response measured in certain ensembles of metallic nanoparticles has been assigned 
to orbital effects of conduction electrons, 
the spin-orbit coupling has been pointed out as a possible origin of the anomalously large diamagnetic response observed in other cases. Such a relativistic effect, arising from the inhomogeneous electrostatic potential seen by the conduction electrons, might originate from the host ionic lattice, impurities, or the self-consistent confining potential. Here we theoretically investigate the effect of the spin-orbit coupling arising from the confining potential, quantifying its contribution to the zero-field magnetic susceptibility and gauging it against the ones generated by other weakly-relativistic corrections. Two ideal geometries are considered in detail, the sphere and the half-sphere, focusing on the expected increased role of the spin-orbit coupling upon a symmetry reduction, and the application of these results to actual metallic nanoparticles is discussed. The matrix elements of the different weakly-relativistic corrections are obtained and incorporated in a perturbative treatment of the magnetic field, leading to tractable semi-analytical and semiclassical expressions for the case of the sphere, while a numerical treatment becomes necessary for the half-sphere. The correction to the zero-field susceptibility arising from the spin-orbit coupling in a single sphere is quite small, and it is dominated by the weakly-relativistic kinetic energy correction, which in turn remains considerably smaller than the typical values of the nonrelativistic zero-field susceptibility. 
Moreover, the spin-orbit contribution to the average response for ensembles of nanoparticles with a large size dispersion 
is shown to vanish.
The symmetry reduction in going from the single sphere to the half-sphere does not
translate into a significant increase of the spin-orbit contribution to the 
zero-field susceptibility.
\end{abstract}

\maketitle

\section{Introduction}
\label{sec:intro}

The spin-orbit coupling (SOC) is a relativistic effect having a decisive role in certain properties of unconfined and confined condensed matter systems. A celebrated example is the spin-orbit driven change from weak localization to anti-weak localization observed in the electronic transport through metallic films \cite{bergmann1984} or ballistic quantum dots at a semiconductor heterojunction \cite{marcus2003,zaitsev2005}. In ferromagnetic materials the SOC underlines the phenomena of magnetic anisotropy and the anomalous Hall effect \cite{nagaosa2010}. The domain of spintronics addresses numerous cases where the SOC influences spin dynamics and spin relaxation \cite{fabian2003,eng-ras-hal,intronati_2012}. At the fundamental level, the SOC is important because it alters the symmetry properties of the electronic Hamiltonian. As a consequence, the statistics of energy levels in time-reversal symmetric (chaotic or disordered) confined systems changes from orthogonal to symplectic (Gaussian or circular) distributions when going from vanishing to strong SOC \cite{taniguchi1994,alhassid2000}. 

The tunneling resonances of disordered metal nanoparticles and the magnetic response of an ensemble of metallic nanoparticles are two examples of physical properties depending on the level statistics of a confined system, and thus on the SOC. 

In the first case, the extracted $g$-factor of the discrete energy levels was found  \cite{salinas1999,davidovic1999} to be below the free-electron value of $g_0=2$. 
Such a reduction could be explained by the fact that the energy eigenstates in the presence of SOC, not being purely spin up or spin down, respond more weakly to an applied magnetic field than pure spin states \cite{kawab70, halpe86_RMP}. The statistical distribution of the $g$-factors has been obtained from random matrix theory, using the spin-orbit scattering rate as a phenomenological parameter in order to describe the  transition between statistical ensembles \cite{brouwer2000,matveev2000}. 

In the second case, the zero-field susceptibility (ZFS) averaged over a nanoparticle ensemble is determined \cite{bouchiat1989,imry1991} by the magnetic field dependence of the variance in the number of energy levels below the chemical potential [see Eq.~\eqref{eq:F2} below]. Therefore, in the disordered or chaotic regimes, it depends on the transition between statistical ensembles driven by the influence on the SOC. In particular, the SOC has been invoked to be responsible for the large diamagnetism measured in an ensemble of gold nanorods \cite{orrit2013,titov2019}. This is a very interesting proposal, since among the anomalous magnetic responses observed in ensembles of metallic nanoparticles, only the paramagnetic behavior \cite{hori99_JPA, nakae00_PhysicaB, hori04_PRB, yamam04_PRL, yamam06, guerr08, guerr08_Nanotechnology, barto12_PRL, agrac17_ACSOmega} has been accounted for \cite{viloria2018orbital}, while no satisfying theory existed to describe the observed
\cite{cresp04_PRL, dutta07_APL, guerr08_Nanotechnology, orrit2013, hori04_PRB}
large diamagnetic response in metallic nanoparticles.\footnote{Superconducting fluctuations that persist beyond the critical temperature were shown to result in a relatively large diamagnetic response \cite{imry2015}, which nevertheless remains one to two orders of magnitude smaller than the one reported in the experiments of Ref.~\cite{orrit2013}.}$^,$\footnote{The most spectacular among the anomalous responses, which is the ferromagnetic behavior \cite{cresp04_PRL, cresp06_PRL, dutta07_APL, donni07_AdvMater, garit08_NL, guerr08_Nanotechnology, guerr08, venta09, donni10_SM, maitr11_CPC, agrac17_ACSOmega, grege12_CPC}, requires to have ferromagnetism at the level of individual nanoparticles or through the interparticle dipolar interaction \cite{viloria2018orbital,percebois2020}, and will not be addressed in this work. (See, e.g., Ref.~\cite{nealo12_Nanoscale} for a discussion of the different magnetic behaviors exhibited by ensembles of gold nanoparticles.)}

In the previously presented cases of the tunneling resonances of disordered metal nanoparticles and the magnetic response of an ensemble of metallic nanoparticles, the strength of the SOC is a key parameter that needs to be determined by microscopic theories. Towards this goal, the main source of SOC must be identified. 
The genesis of the SOC for the conduction electrons of a metallic nanoparticle lies in the existence of an inhomogeneous electrostatic potential, which may have an intrinsic origin (the host ionic lattice) or an extrinsic origin (impurities or the confining potential). Since gold is a heavy atom, SOC plays an important role in its band structure \citep{range12_PRB}, but the effect for the conduction electrons is mainly seen by the Bloch part of the wave function, while the smooth part remains unaffected. This observation is consistent with the $g$-factor $g_{\rm Au}=2.1$ measured by electron spin resonance in macroscopic gold samples \cite{monod1977} and the fact that the spin diffusion length in gold is typically 
limited by extrinsic effects \cite{isasa2015}. Impurities have been invoked to be responsible for the SOC of Ag nanoparticles intentionally doped with Au \cite{salinas1999}, but they are expected to play a lesser role in ballistic nanoparticles where the magnetic susceptibility has been measured. In this last 
setup,  the electronic confinement remains as the chief source of SOC. The effect of this latter mechanism in the ZFS is the goal of this paper, where we analyze model systems of noninteracting electrons with different kinds of confinement, making the link with experimentally relevant cases of metallic nanoparticles. 

The SOC yields a contribution to the fine-structure of atomic spectra which is of the 
same order as those arising from other weakly-relativistic corrections, namely the 
kinetic energy and Darwin terms \cite{schwabl1997aqm}. Hence, we contrast the SOC with the previously cited weakly-relativistic corrections, as well as with the angular
magneto-electric coupling relevant at finite magnetic field, for a confined electron gas
in the limit of Fermi velocities $v_\mathrm{F}$ much smaller than the speed of light in vacuum $c$. 

The ZFS of a nonrelativistic, three dimensional, unconfined (bulk) degenerate electron gas is $\chi_\mathrm{b}^{(\rm nr)}=\chi_\mathrm{L}+\chi_\mathrm{P}$,  resulting from the combined effect of the diamagnetic Landau susceptibility \cite{landau}
\begin{equation}
\label{eq:Landau_sus}
 \chi_\mathrm{L} = -\frac{1}{12\pi^2} \, \frac{e^2k_\mathrm{F}}{m^{*}c^2} \,    , 
\end{equation}
arising from the orbital motion, and the paramagnetic Pauli susceptibility
\begin{equation}
\label{eq:Pauli_sus}
\chi_\mathrm{P} = 3 \left|\chi_\mathrm{L}\right| \,    ,
\end{equation}
originating from the Zeeman effect over the electron spin. We note $-e$ the electron charge, $k_\mathrm{F}$ the Fermi wave vector, and $m^{*}$ the effective mass. In metals, the difference between $m^{*}$ and the free electron mass $m$ is very small (i.e., $m^{*}_{\rm Au}=1.1\,m$) and therefore we will identify the two electron masses.\footnote{The situation considerably changes when going from the case of metals to that of semiconductors, where such an identification is not valid and Eq.~\eqref{eq:Pauli_sus} does not hold.} Throughout the paper, we use cgs units, which for the case of gold result in a magnetic susceptibility  $\chi_\mathrm{L}=-2.9\times10^{-7}$. The ZFS of macroscopic gold samples is 
$\chi_{\rm Au} \simeq - 9.3 \; |\chi_\mathrm{L}|$, as the contribution $\chi_\mathrm{b}^{(\rm nr)}$ from the conduction electrons is dominated by the Larmor diamagnetic response arising from the closed-shell ion-core electrons \cite{suzuki2012}. 

The geometrical confinement of a metal to the nano- or micro-scales does not affect the response of the core electrons, while the effect on conduction electrons might be important, leading to a diamagnetic or paramagnetic ZFS with typical values considerably larger than $|\chi_\mathrm{L}|$, and the Larmor susceptibility as well, depending on the temperature, the Fermi wave vector, and the size of the containing box. The enhancement of the magnetic response is a consequence of the orbital motion of the conduction electrons \cite{viloria2018orbital}, and can be estimated with the help of semiclassical expansions (see Appendix \ref{app:sc}).

In the relativistic case the diamagnetic and paramagnetic contributions cannot be separated. The bulk ZFS of a weakly-relativistic electron gas is given, in the  limit $v_\mathrm{F} \ll c$, by \cite{rukhadze_1960}\footnote{Other values of the ZFS have been proposed in the literature \cite{chudnovsky1981}. However, the result \eqref{eq:weakly-rela_sus} is easily obtainable in the grand canonical ensemble by using the relativistic Landau levels, and it is in agreement with the numerical calculations that we present in Sec.~\ref{sec:kedc}.} 
\begin{equation}
\label{eq:weakly-rela_sus}
\chi_\mathrm{b}^{(\rm wr)} = \left[2-\frac{1}{3}\left(\frac{v_\mathrm{F}}{c}\right)^2\right]  \left| \chi_\mathrm{L} \right|    \,    .
\end{equation}
Thus, in normal metals the weakly-relativistic correction with respect to $\chi_\mathrm{b}^{(\rm nr)}$ is very small. Given the dramatic increase of the ZFS with respect to $\chi_\mathrm{L}$ induced by the electronic confinement in the nonrelativistic case, we might ask if a similar effect occurs for the weakly-relativistic susceptibility, and in particular for the contribution arising from the SOC. This is the goal of this work, where we analyze the case of a model system and apply it to metallic nanoparticles. 

Our paper is organized as follows: 
In Sec.~\ref{sec:msffs} we recall the thermodynamical quantum-mechanical and semiclassical formalisms which enable us to obtain the ZFS of a confined system.
In Sec.~\ref{sec:wrd} we present our model which we employ in order 
to assess the relevance of weakly-relativistic effects for the ZFS of spherical (Sec.~\ref{sec:sn})
and half-spherical nanoparticles (Sec.~\ref{sec:hsn}). 
We conclude in Sec.~\ref{sec:ccl}.
Six appendixes presenting some of the details of our calculations, as well as alternative semiclassical derivations, complete the manuscript.

\section{Magnetic response of a finite fermion system}
\label{sec:msffs}

The magnetic moment of a system of noninteracting fermions in a volume $\mathcal{V}$, at a temperature $T$, and with a chemical potential $\mu$ is given by \cite{ruite91_PRL,leuwe93_PhD}
\begin{align}
\label{eq:magnetization}
\mathcal{M}&=-\frac{\partial \Omega}{\partial H} 
\nonumber\\
&=-\sum_{\{\lambda\}}  f_{\mu}(E_{\lambda}) \, \frac{\partial E_\lambda}{\partial B} \, ,
\end{align}
with
\begin{equation}
\label{eq:Omega}
\Omega(\mu, T, B)=-
k_\mathrm{B}T\int_0^\infty\mathrm{d}E\,\varrho(E, B)
\ln{\left(1+\mathrm{e}^{(\mu-E)/k_{\rm B}T}\right)}
\end{equation}
the thermodynamic potential associated with the grand canonical ensemble (GCE),
$f_{\mu}(E)=\{\exp{([E-\mu]/k_{\rm B}T)}+1\}^{-1}$ the Fermi-Dirac distribution, and $k_{\rm B}$ the Boltzmann constant. The eigenenergies $E_{\lambda}$ of the  system are characterized by the set of quantum numbers $\{\lambda\}$, while
$\varrho(E, B)=\sum_{\{\lambda\}}  \delta(E-E_{\lambda})$ is the field-dependent single-particle density of states (DOS).
Here, $\delta(\zeta)$ denotes the Dirac delta function. In the regime of a small magnetic response of the system that we are interested in, the magnitude of the magnetic field $H$ or the magnetic induction $B$ can be indistinctly used when taking the derivatives in Eq.~\eqref{eq:magnetization}.

The ZFS is obtained from Eq.~\eqref{eq:magnetization} as
\begin{align}
\label{def:ZFS}
	\mathcal{\chi} &= \left.\frac{1}{\mathcal{V}}\frac{\partial\mathcal{M}}{\partial H}\right|_{H=0} \nonumber \\
	&=
	-\left.\frac{1}{\mathcal{V}}\sum_{\{\lambda\}}\left[f'_{\mu}(E_{\lambda})\left(\frac{\partial E_\lambda}{\partial B}\right)^2+ f_{\mu}(E_{\lambda})\frac{\partial^2 E_\lambda}{\partial B^2}\right]\right|_{B=0}.
\end{align}
Note that the unconstrained ZFS of the bulk, discussed in the introduction, follows from Eq.~\eqref{def:ZFS} while taking the limit $\mathcal{V} \rightarrow \infty$. This last procedure is not exempted of some subtleties \cite{leeuw21_JP,richt96_PhysRep, ullmo95_PRL}.

If the fermion system is not able to exchange particles with a reservoir, and then has a fixed number of particles $N$ (as is the case of a metallic nanoparticle), the free energy $F$, associated with the canonical ensemble (CE),  should be used instead of the thermodynamic potential $\Omega$ [cf.\ Eq.~\eqref{eq:Omega}] in the definitions \eqref{eq:magnetization} and \eqref{def:ZFS} of the $N$-fixed magnetic moment and ZFS, $\mathcal{M}_N$ and $\chi_N$, respectively. For large $N$ the difference between $\chi$, evaluated in the GCE at the Fermi energy $E_\mathrm{F}$, and $\chi_N$ is generally very small. However, such a correction becomes essential whenever the GCE magnetic response vanishes, as is the case of the average ZFS over an ensemble of nanoparticles with a large size dispersion \cite{viloria2018orbital}.

The above corrections can be readily incorporated in a semiclassical formalism in which the DOS is decomposed into a smooth (Weyl) and an oscillating (in energy or over the nanoparticle ensemble) part as 
\begin{equation}
\label{trace3D}
\varrho(E, B)={\bar \varrho}(E,B)+\varrho^\mathrm{osc}(E,B) \, .
\end{equation}
The trace formula expresses $\varrho^\mathrm{osc}(E,B)$ as a sum running over the periodic trajectories $\xi$ of the system,
\begin{equation}
\label{trace3Dp}
\varrho^\mathrm{osc}(E, B)= \sum_{\xi} \varrho_{\xi}(E,B) \, ,
\end{equation}
where the specific form of the contributions $\varrho_{\xi}(E,B)$ depends on whether $\xi$ represents  isolated periodic orbits or degenerate families of periodic orbits \cite{gutzwiller1990}. 
For temperatures such that $k_\mathrm{B}T$ is larger than the typical level 
spacing, the free energy can be approximated as \cite{imry1991, richt96_PhysRep}
\begin{equation}
\label{eq:F}
F(N,T,B)\simeq {\bar F} +\Delta F^{(1)}+\Delta F^{(2)},
\end{equation}
where 
\begin{subequations}
\begin{align}
\label{eq:F0}
{\bar F} &= {\bar \Omega}({\bar \mu},T,B)+{\bar \mu} N  \, ,
\\
\label{eq:F1}
\Delta F^{(1)} &= \Omega^\mathrm{osc}({\bar \mu},T,B) \, ,
\\
\label{eq:F2}
\Delta
F^{(2)} &= \frac{1}{2{\bar \varrho}({\bar \mu},0)}
\left[\int_0^\infty\mathrm{d}E\,\varrho^\mathrm{osc}(E,B) \, f_{\bar \mu}(E)\right]^2 \, .
\end{align}
\end{subequations}
In the equations above, ${\bar \Omega}$ and $\Omega^\mathrm{osc}$ are obtained, respectively, when ${\bar \varrho}$ and $\varrho^\mathrm{osc}$ are used in Eq.~\eqref{eq:Omega} instead of $\varrho$. The mean chemical potential ${\bar \mu}$  is determined in such a way as to ensure accommodating the $N$ electrons while using the mean density of states ${\bar \varrho}(E,0)$, that is, 
\begin{equation}
\label{eq:Ncoserv}
N = \int_0^\infty\mathrm{d}E \, {\bar \varrho}(E,0) \, f_{\bar \mu}(E) \, .
\end{equation}

When only the orbital motion is considered, the term ${\bar F}$ of Eq.~\eqref{eq:F0} leads to the Landau susceptibility presented in Eq.~\eqref{eq:Landau_sus}. The calculation of $\Omega^\mathrm{osc}$ in Eq.~\eqref{eq:F1} involves the energy integration of rapidly oscillating functions of $E$ (appearing through the classical action of the periodic orbits) that results in \cite{richt96_PhysRep}
\begin{equation}
\label{eq:chi1}
\chi^\mathrm{osc} = -\left.\frac{1}{\mathcal{V}}\sum_{\xi} R\left(\tau_{\xi}/\tau_{T}\right) 
\left(\frac{\hbar}{\tau_{\xi}} \right)^2 \, \frac{\partial^2 \varrho^\mathrm{osc}(E_\mathrm{F},B)}{\partial B^2}\right|_{B=0} \, ,
\end{equation}
where $\hbar$ is Planck's (reduced constant), $\tau_{\xi}$ is the period of the classical orbit, and $\tau_{T}=\hbar/\pi k_{\rm B}T$. The thermal factor
\begin{equation}
\label{eq:R_T}
R(\zeta)=\frac{\zeta}{\sinh{(\zeta)}} 
\end{equation}
exponentially cuts off the long trajectories of the semiclassical expansion.
The two previously discussed contributions to $\chi_N$, arising from ${\bar F}$ and $\Delta F^{(1)}$, give together a good account of the ZFS for an individual nanoparticle \cite{viloria2018orbital}, when compared to a quantum-mechanical perturbative (in the magnetic field $B$) calculation \cite{ruite91_PRL,leuwe93_PhD}. The magnetic field dependence of ${\bar F}$ is much weaker than that of $\Delta F^{(1)}$ (which is given by the magnetic flux enclosed by the periodic orbits). Thus, $\chi^\mathrm{osc}$ (related with the confinement) dominates over the bulk contribution \eqref{eq:Landau_sus}. The finite-$N$ correction arising from $\Delta F^{(2)}$ in  Eq.~\eqref{eq:F2} becomes crucial when considering an ensemble of nanoparticles, and it is important in order to assess the relevance of the relativistic corrections that we study in this work.

The above formulation has been applied to the problems of
the persistent currents in metallic rings \cite{bouchiat1989, imry1991, schmi91, felix91, altsh88_PRL} and the 
orbital magnetism in low-dimensional ballistic 
systems \cite{richt96_PhysRep, ullmo95_PRL} for weakly-interacting Landau quasiparticles, as well as including the particularly relevant Cooper-like electronic 
correlations \cite{ambeg90_PRL, ullmo98_PRL}.
In the absence of spin-orbit interaction, the DOS oscillations characterized by $\varrho^\mathrm{osc}$ [i.e., Eq.~\eqref{trace3Dp}]
exhibit a diminishing amplitude as the magnetic field is turned on, in line with the enhanced spectral rigidity
associated with the transition from orthogonal to unitary Gaussian ensembles describing chaotic or disordered systems. 
The dependence of measurable physical effects on the transition between statistical ensembles, discussed in the introduction, appears in the case of the average response of a large number of nanostructures, where the contribution arising from $\Delta F^{(1)}$ vanishes, and thus 
$\Delta F^{(2)}$ provides the leading response. The average ZFS resulting from 
Eqs.~\eqref{eq:magnetization}, \eqref{def:ZFS}, and \eqref{eq:F2} is then expected to be paramagnetic. On the one hand, the previous reasoning is confirmed by the paramagnetic ZFS measured in ensembles of $10^5$ phase-coherent, ballistic squares lithographically defined on a high 
mobility GaAs heterojunction \cite{levy93_physicaB}. On the other hand, the orbital response of 
$10^5$ microscopic silver rings was found to be diamagnetic \cite{deblo02_PRL}. 
The resulting discrepancy with the previous theory has not been settled, and points to the possible 
role of attractive electron-electron interaction above the superconducting transition temperature, 
as proposed in Refs.~\cite{bary08_PRL, imry2015}.\footnote{For a complete account of theory 
and experiments on the persistent current of normal metal rings, see Ref.~\cite{Shanks}.}

For a system with a strong SOC, the magnetic field drives the transition from symplectic 
to unitary Gaussian ensembles. From the larger spectral rigidity of the former, one might conclude
that there is an SOC-induced inversion of the average magnetic response. However, as pointed out in Ref.~\cite{titov2019}, such a reasoning does not always hold, since in the presence of a strong 
SOC the breaking of the Kramers degeneracy occurs at magnetic fields much smaller than those
lifting the Zeeman splitting in the absence of SOC. While Ref.~\cite{titov2019} predicted, 
within a noninteracting model, a switch from paramagnetic to diamagnetic ZFS in the case of 
metallic nanoparticles, the average persistent current in disordered wires was shown to retain 
its paramagnetic character even if a strong SOC is present for noninteracting electrons \cite{altsh88_PRL, mathu91_PRB}, 
as well as in the case of a repulsive electron-electron interaction \cite{schmi91, ambeg90_PRL}.
The nontrivial effect of SOC and the angular magneto-electric coupling for 
the persistent current in conducting rings has also been analyzed in terms of a Berry phase
\cite{Aronov1993}. The persistent currents in ballistic mesoscopic rings wish Rashba SOC 
have been studied \cite{Splettstoesser2003}. The coupling strength of the previous mechanism, 
often taken as a phenomenological parameter, has been linked with measurable features
of the flux dependence of the persistent currents.


\section{Weakly-relativistic description}
\label{sec:wrd}

The weakly-relativistic Hamiltonian for an electron subject to the action of static electric and magnetic induction fields, respectively 
$\mathbf{E}(\mathbf{r})=-\nabla \phi(\mathbf{r})$ and $\mathbf{B}(\mathbf{r})=\nabla\times\mathbf{A}(\mathbf{r})$  (with $\phi$ and $\mathbf{A}$ the scalar and vector potentials, respectively), is given as a $2\times 2$ matrix by \cite{foldy_1950,schwabl1997aqm}
\begin{equation}
\label{eq:Hamiltonian}
\mathcal{H}^{(\rm wr)}=\mathcal{H}^{(\rm nr)}+ \Delta \mathcal{H} \, .
\end{equation}
The nonrelativistic Hamiltonian $\mathcal{H}^{(\rm nr)}$ is
\begin{equation}
\label{eq:Hamiltonian_nr}
\mathcal{H}^{(\rm nr)}=\mathcal{H}^{(\rm orb)}+\mathcal{H}^{(\rm Z)} \, ,
\end{equation}
with
\begin{subequations}
\begin{align} 
\mathcal{H}^{(\rm orb)}&=
\frac{1}{2m}\left[\mathbf{p}+\frac{e}{c}\mathbf{A}(\mathbf{r})\right]^2 - e \phi(\mathbf{r})
 \, ,
\label{eq:H_0}
\\ 
\mathcal{H}^{(\rm Z)}&=\frac{e\hbar}{2mc } \, \boldsymbol{\sigma} \cdot \mathbf{B}(\mathbf{r}) \, ,
\label{eq:H_Zeeman}
\end{align}
\end{subequations}
representing, respectively, the spin-independent  Hamiltonian associated with the orbital motion (orb), and the Zeeman energy of the spin in the magnetic field (Z). We note $\mathbf{p}$ the electron momentum and $\boldsymbol{\sigma}$ the vector of Pauli matrices, related with the electron spin angular momentum operator by $\mathbf{S} = (\hbar/2) \boldsymbol{\sigma}$. The weakly-relativistic correction  $\Delta \mathcal{H}$ 
in Eq.~\eqref{eq:Hamiltonian} writes
\begin{equation}
\label{eq:Hamiltonian_Delta}
\Delta \mathcal{H}=\mathcal{H}^{(\rm k)}+\mathcal{H}^{(\rm so-ame)} +\mathcal{H}^{(\rm D)}+\mathcal{H}^{(\rm r)} \, ,
\end{equation}
with
\begin{subequations}
\begin{align}
\mathcal{H}^{(\rm k)}&= -\frac{1}{8m^3c^2} \, \left\{  \boldsymbol{\sigma}\cdot\left[\mathbf{p}+\frac{e}{c}\mathbf{A}(\mathbf{r})\right] \right\}^4 \, ,
\label{eq:H_k}
\\  
\mathcal{H}^{(\rm so-ame)}&=\frac{e\hbar}{4m^2c^2} \, \boldsymbol{\sigma} \cdot \left\{\mathbf{E}(\mathbf{r}) \times \left[\mathbf{p}+\frac{e}{c}\mathbf{A}(\mathbf{r})\right]\right\} \, ,
\label{eq:H_soame}
\\ 
\mathcal{H}^{(\rm D)} &= \frac{e\hbar^2}{8m^2c^2} \,
\nabla \cdot \mathbf{E}(\mathbf{r}) \, , 
\label{eq:H_Darwin}
\\
\mathcal{H}^{(\rm r)} &= \frac{\mathrm{i} e\hbar^2}{8m^2c^2} \, \boldsymbol{\sigma} \cdot \left[\nabla \times \mathbf{E}(\mathbf{r}) \right] \, .
\label{eq:H_r}
\end{align}
\end{subequations}
$\mathcal{H}^{(\rm k)}$ contains the first relativistic correction to the kinetic energy, $\mathcal{H}^{(\rm so-ame)}$ combines the spin-orbit (so) and the angular magneto-electric (ame) \cite{mondal_PRB2015} couplings,  $\mathcal{H}^{(\rm D)}$ stands for the Darwin term responsible for the zitterbewegung effect in atomic physics, while the last term $\mathcal{H}^{(\rm r)}$ with the curl of $\mathbf{E}$ vanishes for conservative potentials.

Our idealized nanoparticle is defined by a confining potential $V(\mathbf{r})=-e \phi(\mathbf{r})$ acting on noninteracting electrons. We assume an applied magnetic field along the $z$-direction, and thus a magnetic induction $\mathbf{B} =  B\, {\bf {\hat e}}_z$  that in the symmetric gauge can be represented by $\mathbf{A}(\mathbf{r})=\frac 12 \mathbf{B} \times \mathbf{r}$. Under the previous hypothesis the different components of
the Hamiltonian \eqref{eq:Hamiltonian} admit simpler expressions that we describe below. The nonrelativistic spinless term \eqref{eq:H_0} describing the electron orbital motion can be  written as
\begin{equation}
\mathcal{H}^{(\rm orb)}=
\mathcal{H}^{(0)}+\mathcal{H}^{(\rm para)}+\mathcal{H}^{(\rm dia)}
 \, ,
\end{equation}
with
\begin{subequations}
\begin{align} 
\mathcal{H}^{(0)}&=\frac{\mathbf{p}^2}{2m} + V(\mathbf{r}) \, ,
\label{eq:H_00}
\\ 
\mathcal{H}^{(\rm para)}&= \frac{\omega_\mathrm{c}}{2} \, L_z  ,
\label{eq:H_para}
\\ 
\mathcal{H}^{(\rm dia)}&= \frac{m \omega_\mathrm{c}^2}{8} \left(x^2+y^2 \right) \, .
\label{eq:H_dia}
\end{align}
\end{subequations}
Here, $\mathcal{H}^{(0)}$ is the zero-field Hamiltonian describing a spinless particle in the confining box, while the terms $\mathcal{H}^{(\rm para)}$ and $\mathcal{H}^{(\rm dia)}$ represent paramagnetic and diamagnetic contributions, respectively, depending of the $B$-field  through the cyclotron frequency $\omega_\mathrm{c}=eB/m^{*}c$ (as we discussed in Sec.~\ref{sec:intro} we will identify $m^{*}$ and $m$). We denote $\mathbf{L}$ the orbital angular momentum of the electron and $L_z$ its $z$ component.

Denoting $\mu_\mathrm{B}=e\hbar/2mc$ the Bohr magneton, it is convenient to group the paramagnetic \eqref{eq:H_para} and Zeeman \eqref{eq:H_Zeeman} components in a term representing the coupling of the total magnetic moment
\begin{equation}
\label{eq:magmoml}
\boldsymbol{\mu} = -\frac{\mu_\mathrm{B}}{\hbar} \left(\mathbf{L}+g_0 \mathbf{S}\right)
\end{equation}
to the magnetic induction, leading to
\begin{align}
\label{eq:Hmagmoml}
\mathcal{H}^{(\mu)} &= \mathcal{H}^{(\rm para)} + \mathcal{H}^{(\rm Z)} 
\nonumber\\
&= - \boldsymbol{\mu} \cdot \mathbf{B} \, .
\end{align}
In the ideal system that we are describing, we use the g-factor $g_0=2$. 

The correction \eqref{eq:H_soame} takes the form
\begin{equation}
\label{eq:H_soame2}
\mathcal{H}^{(\rm so-ame)}=\mathcal{H}^{(\rm so)}+ \mathcal{H}^{(\rm ame)} \, ,
\end{equation}
with
\begin{subequations}
\label{eq:H_soame3}
\begin{align} 
\mathcal{H}^{(\rm so)}& = \frac{1}{2m^2c^2} \, \mathbf{S} \cdot \left[\nabla V(\mathbf{r}) \times \mathbf{p}\right] \, ,
\label{eq:H_so3}
\\ 
\mathcal{H}^{(\rm ame)}&= \frac{e}{4m^2c^3} \, \mathbf{S} \cdot \left[\nabla V(\mathbf{r}) \times \left( \mathbf{B} \times \mathbf{r} \right) \right] \, ,
\label{eq:H_ame}
\end{align}
\end{subequations}
while the Darwin term \eqref{eq:H_Darwin} can be written as
\begin{equation}
\mathcal{H}^{(\rm D)} = \frac{\hbar^2}{8m^2c^2} \, \nabla^{2} V(\mathbf{r}) \, .
\label{eq:H_Darwin3}
\end{equation}

Our main interest is the effect of the SOC correction \eqref{eq:H_so3} on the magnetic response of different kinds of nanoparticles. In order to assess its relevance, we need to also examine the role of the other weakly-relativistic corrections.

\section{Spherical nanoparticles}
\label{sec:sn}

\subsection{Nanoparticle modeling}
\label{sec:model}

We here consider a spherically-symmetric confinement defined by the potential 
\begin{equation}
\label{eq:confining_potential}
V(r) = V_0 \, \Theta(r-a) \, ,
\end{equation}
where we adopt the spherical coordinates $(r,\theta,\varphi)$ with $r=|\mathbf{r}|$. We denote $\Theta(\zeta)$ the Heaviside step function, $V_0$ the height of the confining wall, and $a$ the nanoparticle radius. The spherical symmetry of the problem allows one to further simplify the expressions of Eq.~\eqref{eq:H_soame3} for the spin-orbit and angular magneto-electric couplings as
\begin{subequations}
\label{eq:H_soame4}
\begin{align} 
\mathcal{H}^{(\rm so)}& = \frac{1}{2m^2c^2} \, \frac{1}{r} \left(\frac{{\rm d}V}{{\rm d}r} \right) \mathbf{S} \cdot \mathbf{L} \, ,
\label{eq:H_so2}
\\ 
\mathcal{H}^{(\rm ame)}&= - \frac{\mu_\mathrm{B} B}{2mc^2 \hbar} \, r \left(\frac{{\rm d}V}{{\rm d}r} \right) \sin{(\theta)} \
\mathbf{S} \cdot {\bf {\hat e}}_{\theta} \, ,
\label{eq:H_ame2}
\end{align}
\end{subequations}
together with that of Eq.~\eqref{eq:H_Darwin3} for the Darwin term, which can be written as
\begin{equation}
\mathcal{H}^{(\rm D)} = \frac{\hbar^2}{8m^2c^2} \, \frac{1}{r^2} \,  \frac{{\rm d}}{{\rm d}r}  \left(r^2  \, \frac{{\rm d}V}{{\rm d}r}  \right) \, .
\label{eq:H_Darwin2}
\end{equation}

The previously exposed model of free electrons contained in a hard-wall box constitutes a first approximation for describing noble-metal nanoparticles. Within this scheme, the jellium approximation is adopted for the description of the conduction electrons and $V(r)$ in Eq.~\eqref{eq:confining_potential} should be understood as the self-consistent electrostatic potential resulting from the confinement due to the positive background from the ionic lattice, together with the mean-field treatment of the electron-electron interaction. Density functional theory calculations~\cite{weick05_PRB, weick06_PRB} indicate that, for not too small nanoparticles and in the absence of a magnetic field, the spherical jellium model \eqref{eq:confining_potential} is a good approximation for the confining potential, 
where $V_0=E_\mathrm{F}+W$, with $W$ the work function of the nanoparticle.

The advantages and disadvantages of the jellium model have been thoroughly discussed in the literature concerning the electronic properties of metallic nanoparticles and the physics of surface plasmon resonances \cite{brack93_RMP,molina2002}. For very small metallic nanoparticles (or clusters) the corrections to the ionic structure have been approached by the use of pseudopotential perturbation theory as a multipole expansion \cite{ekardt1994}
and the effect of surface irregularities and the underlying crystalline lattice has been addressed with numerical 
methods \cite{bucher1990, pavloff1992}. The influence of smooth disorder in low-dimensional systems has been addressed with the help 
of semiclassical methods \cite{richter1996, richter96_JMP}.
We will leave aside the case of very small nanoparticles while neglecting the effect of surface and/or bulk disorder, and we will then not be concerned with such corrections. 

The finite height $V_0$ of the confining wall is responsible for the spill-out effect describing the nonzero probability to find an electron density outside the nanoparticle \cite{weick06_PRB}. The proper description of the spill-out, as well as the lifetime of the surface plasmon resonance necessitates to go beyond the discontinuous form  \eqref{eq:confining_potential} of the confining potential, including the abruptness of the potential jump at the nanoparticle surface. In the case of the ZFS, the abruptness of the potential jump is not a crucial parameter, and moreover, the careful analysis of the wave-function behavior close to a potential discontinuity, that we perform in the sequel, shows that the precise value of $V_0$ is not determinant for the nonrelativistic ZFS, nor for the weakly-relativistic corrections (with the exception of the Darwin contribution discussed in Appendix \ref{app:mebsp}), justifying the limit of $V_0 \rightarrow \infty$  commonly adopted in the nonrelativistic case \cite{ruite91_PRL,ruite93_MPLB, leuwe93_PhD,viloria2018orbital}.  

When going from the description of the ideal model system to that of a gold nanoparticle, we should in principle use $g_{\rm Au}$ instead of $g_{0}$ in the expression \eqref{eq:magmoml} for the total magnetic moment. However, the small difference between $g_{\rm Au}$ and $g_{0}$, like the one between $m^{*}$ and $m$, only induce very small corrections. 
Another effect to take into account when treating metallic nanoparticles is the Larmor diamagnetic response of the core electrons, which are not considered in the idealized model. While the effect of the core electrons on the ZFS is not altered by the confinement, its contribution must be confronted with the corrections under study in order to gauge the relevance of the latter.

\subsection{Nonrelativistic susceptibility}
\label{sec:nrssg}

The magnetization and the ZFS in the nonrelativistic case have been calculated in the case of a spherical geometry by quantum perturbation theory \cite{ruite91_PRL,leuwe93_PhD}, as well as through the use of semiclassical expansions \cite{viloria2018orbital} for the DOS. We show in Appendix \ref{app:sc} the connection between the two approaches. The quantum procedure starts from the eigenstates of the unperturbed ($B=0$) Hamiltonian $\mathcal{H}^{(0)}$ of Eq.~\eqref{eq:H_00}, describing nonrelativistic electrons in a spherical potential (of radius $a$), which are characterized in the product basis by the set $\{\lambda\}=\{n,l,m_z,m_s\}$ of quantum numbers, with $n>0$ the principal quantum number, $l \geqslant 0$ the azimuthal quantum number, $m_z \in [-l,l]$ the magnetic quantum number, and $m_s=\pm 1/2$ associated with the spin component along the $z$ direction. The corresponding wave functions are given by the two-component spinors
\begin{subequations}
\label{eq:nlmzmsud}
\begin{align} 
\Psi_{n,l,m_z,+\frac{1}{2}}^{(0)}(\mathbf{r})&=\psi_{n,l,m_z}^{(0)}(\mathbf{r})\begin{bmatrix} 1\\ 0\end{bmatrix} \, ,
\label{eq:nlmzmsu}
\\ 
\Psi_{n,l,m_z,-\frac{1}{2}}^{(0)}(\mathbf{r})&=\psi_{n,l,m_z}^{(0)}(\mathbf{r})\begin{bmatrix} 0\\ 1\end{bmatrix} \, ,
\label{eq:nlmzmsd}
\end{align}
\end{subequations}
with the orbital wave function
\begin{equation}
\label{eq:orbital}
\psi_{n,l,m_z}^{(0)}(\mathbf{r})=R_{n,l}(r) \, Y^{m_z}_l(\vartheta)\; .
\end{equation}
We note $Y_{l}^{m_z}(\vartheta)$ the spherical harmonic of degree $l$ and order $m_z$ as a function of the solid angle $\vartheta=(\theta,\varphi)$, while $R_{n,l}(r)$ stands for the associated radial wave function and $E_{n,l}=E_0 (k_{n,l}\, a)^2$ is the  corresponding eigenenergy, with $E_0=\hbar^2/2m a^2$. 

For the confining potential \eqref{eq:confining_potential}, $R_{n,l}$ can be expressed in terms of Bessel functions (see Appendix \ref{app:mebsp}). In the limiting case of a hard-wall potential ($V_0 \rightarrow \infty$), according to 
Eq.~\eqref{eq:knl}, the eigenenergy of the $m_z$ and spin-degenerate states characterized by the quantum numbers $l$ and $n$ is 
\begin{equation}
\label{eq:E^0}
E^{(0)}_{n,l}=E_0\,  \zeta^2_{n,l}\, ,
\end{equation}
where $\zeta_{n,l}$ is the $n$th root of the spherical Bessel function of the first kind $j_l(\zeta)$, while
\begin{equation}
\label{eq:rwfs}
R_{n,l}(r)=\sqrt{\frac{2}{a^3}} \, \frac{j_l(\zeta_{n,l} \, r/a)}{|j_{l+1}(\zeta_{n,l})|} \, .
\end{equation}

The action of a magnetic field through the term $\mathcal{H}^{(\mu)}$ of Eq.~\eqref{eq:Hmagmoml} is, one the one hand, to break the above-mentioned degeneracy without giving rise to a new basis of eigenstates. On the other hand, the states $\Psi_{n,l,m_z,m_s}^{(0)}$ [cf.\ Eq.~\eqref{eq:nlmzmsud}] are no longer eigenstates once the diamagnetic term $\mathcal{H}^{(\rm dia)}$ of Eq.~\eqref{eq:H_dia} is considered under a finite magnetic field. The matrix elements of $\mathcal{H}^{(\rm dia)}$ in the product basis are given in Appendix \ref{app:mes}. Since they are quadratic in $B$, we can treat $\mathcal{H}^{(\rm dia)}$ by first-order perturbation theory. Thus, up to terms of order $B^2$, the perturbed energies are
\begin{equation}
E_{n,l,m_z,m_s}^{(\rm nr)}=E_{n,l}^{(0)} + \delta E_{n,l,m_z,m_s}^{(\rm nr)} \, ,
\label{eq:Enr}
\end{equation}
where $\delta E_{n,l,m_z,m_s}^{(\rm nr)}$ is the magnetic-field correction, which reads
\begin{equation}
\delta E_{n,l,m_z,m_s}^{(\rm nr)} = E^{(\mu)}_{m_z,m_s}+E^{(\rm dia)}_{n,l,m_z} \, .
\label{eq:dEnr}
\end{equation}
Consistently with Eq.~\eqref{eq:Hmagmoml}, we denote
\begin{equation}
E^{(\mu)}_{m_z,m_s}=E^{(\rm para)}_{m_z}+E^{(\rm Z)}_{m_s} \, ,
\label{eq:Enr_mu}
\end{equation}
with
\begin{subequations}
\label{eq:Enrcomponents}
\begin{align}
E^{(\rm para)}_{m_z} &= \frac{\hbar\omega_\mathrm{c}}{2} \, m_z \, ,
\label{eq:Epara}
\\
E^{(\rm Z)}_{m_s} &= g_0 \, \mu_\mathrm{B} \, m_s \, B \, .
\label{eq:EZeeman}
\end{align}
\end{subequations}
For the case of a hard-wall confinement \cite{ruite91_PRL,leuwe93_PhD},
\begin{equation}
E^{(\rm dia)}_{n,l,m_z} = \frac{m\omega_\mathrm{c}^2a^2}{8} \, \mathcal{R}_{n,l}\, \mathcal{Y}_l^{m_z}
\, ,
\label{eq:Edia}
\end{equation}
with
\begin{subequations}
\begin{align}
\label{eq:R}
  \mathcal{R}_{n,l} &= \frac{1}{3}\left[1+\frac{2}{\zeta_{nl}^2} \left(l-\frac{1}{2}\right)\left(l+\frac{3}{2}\right) \right] \, ,
\\
\label{eq:Y}
  \mathcal{Y}_l^{m_z} &=  \frac 12\left[ \frac{m_z^2 + l(l+1) - 1 }{(l-1/2)(l+3/2)}\right]  \, .
\end{align}
\end{subequations}

According to Eq.~\eqref{def:ZFS}, the nonrelativistic ZFS is determined from the following parameters directly obtained from Eqs.~\eqref{eq:Enrcomponents} and \eqref{eq:Edia}, 
\begin{subequations}
\label{eq:energiesandderivativesnr}
\begin{align}
\left. E_{n,l,m_z,m_s}^{(\rm nr)} \right|_{B=0} & =  E_{n,l}^{(0)}   \, , \\
\left. \frac{\partial E_{n,l,m_z,m_s}^{(\rm nr)}}{\partial B}\right|_{B=0} & =  \mu_{\rm B} \left(m_z+2 m_s \right) \, , \\
 \left.\frac{\partial^2 E_{n,l,m_z,m_s}^{(\rm nr)}}{\partial B^2}\right|_{B=0} &   =  \frac{\mu_{\rm B}^2}{2E_0}  \, \mathcal{R}_{n,l} \,  \mathcal{Y}_l^{m_z}  \, .
\end{align}
\end{subequations}

Treating separately each of the field-dependent energy contributions \eqref{eq:Enrcomponents}, the nonrelativistic ZFS following from Eqs.~\eqref{def:ZFS} and \eqref{eq:energiesandderivativesnr} can be written as
\begin{equation}
\label{eq:chi_nr}
\chi^{(\rm nr)}=\chi^{(\rm Z)} + \chi^{(\rm orb)} \, ,
\end{equation}
in terms of a spin-dependent susceptibility $\chi^{(\rm Z)}$ and the orbital component $ \chi^{(\rm orb)}$. The latter admits the decomposition
\begin{equation}
\label{eq:chi_orb}
\chi^{(\rm orb)}=\chi^{(\rm para)}+\chi^{(\rm dia)} \, .
\end{equation}

Performing the $m_z$ and $m_s$ sums in Eq.~\eqref{def:ZFS}, we have
\begin{equation}
\label{eq:chi_Zeeman}
\chi^{(\rm Z)} = - \frac{2 \mu_\mathrm{B}^2}{\mathcal{V}} \sum_{l=0}^\infty (2l+1)  \sum_{n=1}^\infty \, f'_{\bar \mu_{0}}(E_{nl}^{(0)}) \, ,
\end{equation}
as well as \cite{ruite91_PRL,leuwe93_PhD}
\begin{subequations}
\label{eq:chi_paraanddia}
\begin{align}
\label{eq:chi_para}
\frac{\chi^{(\rm para)}}{|\chi_\mathrm{L}|} &= - \frac{3\pi E_0}{k_\mathrm{F}a}\sum_{l=0}^\infty l(l+1) (2l+1) \sum_{n=1}^\infty \, f'_{\bar \mu_{0}}(E_{nl}^{(0)}) \, ,
\\
\label{eq:chi_dia}
\frac{\chi^{(\rm dia)}}{|\chi_\mathrm{L}|} &= - \frac{3\pi}{k_\mathrm{F}a} \sum_{l=0}^\infty(2l+1) \sum_{n=1}^\infty\, \mathcal{R}_{n,l} \, f_{\bar \mu_{0}}(E_{nl}^{(0)}) \, .
\end{align}
\end{subequations}
The mean chemical potential $\bar \mu_{0}$ is associated with the spectrum of the Hamiltonian $\mathcal{H}^{(0)}$ presented in Eq.~\eqref{eq:H_00}.
Trading the sum over the principal quantum number $n$ by an integral over energy allows to recast  Eq.~\eqref{eq:chi_Zeeman} as
\begin{equation}
\label{eq:chi_Zeemanap}
\frac{\chi^{(\rm Z)}}{|\chi_\mathrm{L}|} =  - \frac{9\pi E_0}{k_\mathrm{F}a}  \sum_{l=0}^\infty (2l+1) \int_{0}^{\infty} {\rm d}E \,  \varrho_l(E) \, f'_{\bar \mu_{0}}(E) \, ,
\end{equation}
where we have introduced the $l$-fixed density of states $\varrho_l(E)$ corresponding to the radial problem, itself related to the zero-field DOS by
\begin{equation}
\label{eq:reldosldos}
\varrho(E,0) = 2 \sum_{l=0}^{l_{\rm max}} \left(2l+1 \right)\varrho_l(E) \, ,
\end{equation}
with $l_{\rm max} $  given by  Eq.~\eqref{eq:lmax}. The prefactor of $2$ in the equation above takes into account the spin degeneracy, as we follow the standard convention of using a spinless 
one-dimensional $\varrho_l(E)$ and a spinful three-dimensional $\varrho(E,B)$. 

For a degenerate electron gas, where $E_\mathrm{F} \gg k_{\rm B}T$, we use $f'_{\mu}(E)=-\delta(E-\mu)$, and thus Eq.~\eqref{eq:chi_Zeemanap} leads to the standard result of the spin-dependent  susceptibility \cite{AM}
\begin{equation}
\label{eq:chi_Zeeman2}
\chi^{(\rm Z)} = \frac{\mu_\mathrm{B}^2}{\mathcal{V}} \, \varrho(E_\mathrm{F},0) \, .
\end{equation}
In the unconstrained case of $a \rightarrow \infty$, the use of the DOS per unit volume for the zero-field, three-dimensional, free electron gas  $g^{(\mathrm{3D})}(E)=(m/\pi^2 \hbar^2) \sqrt{2mE/\hbar^2}$ in Eq.~\eqref{eq:chi_Zeeman2}, results in the form \eqref{eq:Pauli_sus} of the Pauli susceptibility. In the constrained case of a finite $a$, the separation \eqref{trace3D} and the fact that ${\bar \varrho}(E,0)= g^{(\mathrm{3D})}(E) {\mathcal{V}}$ result in a mean spin-dependent susceptibility ${\bar \chi}^{(\rm Z)}=\chi_\mathrm{P}$. Thus, the confinement only adds a small contribution to the bulk susceptibility, which is associated with $\varrho^{\rm osc}(E,0)$ \cite{oppen94_PRB}.

\begin{figure*}[tb]
\begin{center}
\includegraphics[width=.86\linewidth]{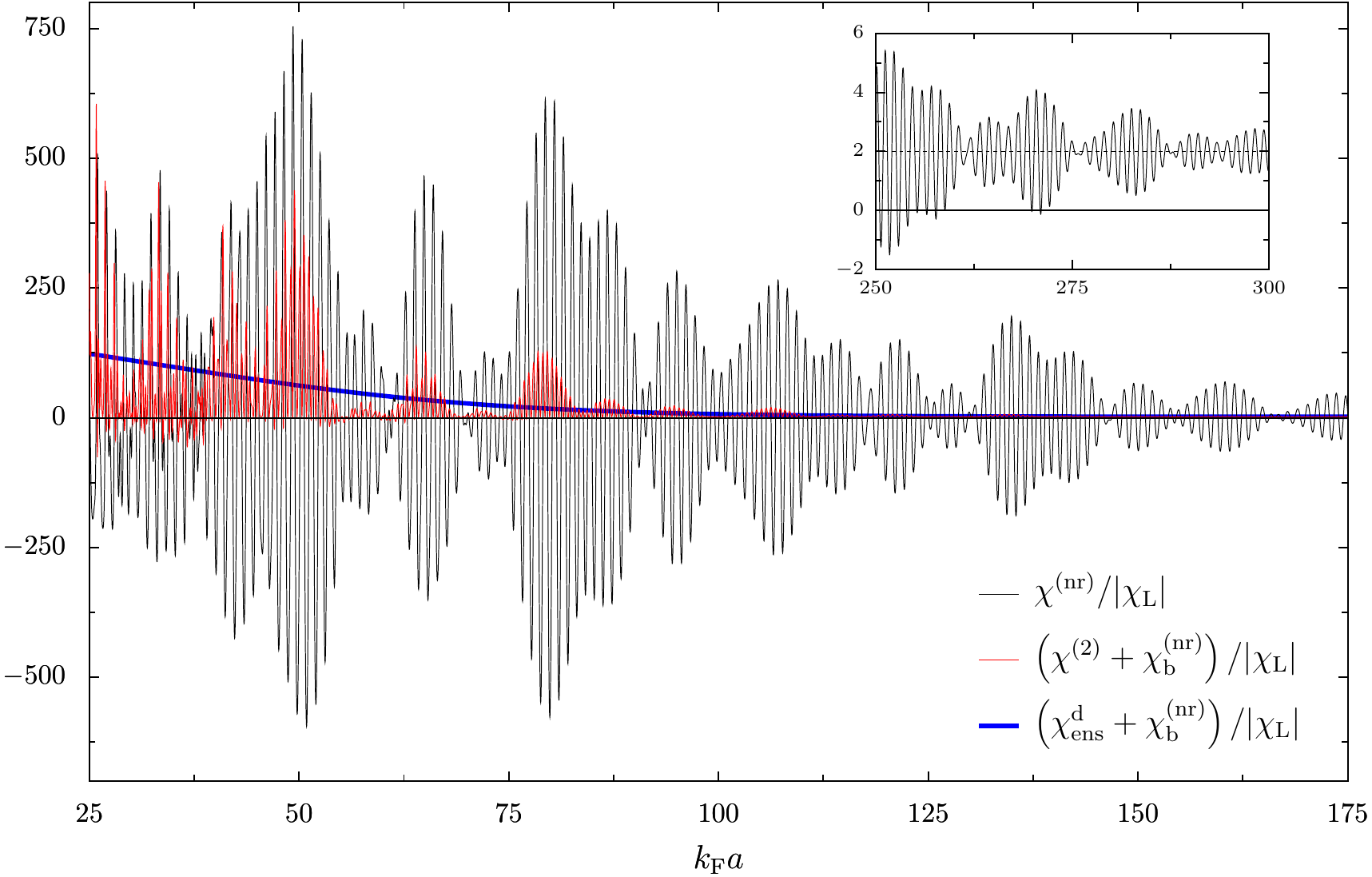}
\caption{\label{fig:chinr} 
In black: nonrelativistic ZFS $\chi^{(\rm nr)}$ for $k_\mathrm{B}T /{\bar \mu}_0 = 5 \times 10^{-3}$ (corresponding to room temperature in the case of gold) obtained from Eqs.~\eqref{eq:chi_Zeeman}--\eqref{eq:chi_paraanddia} (in units of the absolute value of the Landau susceptibility $\chi_\mathrm{L}$), as a function of the nanoparticle radius $a$ (scaled with the Fermi wave vector $k_\mathrm{F}$). 
In red: 
$\chi^{(2)}+\chi_{\rm b}^{\rm (nr)}$, with $\chi^{(2)}$ obtained from the finite-$N$ correction \eqref{eq:F2} to the free energy and the oscillating part of the $B$-dependent density of states $\varrho^\mathrm{osc}(E,B)$.
In blue: ZFS of an ensemble of nanoparticles with an important size dispersion, $\chi_\mathrm{ens}^{\rm d}+\chi_{\rm b}^{\rm (nr)}$, 
where $\chi_\mathrm{ens}^{\rm d}$ is obtained by taking the average of $\chi^{(2)}$ over a Gaussian probability distribution (in size). 
Inset: Corresponding ZFSs for larger nanoparticle radii, showing the approach to the bulk value given by Eqs.~\eqref{eq:Landau_sus} and \eqref{eq:Pauli_sus} indicated by a dashed line.}
\end{center}
\end{figure*}

In the constrained case, the numerical implementation of 
Eqs.~\eqref{eq:chi_Zeeman} and \eqref{eq:chi_paraanddia} leads to the nonrelativistic ZFS $\chi^{(\rm nr)}$ presented in Fig.~\ref{fig:chinr} (in black) as a function of $k_\mathrm{F}a$. In the shown interval, these numerical results are almost indistinguishable from those in which $\chi^{(\rm nr)}_{\rm b}=2|\chi_\mathrm{L}|$ is added to the semiclassical ZFS $\chi^{(\rm orb)-osc}$ of Eq.~\eqref{eq:chi_1}. 
The oscillations of $\chi^{(\mathrm{nr})}$ as a function of $k_\mathrm{F}a$ are typically
much larger than $|\chi_\mathrm{L}|$ and can be understood as a shell structure \cite{fraue98_PRB}.
The suppression of these oscillations for large sizes can be understood, at the semiclassical level, by the thermal damping \eqref{eq:R_T} acting on the contribution
of each family of classical periodic orbits. 
For very large $a$ (inset), the oscillations of $\chi^{(\rm nr)}$ 
are quite reduced, and we can see that they are 
around the bulk ZFS $\chi^{(\rm nr)}_{\rm b}$, given by Eqs.~\eqref{eq:Landau_sus} and \eqref{eq:Pauli_sus}, as the confinement becomes irrelevant in such a limit. The typical values of $\chi^{(\rm nr)}$ are important in order to assess the relevance of the relativistic corrections to be calculated in the sequel. Also for comparison purposes, we show in Fig.~\ref{fig:chinr} (in red) the ZFS $\chi^{(2)}+\chi^{(\rm nr)}_{\rm b}$ arising from finite-$N$ corrections to the free energy $\eqref{eq:F2}$ using the semiclassical expression for the oscillating part of the density of states $\varrho^\mathrm{osc}(E,B)$ given by Eqs.~\eqref{eq:rho_osc}, \eqref{eq:rho_osc2}, and \eqref{eq:mf}, as detailed in Ref.~\cite{viloria2018orbital}.\footnote{Up to the constant term $\chi^{(\rm nr)}_{\rm b}$, Fig.~\ref{fig:chinr} presents equivalent results to those of Fig.~9(b) of Ref.~\cite{viloria2018orbital}, correcting a factor of $\pi/4$ that erroneously overrated the contribution $\chi^{(2)}$.} In addition, we show in Fig.~\ref{fig:chinr} (blue line) the ZFS
$\chi_\mathrm{ens}^\mathrm{d}+\chi^{(\rm nr)}_{\rm b}$
of an ensemble of metallic nanoparticles 
with an important size dispersion. Here, $\chi_\mathrm{ens}^\mathrm{d}$ is the ensemble average of $\chi^{(2)}$ over a Gaussian probability distribution of the size parameter $a$ (cf.\ Eq.~(30) in Ref.~\cite{viloria2018orbital}).

\subsection{Kinetic correction}
\label{sec:kedc}

The zero-field Dirac equation in a spherical potential box admits an exact solution \cite{greiner1990relativistic,Alberto_1998}. While the inclusion of an infinitesimal magnetic field allowing to address the ZFS can in principle be implemented as a perturbation \cite{mikhailov_1968}, it is simpler to proceed from the weakly-relativistic Hamiltonian \eqref{eq:Hamiltonian}. The kinetic term \eqref{eq:H_k} can be written as $\mathcal{H}^{(\rm k)}= - \left\{ \mathcal{H}^{(\rm nr)} + e\phi \right\}^2/2mc^2$. Therefore, in the case of a hard-wall confinement, and up to quadratic terms in $B$, it leads to the energy correction 
\begin{widetext}
\begin{align}
\label{eq:keshift}
E_{n,l,m_z,m_s}^{(\rm k)} &= - \frac{1}{2mc^2} \, \langle \Psi_{n,l,m_z,m_s}^{(0)} | \left( \frac{\mathbf{p}^{2}}{2m} \right)^{2} + \frac{\mathbf{p}^{2}}{m} \, \mathcal{H}^{(\mu)} + \left(\mathcal{H}^{(\mu)}\right)^{2} + \frac{\mathbf{p}^{2}}{2m} \, \mathcal{H}^{(\rm dia)} + \mathcal{H}^{(\rm dia)} \, \frac{\mathbf{p}^{2}}{2m} |  \Psi_{n,l,m_z,m_s}^{(0)}  \rangle 
\nonumber\\
&= - \frac{\left( E_{n,l}^{(0)} \right)^2 }{2mc^2} + \delta E_{n,l,m_z,m_s}^{(\rm k)} \, ,
\end{align}
where the $B$-dependent component is given by
\begin{equation}
\label{eq:keshift_delta}
\delta E_{n,l,m_z,m_s}^{(\rm k)} = - \frac{1}{2mc^2} \left[2 E_{n,l}^{(0)} \, E^{(\mu)}_{m_z,m_s} + \left(E^{(\mu)}_{m_z,m_s}\right)^2
+ 2 E_{n,l}^{(0)} \, E^{(\rm dia)}_{n,l,m_z}  \right] \, ,
\end{equation}
with $E_{n,l}^{(0)}$, $E^{(\mu)}_{m_z,m_s}$, and $E^{(\rm dia)}_{n,l,m_z}$ given, respectively, by Eqs.~\eqref{eq:E^0}, \eqref{eq:Enr_mu}, and \eqref{eq:Edia}. Since the product basis is constituted of eigenvectors of $\mathbf{p}^{4}$ and $\mathcal{H}^{(\mu)}$, the off-diagonal matrix elements of $\mathcal{H}^{(\rm k)}$ are of quadratic order in $B$, and thus do not need to be considered. 
\end{widetext}

Even if the modification of the nonrelativistic ZFS $\chi^{(\rm nr)}$ due to the kinetic correction \eqref{eq:keshift} does not have a physical meaning by itself, it is nevertheless interesting to calculate it in view of weighting its importance against the other weakly-relativistic modifications. Moreover, in the case of a very large radius $a$, where the role of the confining potential $V(r)$ should become irrelevant, the ZFS $\chi^{(\rm nr-k)}$ taking only into account the modification of $\chi^{(\rm nr)}$ due to the correction \eqref{eq:keshift}, can be compared with the bulk weakly-relativistic ZFS $\chi_\mathrm{b}^{(\rm wr)}$ of Eq.~\eqref{eq:weakly-rela_sus}.

An important aspect of the correction \eqref{eq:keshift} is that it induces at $B=0$ an energy shift of all levels. In particular, within the grand canonical ensemble, the shift of the Fermi level translates into a renormalization 
\begin{equation}
\label{eq:deltamuk}
\Delta \mu^{(\rm k)} \simeq - \frac{\bar{\mu}_{0}^2}{2m c^2}
\end{equation}
of the zero-field nonrelativistic mean chemical potential $\bar{\mu}_{0}$. 

The kinetic correction \eqref{eq:keshift} results in the eigenenergies  $E_{n,l,m_z,m_s}^{(\rm nr-k)}=E_{n,l,m_z,m_s}^{(\rm nr)}+E_{n,l,m_z,m_s}^{(\rm k)}$ from which the ZFS $\chi^{(\rm nr-k)}$ can be obtained by using the parameters
\begin{subequations}
\label{eq:energiesandderivativeskc}
\begin{align}
\label{eq:energiesandderivativeskca}
\left. E_{n,l,m_z,m_s}^{(\rm nr-k)} \right|_{B=0} & =  E_{n,l}^{(0)} \left[1-\frac{E_{n,l}^{(0)}}{2mc^2}\right] \, , \\
\label{eq:energiesandderivativeskcb}
\left. \frac{\partial E_{n,l,m_z,m_s}^{(\rm nr-k)} }{\partial B} \right|_{B=0} & =  \mu_{\rm B} \left(m_z+2 m_s \right) 
\left[1-\frac{E_{n,l}^{(0)}}{mc^2}\right]
\, , \\
\label{eq:energiesandderivativeskcc}
 \left. \frac{\partial^2 E_{n,l,m_z,m_s}^{(\rm nr-k)} }{\partial B^2} \right|_{B=0} &   =  
\mu_{\rm B}^2\Bigg(\frac{1}{2E_0}\mathcal{R}_{n,l}\mathcal{Y}_l^{m_z}\left[1-\frac{E^{(0)}_{n,l}}{mc^2}\right]
\nonumber\\
&\hspace{.5cm}-\frac{1}{mc^2}(m_z+2m_s)^2\Bigg)
 \, .
\end{align}
\end{subequations}

\begin{widetext}
Performing the $m_z$ and $m_s$ sums in Eq.~\eqref{def:ZFS}, while working up to linear order in $E_{n,l}^{(0)}/mc^2$, we have
\begin{align}
\frac{ \chi^{(\rm nr-k)}}{|\chi_{\rm L}|}=&-\frac{6\pi \, E_0}{k_{\rm F}a}\sum_{l=0}^{\infty} (l+1/2)\sum_{n=0}^\infty \left\{[l(l+1)+3] \, f'_{\bar \mu}\!\left(E_{n,l}^{(0)}\left[1-\frac{E_{n,l}^{(0)}}{2mc^2}\right]\right) \left[1-2\frac{E^{(0)}_{n,l}}{mc^2}\right] \right.\nonumber\\
		 &\left.+ \frac{\mathcal{R}_{n,l}}{E_0} \, f_{\bar \mu}\!\left(E_{n,l}^{(0)}\left[1-\frac{E_{n,l}^{(0)}}{2mc^2}\right]\right)\left[1-\frac{E^{(0)}_{n,l}}{mc^2}\right]-\frac{l(l+1)+3}{mc^2} \, f_{\bar \mu}\!\left(E_{n,l}^{(0)}\left[1-\frac{E_{n,l}^{(0)}}{2mc^2}\right]\right) \right\} \, .
\end{align}
	
As discussed above, we are interested in the correction 
\begin{equation}
\Delta \chi^{(\rm k)}  = \chi^{(\rm nr-k)} -  \chi^{(\rm nr)} \, ,
\end{equation}
where $\chi^{(\rm nr)}$ is defined in Eq.~\eqref{eq:chi_nr}, and thus associated with the mean chemical potential $\bar \mu_{0}$ of the nonrelativistic problem, while $\chi ^{(\rm nr-k)}$ is associated with the renormalized mean chemical potential ${\bar \mu}={\bar \mu}_0+\Delta \mu^{(\rm k)}$ [cf.\ Eq.~\eqref{eq:deltamuk}]. We then write 
\begin{align}
\label{eq:finalexpressionchik1}
\frac{\Delta \chi^{\rm (k)}}{|\chi_{\rm L}|}=& - \frac{6\pi \, E_0}{k_{\rm F}a \, m c^2}\sum_{l=0}^{\infty} (l+1/2)\sum_{n=1}^\infty \left\{\left[l(l+1)+3\right] \, f_{\bar{\mu}_0}''(E_{n,l}^{(0)} ) + \frac{ \mathcal{R}_{n,l}}{E_0} \, f_{\bar{\mu}_0}'(E_{n,l}^{(0)}) \right\}\left[-\frac{\left(E_{n,l}^{(0)}\right)^2}{2mc^2} - \Delta \mu^{\rm (k)} \right] \nonumber\\
&+\frac{6\pi \, E_0}{k_{\rm F}a \, mc^2}\sum_{l=0}^{\infty} (l+1/2)\sum_{n=1}^{\infty}\left\{2[l(l+1)+3] \, E_{n,l}^{(0)} \, f_{\bar{\mu}_0}'(E^{(0)}_{n,l})+
	 \left( \frac{\mathcal{R}_{n,l}}{E_0} \, E_{n,l}^{(0)}+l(l+1)+3 \right)f_{\bar{\mu}_0}(E^{(0)}_{n,l})\right\}.
\end{align}
\end{widetext}

\begin{figure*}[tb]
\begin{center}
\includegraphics[width=.9\linewidth]{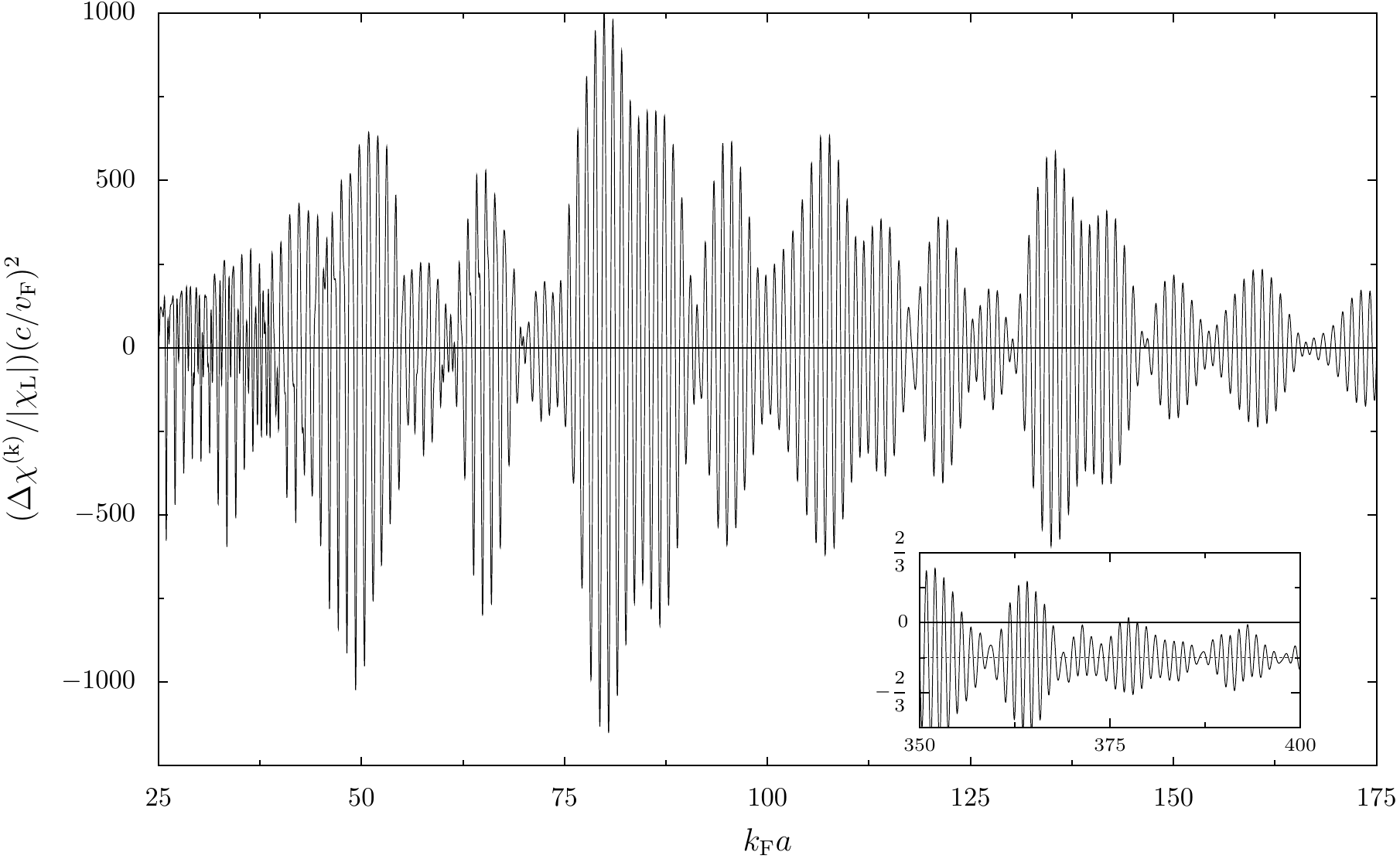}
\caption{\label{fig:chideltak} 
Kinetic correction $\Delta \chi^{(\rm k)}$ to the ZFS for $k_\mathrm{B}T /{\bar \mu}_0 = 5 \times 10^{-3}$ obtained from Eq.~\eqref{eq:finalexpressionchik1} [in units of the absolute value of the Landau susceptibility $\chi_\mathrm{L}$ and multiplied by the scaling factor $(c/v_{\rm F})^2$], as a function of the nanoparticle radius $a$ (scaled with the Fermi wave vector $k_\mathrm{F}$). Inset: $\Delta \chi^{(\rm k)}$ (in the same above-mentioned units as in the main panel) corresponding to larger sizes, showing the approach to the weakly-relativistic bulk value $-1/3$ [cf.\ Eq.~\eqref{eq:weakly-rela_sus}] and indicated by a dashed line.}
\end{center}
\end{figure*}

The first sum in the above equation results from the zero-field component of the correction $E_{n,l,m_z,m_s}^{(\rm k)}$, and thus it could have alternatively been derived by simply implementing in the ZFS expressions \eqref{eq:chi_Zeeman}--\eqref{eq:chi_paraanddia} the shift of $E^{(0)}_{n,l}$ to $E^{(0)}_{n,l}[1-E^{(0)}_{n,l}/2mc^2]$, with the corresponding renormalization $\Delta \mu^{(\rm k)}$ of the chemical potential. The second sum in \eqref{eq:finalexpressionchik1} represents the nontrivial effect of the magnetic-field dependent contribution $\delta E_{n,l,m_z,m_s}^{(\rm k)}$ to the kinetic correction, given by Eq.~\eqref{eq:keshift_delta}.

In Fig.~\ref{fig:chideltak} we present the numerical evaluation of the kinetic correction $\Delta \chi^{\rm (k)}$ to the nonrelativistic ZFS, according to Eq.~\eqref{eq:finalexpressionchik1}, as a function of $k_{\rm F}a$. For large $a$ (inset), the values of $\Delta \chi^{\rm (k)}$ oscillate around $-1/3 \, |\chi_{\rm L}| \, (v_{\rm F}/c)^2$, consistently with the result \eqref{eq:weakly-rela_sus}  for the bulk ZFS $\chi^{\rm (wr)}_{\rm b}$, since in such a limit the confinement becomes irrelevant and the kinetic correction is the only weakly-relativistic effect that needs to be taken into account. The oscillations as a function of $k_{\rm F}a$ are much larger than the bulk value, but remain considerably smaller than the typical values of $\chi^{(\rm nr)}$ exhibited in Fig.~\ref{fig:chinr} (and even of the finite-$N$ correction characterizing the response of an ensemble of nanoparticles, see the blue line in Fig.~\ref{fig:chinr}).

The correction $\Delta \chi^{\rm (k)}$ of Eq.~\eqref{eq:finalexpressionchik1} admits a one-dimensional semiclassical treatment, analogous to that of the nonrelativistic ZFS treated in Appendix \ref{app:sc}. In particular, the smooth part of $\Delta \chi^{\rm (k)}$ can be evaluated along the lines of Eqs.~\eqref{eq:chi_para_diaps}--\eqref{eq:chi_diapss}, and we find in the leading order in $k_\mathrm{F}a$, that the smooth part of the first sum behaves as $1/5 \, (k_{\rm F}a)^2 \, \left(v_{\rm F}/c\right)^2$, while the second sum cancels the previous contribution. The next-leading term of the smooth part of $\Delta \chi^{\rm (k)}$ is of order $(k_\mathrm{F}a)^0$, but for the same reasons discussed in  Appendix \ref{app:sc}, the correct asymptotic value $\chi^{\rm (wr)}_{\rm b}$ of the bulk yielded by the numerical calculation, is not recovered by our semiclassical approach.

\subsection{Spin-orbit coupling in a spherical nanoparticle}
\label{sec:socsn}

While the product eigenbasis of $\mathcal{H}^{(0)}$ used in Sec.~\ref{sec:nrssg}, and characterized by the set $\{\lambda\}=\{n,l,m_z,m_s\}$,  remained the appropriate one once the term $\mathcal{H}^{(\mu)}$ of Eq.~\eqref{eq:Hmagmoml} was taken into account, the inclusion of the SOC term $\mathcal{H}^{(\rm so)}$ of Eq.~\eqref{eq:H_so2} makes it more convenient to change to the eigenbasis of the total angular momentum $\mathbf{J}=\mathbf{L}+\mathbf{S}$, characterized by the set $\{\tilde\lambda\}=\{n,j,m_j,l\}$. The rule of addition of angular momentum results in $[l]\otimes[1/2]=[l-1/2,l]\oplus [l+1/2,l]$, where the left-hand side is the tensor product of the orbital angular momentum $l \ne 0$ subspace with that of the spin $1/2$ and the  right-hand side represents the direct sum of the subspaces with total angular momenta $j=l \mp 1/2$ and orbital angular momentum $l$. For radially symmetric electrostatic potentials the separability between radial and angular coordinates allows to write the eigenstates of the coupled basis in terms of the spinors 
\cite{greiner1990relativistic}
\begin{equation}
\Phi_{n,j,m_j}^{(\pm)}(\mathbf{r})=R_{n,j\pm1/2}(r) \, {\Upsilon}^{(\pm)}_{j,m_j}(\vartheta) \, ,
\end{equation}
where we have defined the spinor spherical harmonics
\begin{align}\label{eq:spinsh}
	{\Upsilon}^{(\pm)}_{j,m_j}(\vartheta)=\;&\frac{1}{\sqrt{2(j \pm 1/2)+1}}
	\nonumber\\&\times\begin{bmatrix}
	 \mp\sqrt{j \pm 1/2 \mp m_j+1/2 } \, Y_{j \pm 1/2}^{m_j-1/2}(\vartheta)\\[.25truecm] 
	\sqrt{j \pm 1/2 \pm m_j+1/2 } \, Y_{j \pm 1/2}^{m_j+1/2}(\vartheta)
	  \end{bmatrix} 
\end{align}
in terms of the usual spherical harmonics $Y_l^{m_z}$ introduced in Eq.~\eqref{eq:orbital}.
The label $(\pm)$ corresponds to $l=j \pm 1/2$, and sets the parity $(-1)^{j \pm 1/2}$ of the state. The associated eigenenergies of $\mathcal{H}^{(0)}$ in the case of a  spherical confining potential are ${\cal E}_{n,j,(\pm)}^{(0)}=E_{n,j \pm 1/2}^{(0)}$, and therefore there is degeneracy between the $2j\!+\!1$ dimensional subspace $\{n,j,(+)\}$ and the $2j\!+\!3$ dimensional subspace $\{n,j+1,(-)\}$, as they are characterized by the same quantum number $l$.

The coupled basis remains an  eigenbasis of the subspace $\left\{n,j,(+)\right\} \oplus \{n,j+1,(-)\}$ once $\mathcal{H}^{(\rm so)}$ is taken into account, while the eigenenergies change according to
\begin{align}
{\cal E}_{n,j,(\pm)}^{({\rm so})} &= \langle \Phi_{n,j,m_j}^{(\pm)} |\mathcal{H}^{(\rm so)}|  \Phi_{n,j,m_j}^{(\pm)} \rangle \nonumber\\
&=\frac{\hbar^2}{4m^2c^2a^2}\left[\mp\left(j+\frac{1}{2}\right)-1\right]  I^{(\rm so)}_{n, j \pm 1/2} \, ,
\end{align}
with the radial matrix element
\begin{equation}
\label{eq:rme}
I^{(\rm so)}_{n,l}=a^2 \int_{0}^{\infty} \mathrm{d}r  \, r \left[ R_{n,l}(r) \right]^2  V^{\prime}(r)     \, .
\end{equation}
For the potential \eqref{eq:confining_potential}, we have 
\begin{equation}
\label{eq:rme2}
I^{(\rm so)}_{n,l}=V_{0} \, a^3 \left[ R_{n,l}(a) \right]^2     \, .
\end{equation}
In the limit where the confining potential approaches a hard wall, the product $V_{0}   \left[ R_{n,l}(a) \right]^2$ remains finite \cite{yanno92_AnnPhys}. Using the limiting expressions \eqref{eq:knlandnorma} and \eqref{eq:jl} we obtain
\begin{equation}
\label{eq:I_SO}
	I^{(\rm so)}_{n,l}=\frac{\hbar^2}{ma^2} \, \zeta_{n,l}^2 \, ,
\end{equation}
and therefore
\begin{equation}
\label{eq:esoj}
{\cal E}_{n,j,(\pm)}^{({\rm so})} = \left[\mp\left(j+\frac{1}{2}\right)-1\right] \, \zeta_{n,j\pm 1/2}^2 \, \frac{E_0^2}{mc^2} \, ,
\end{equation}
independently of $V_0$.

Obviously, ${\cal E}_{n,1/2,(-)}^{({\rm so})}=0$, since the spherical symmetry of the $s$-states ($l=0$) renders the SOC ineffective.
The degeneracy between the subspaces $\{n,j,(+)\}$ and $\{n,j+1,(-)\}$ for $B = 0$ is broken by $\mathcal{H}^{(\rm so)}$. For $B \ne 0$, the remaining degeneracy within each subspace is lifted according to the different values of $m_j$.

\subsection{Perturbative treatment of the magnetic field}
\label{sec:ptmf}
Once the term $\mathcal{H}^{(\mu)}$ of Eq.~\eqref{eq:Hmagmoml} is taken into account, $m_j$ is still a good quantum number, but the coupled basis is no longer an eigenbasis  of the subspace $\left\{n,j,(+)\right\} \oplus \{n,j+1,(-)\}$. Therefore, in order to treat the terms $\mathcal{H}^{(\mu)}$, $\mathcal{H}^{(\rm dia)}$, $\mathcal{H}^{(\rm k)}$, $\mathcal{H}^{(\rm ame)}$ in the two lowest orders in $B$, the perturbative approach in magnetic field of Sec.~\ref{sec:nrssg} yielding the nonrelativistic ZFS has to be extended using the decomposition in subspaces of fixed $m_j$. These are represented by
\begin{align}
\label{eq:decomposition_spaces}
\left\{n,j,(+)\right\} \oplus& \{n,j+1,(-)\} = {\cal S}_{n,j+1/2,({\rm d})}^{\rm p} \nonumber\\ 
&\oplus \cup_{m_j=-j}^{m_j=+j}{\cal S}_{n,j+1/2,m_j}^{\rm e}
\oplus {\cal S}_{n,j+1/2,({\rm u})}^{\rm p} 
\end{align}
in terms of the  down ($\rm d$) and up ($\rm u$) one-dimensional subspaces 
\begin{equation}
\label{eq:product_spaces}
{\cal S}_{n,l,({\rm d/u})}^{\rm p} = \left\{n,l+1/2,(-)\right\}\big|_{m_j=\mp(l+1/2)} \, ,
\end{equation}
and the two-dimensional subspaces ${\cal S}_{n,l,m_j}^{\rm e}$ subtended by the vectors $|\Phi_{n,l \mp 1/2,m_j}^{(\pm)}\rangle$ of the coupled basis, associated with the quantum numbers $\{n,l \mp 1/2,m_j,l\}$ with  $l\ne 0$ and  $|m_j| \leqslant l-1/2$. The labels $\rm p$ ($\rm e$) stand for ``product" (``entangled"), characterizing the one (two)-dimensional subspaces where the coupled basis does (does not) coincide with the product basis. The choice of using the index $l$ (instead of $j$) in order to label the subspaces is motivated in view of the book-keeping for the sum over states. We notice that the definition \eqref{eq:product_spaces} is also valid for the subspaces with $l=0$, which are not considered in the decomposition \eqref{eq:decomposition_spaces}, but should be included when taking the sum over states yielding a thermodynamic quantity like the ZFS.

The Hamiltonian \eqref{eq:Hamiltonian}, restricted to the subspaces ${\cal S}_{n,l,m_j}^{\rm e}$, can be expressed by the $2 \times 2$ matrix 
\begin{equation}
\label{eq:redHam}
\mathcal{H}_{n,l,m_j}=\begin{pmatrix}
E_{n,l,m_j}^{(+)} &-\mu_{\rm B}B \, \aleph_{l,m_j} \\
-\mu_{\rm B}B \, \aleph_{l,m_j} & E_{m,l,m_j}^{(-)}
\end{pmatrix} \, .
\end{equation}
In the diagonal matrix elements $E_{n,l,m_j}^{(\pm)}$ we separate the field-independent and the field-dependent contributions as
\begin{equation}
E_{n,l,m_j}^{(\pm)} = E_{n,l,(\pm)}^{(0)} +  \delta E_{n,l,m_j}^{(\pm)}  \, ,
\label{eq:Erel}
\end{equation}
with
\begin{equation}
E_{n,l,(\pm)}^{(0)} = E_{n,l}^{(0)}\left[1-\frac{E_{n,l}^{(0)}}{2mc^2}\right] + E^{(\rm D)}_{n,l} + E_{n,l,(\pm)}^{({\rm so})} \, .
\label{eq:ErelB0}
\end{equation}
The first term in the right-hand side of the above equation represents the $B=0 $ nonrelativistic eigenvalue together with its kinetic energy correction, as expressed in Eq.~\eqref{eq:energiesandderivativeskca}. The second term is the Darwin correction, given for the case of $V_0 \gg E_0$ by Eq.~\eqref{eq:ED}. Recalling Eq.~\eqref{eq:esoj}, it is convenient to express the SOC contribution as
\begin{align}
\label{eq:EpmSO}
E_{n,l,(\pm)}^{({\rm so})} &= {\cal E}_{n,l \mp 1/2,(\pm)}^{({\rm so})} 
\nonumber\\
&= 
\left[\mp\left(l+\frac{1}{2}\right)-\frac{1}{2}\right]  \zeta_{n,l}^{2} \, \frac{E_0^2}{mc^2} \, .
\end{align}

The spin splitting in the subspace ${\cal S}_{n,l,m_j}^{\rm e}$ is
\begin{align} 
\label{eq:criteriumSO}
\Delta_{n,l} & = E_{n,l,(-)}^{(0)}-E_{n,l,(+)}^{(0)}  
\nonumber\\
&=
E_{n,l,(-)}^{({\rm so})} - E_{n,l,(+)}^{({\rm so})}
\nonumber\\
& = 2 \left(l+\frac{1}{2}\right)  \zeta_{n,l}^{2} \, \frac{E_0^2}{mc^2} \, .
\end{align} 
In reducing the Hamiltonian \eqref{eq:Hamiltonian} to its $2 \times 2$ form \eqref{eq:redHam}, we are assuming an infinitesimal field $B$, and neglecting the coupling between the subspaces ${\cal S}_{n,l,m_j}^{\rm e}$ and ${\cal S}_{n',l,m_j}^{\rm e}$. This last approximation requires $\Delta_{n,l} \ll E_{n+1,l}^{(0)} - E_{n,l}^{(0)} \simeq 2 \pi  \zeta_{n,l} E_0 $, where we have used the asymptotic form of the zeros of $j_l(\zeta)$. The previous condition translates into $(l+1/2)(v_\mathrm{F}/c)^2 \ll 2\pi  k_\mathrm{F}  a$, which is always verified in the weakly-relativistic limit since $l_{\rm max} + 1/2 \simeq k_\mathrm{F} a$ (see 
Appendix \ref{app:sc}). Nevertheless, our perturbative treatment of $\mathcal{H}^{(\rm so)}$ can become problematic in cases of quasi-degeneracies between eigenstates  $E_{n,l}^{(0)}$ and $E_{n',l'}^{(0)}$ corresponding to different quantum numbers.
We remark that $\mathcal{H}^{(\rm so)}$ does not induce a renormalization of the chemical potential, since the shift $-E_0 E_{n,l}^{(0)} /mc^2$ associated with each of the subspaces ${\cal S}_{n,l,m_j}^{\rm e}$ is compensated by the correction $l E_0 E_{n,l}^{(0)} /mc^2$ characterizing each of the two subspaces ${\cal S}_{n,l,({\rm d/u})}^{\rm p}$.

The $B$-dependent components of the diagonal matrix elements \eqref{eq:Erel} are given by
\begin{align}
\delta E_{n,l,m_j}^{(\pm)} =& \;  E_{l,m_j,(\pm)}^{({\mu})} + E^{(\rm dia)}_{n,l,m_j,(\pm)} 
\nonumber \\
& +\delta E_{n,l,m_j,(\pm)}^{(\rm k)} + E^{(\rm ame)}_{n,l,m_j,(\pm)} \, .
\label{eq:deltaErel}
\end{align} 
The first term can be obtained with the aid of the Wigner--Eckart theorem, which allows us to write  
\begin{align}
\label{eq:dmehmu}
{\cal E}_{j,m_j,(\pm)}^{({\mu})} &= \langle \Phi_{n,j,m_j}^{(\pm)} |\mathcal{H}^{(\mu)}|  \Phi_{n,j,m_j}^{(\pm)} \rangle 
\nonumber\\
&=  \frac{j+1/2}{j+1/2 \pm 1/2}\, \mu_{\rm B}  B  m_j \, ,
\end{align}
independently of $n$, and therefore
\begin{align}
\label{eq:Emucp}
E_{l,m_j,(\pm)}^{({\mu})} &= {\cal E}_{l \mp 1/2,m_j,(\pm)}^{({\mu})} 
\nonumber\\
&= \left(1 \mp \frac{1/2}{l+1/2}\right) \mu_{\rm B}  B  m_j \, .
\end{align}

The diamagnetic contributions in Eq.~\eqref{eq:deltaErel} follow from the matrix elements
\begin{align} 
\label{eq:media}
{\cal E}_{n,j,m_j,(\pm)}^{({\rm dia})} & =  \langle \Phi_{n,j,m_j}^{(\pm)} |\mathcal{H}^{(\rm dia)}| \Phi_{n,j,m_j}^{(\pm)} \rangle \nonumber \\ 
 & = 
\left[1+\frac{m_j^2}{j(j+1)}\right] \frac{\mu_{\rm B}^2B^2}{8E_0} \, \mathcal{R}_{n,j \pm 1/2}  \, ,
\end{align} 
where $\mathcal{R}_{n,l}$ has been defined in Eq.~\eqref{eq:R}. Thus,
\begin{align}
\label{eq:media2}
E_{n,l,m_j,(\pm)}^{({\rm dia})} & =  {\cal E}_{n,l \mp 1/2,m_j,(\pm)}^{({\rm dia})} 
\nonumber \\ 
& =  \left[1+\frac{m_j^2}{(l \mp 1/2)(l \mp 1/2 +1)}\right] \frac{\mu_{\rm B}^2B^2}{8E_0} \, \mathcal{R}_{n,l}  \, .
\end{align}

The diagonal matrix elements of $\mathcal{H}^{(\rm k)}$ in the subspace ${\cal S}_{n,l,m_j}^{\rm e}$ can be obtained, up to quadratic order in $B$, by a similar expression to that of Eq.~\eqref{eq:keshift}, where the vectors $|\Psi_{n,l,m_z,m_s}^{(0)}  \rangle $ of the product basis  have to be replaced by the ones of the coupled basis, $|  \Phi_{n,j,m_j}^{(\pm)} \rangle$, leading to
\begin{equation}
\label{eq:keshiftcb}
E_{n,l,m_j,(\pm)}^{(\rm k)} = - \frac{\left( E_{n,l}^{(0)} \right)^2 }{2mc^2} + \delta E_{n,l,m_j,(\pm)}^{(\rm k)} \, ,
\end{equation}
with
\begin{align}
\label{eq:deltaekcp}
\delta E_{n,l,m_j,(\pm)}^{(\rm k)} =& - \frac{E_{n,l}^{(0)} }{mc^2} \left( E^{(\mu)}_{l,m_j,(\pm)} +  \, E^{(\rm dia)}_{n,l,m_j,(\pm)}  \right) 
\nonumber\\
&- \frac{\left(\mu_{\rm B}B \right)^2}{2mc^2} \left[m_j^2\left(1 \mp \frac{1}{l+1/2} \right) + \frac{1}{4}  \right]
\, .
\end{align}

The last contribution to $\delta E_{n,l,m_j}^{(\pm)}$ in Eq.~\eqref{eq:deltaErel} results from the diagonal matrix element of $\mathcal{H}^{(\rm ame)}$ in the coupled basis, which with the help of Eqs.~\eqref{eq:dmeamec},
 \eqref{eq:I_SO}, and \eqref{eq:dmeamecai2}, can be written as
\begin{align}
\label{eq:dmecbame2}
E_{n,l,m_j,(\pm)}^{({\rm ame})} &= {\cal E}_{n,l \mp 1/2,m_j,(\pm)}^{({\rm ame})} 
\nonumber\\
&= \mp \mu_{\rm B}  B \, \frac{E_{n,l}^{(0)}}{2mc^2} \, 
 \frac{m_j (l \mp 1/2 + 1/2)}{(l \mp 1/2)(l \mp 1/2+1)} \, .
\end{align}

We remark that the expression \eqref{eq:Emucp} of $E_{l,m_j,(\pm)}^{({\mu})}$ does not simply follow from the result \eqref{eq:Enr_mu} for $E_{m_z,m_s}^{({\mu})}$, since the latter represents the exact energy shift associated with $\mathcal{H}^{(\mu)}$ in the product basis, while the former is just the first-order perturbative correction in the coupled basis. Similarly, the perturbative correction \eqref{eq:media2} for $E_{n,l,m_j,(\pm)}^{({\rm dia})}$ does not simply follow from the analogous correction \eqref{eq:Edia} for $E^{(\rm dia)}_{n,l,m_z}$, nor does $\delta E_{n,l,m_j,(\pm)}^{(\rm k)}$ in Eq.~\eqref{eq:deltaekcp} from $E_{n,l,m_z,m_s}^{(\rm k)}$ in Eq.~\eqref{eq:keshift}. In the same vein, the angular magneto-electric correction $E_{n,l,m_j,(\pm)}^{({\rm ame})}$ of Eq.~\eqref{eq:dmecbame2} follows from the matrix element ${\cal E}_{n,l \mp 1/2,m_j,(\pm)}^{({\rm ame})}$ in the coupled basis. 

The off-diagonal matrix elements of the restricted Hamiltonian \eqref{eq:redHam} need to be only obtained up to linear order in $B$. Thus, the nonrelativistic component does not have a term associated with $\mathcal{H}^{(\rm dia)}$, but consists only of 
\begin{align}
\label{eq:offdiagonal}
\langle \Phi_{n,l+1/2,m_j}^{(-)} |&\mathcal{H}^{(\mu)}| \Phi_{n,l-1/2,m_j}^{(+)} \rangle =
\nonumber\\
&- \mu_{\rm B}B 
\frac{\sqrt{\left(l+1/2\right)^2-m_j^2} }{2(l+1/2)} \, ,
\end{align} 
independently of $n$.  The only off-diagonal matrix element of $\mathcal{H}^{(\rm k)}$ that we need to consider is
\begin{align}
\label{eq:offdiagonalp2mu}
\langle \Phi_{n,l+1/2,m_j}^{(-)}  |&\left( - \frac{1}{mc^2} \right) \frac{\mathbf{p}^{2}}{2m} \mathcal{H}^{(\mu)} | \Phi_{n,l-1/2,m_j}^{(+)} \rangle =
\nonumber \\
 & \, \mu_{\rm B}B \,  \frac{E_{n,l}^{(0)}}{2mc^2} \, 
\frac{\sqrt{\left(l+1/2\right)^2-m_j^2} }{l+1/2} \, .
\end{align}

The remaining contribution to the off-diagonal matrix element arising from the angular magneto-electric coupling is given by Eqs.~\eqref{eq:odmeame} and 
\eqref{eq:odmeamecai2} as
\begin{align}
\label{eq:odmeamel}
\langle  \Phi_{n,l+1/2,m_j}^{(-)} |& \mathcal{H}^{(\rm ame)} | \Phi_{n,l-1/2,m_j}^{(+)} \rangle =
\nonumber\\
&\,
 -  \mu_{\rm B}B \,  \frac{E_{n,l}^{(0)}}{4mc^2} \, 
\frac{ \, \sqrt{\left(l+1/2\right)^2-m_j^2} }{l+1/2} \, .
\end{align}

The off-diagonal  matrix element of $\mathcal{H}_{n,l,m_j}$ with the form of Eq.~\eqref{eq:redHam} is determined by the $B$-independent dimensionless parameter
\begin{equation}
\aleph_{l,m_j}=  \frac{\sqrt{\left(l+1/2\right)^2-m_j^2}}{2(l+1/2)} 
 \left[ 1 - \frac{E_{n,l}^{(0)}}{2mc^2} \right] \, .
\end{equation}

The diagonalization of $\mathcal{H}_{n,l,m_j}$ yields the $B$-dependent low ($\rm l$) and high ($\rm h$) eigenenergies within the subspaces ${\cal S}_{n,l,m_j}^{\rm e}$, respectively given by
\begin{align}
	E^{(\rm l/h)}_{n,l,m_j}=&\;\frac{E_{n,l,m_j}^{(-)}+E_{n,l,m_j}^{(+)}}{2} 
	\nonumber\\
	&\mp \sqrt{\left(\frac{E_{n,l,m_j}^{(-)}-E_{n,l,m_j}^{(+)}}{2}\right)^2+\mu_{\rm B}^2B^2 \, \aleph^2_{l,m_j}} \, .
\end{align}

In analogy with the nonrelativistic parameters of Eq.~\eqref{eq:energiesandderivativesnr}, the contributions to the ZFS stemming from the ($\rm l$/$\rm h$) eigenstates within the subspaces ${\cal S}_{n,l,m_j}^{\rm e}$ are determined by 
\begin{widetext}
\begin{subequations}
\label{eq:energiesandderivativese}
\begin{align}
\left.E^{(\rm l/h)}_{n,l,m_j}\right|_{B=0} &=  \left.E_{n,l,m_j}^{(\pm)}\right|_{B=0} 
= E_{n,l,(\pm)}^{(0)}  \, , 
\\
 \left.\frac{\partial E^{(\rm l/h)}_{n,l,m_j}}{\partial B}\right|_{B=0} &=  \left.\frac{\partial E^{(\pm)}_{n,l,m_j}}{\partial B}\right|_{B=0} 
= \mu_{\rm B} \, m_j \left\{   \left(1 \mp \frac{1/2}{l+1/2}\right)
\left[1 - \frac{E_{n,l}^{(0)}}{mc^2} \right]
\mp  \frac{E_{n,l}^{(0)}}{2mc^2} \
 \frac{l \mp 1/2 + 1/2}{(l \mp 1/2)(l \mp 1/2+1)} \right\}
  \, , 
\end{align}
\begin{align}
 \left.\frac{\partial^2 E^{(\rm l/h)}_{n,l,m_j}}{\partial B^2}\right|_{B=0} \!
 &=  \left.\frac{\partial^2 E^{(\pm)}_{n,l,m_j}}{\partial B^2}\right|_{B=0} \mp \frac{2\mu_{\rm B}^2 \,\aleph_{l,m_j}^2}{\Delta_{n,l}}  =
\frac{\mu_{\rm B}^2}{4} \left\{\!
\left(1+\frac{m_j^2}{(l \mp 1/2)(l+1\mp 1/2)}\right) 
 \frac{\mathcal{R}_{n,l}}{E_0} \left[1 - \frac{E_{n,l}^{(0)}}{mc^2} \right] \right.
 \nonumber\\
& 
\left.
 \hspace{.5truecm}-\frac{1}{mc^2} \left(4 m_j^2\left(1 \mp \frac{1}{l+1/2} \right) + 1  \right)
\mp \frac{\left(l+1/2\right)^2-m_j^2}{(E_0/mc^2) E_{n,l}^{(0)} \left(l+1/2\right)^3} \left[1 - \frac{E_{n,l}^{(0)}}{2mc^2} \right]^2 \right\} \, .
\end{align}
\end{subequations}

The eigenenergies of the down (up) state of the subspace ${\cal S}_{n,l,({\rm d/u})}^{\rm p}$ can be written as
\begin{equation}
E_{n,l}^{(\rm d/u)} = E_{n,l,(-)}^{(0)}  + \delta E_{n,l}^{(\rm d/u)} \, ,
\label{eq:Ereldu}
\end{equation}
with
\begin{equation}
\delta E_{n,l}^{(\rm d/u)}  = E_{l,\mp (l+1/2),(-)}^{({\mu})} + E^{(\rm dia)}_{n,l,\mp (l+1/2),(-)} 
+
\delta E_{n,l,\mp (l+1/2),(-)}^{(\rm k)} + E_{n,l,\mp (l+1/2),(-)}^{({\rm ame})}
\, .
\label{eq:deltaEreldu}
\end{equation}
Therefore, in addition to the parameters \eqref{eq:energiesandderivativese}, we have to consider the contributions to the ZFS associated with the ($\rm d/\rm u$) states, which follow from  
\begin{subequations}
\label{eq:energiesandderivativespa}
\begin{align}
\left.E^{(\rm d/u)}_{n,l}\right|_{B=0} & = E_{n,l,(-)}^{(0)}  \, , \\
\left. \frac{\partial E^{(\rm d/u)}_{n,l}}{\partial B}\right|_{B=0} & = \mp \mu_{\rm B}   (l+1) \left\{ 1 - \frac{E_{n,l}^{(0)}}{mc^2} \right. 
+ \left.
\frac{E_{n,l}^{(0)}}{2mc^2} \, \frac{1}{l+3/2} \right\}
, \\
\left. \frac{\partial^2 E^{(\rm d/u)}_{n,l}}{\partial B^2}\right|_{B=0} &   =  \mu_{\rm B}^2 \Bigg\{  \frac{l+1}{l+3/2} \, \frac{\mathcal{R}_{n,l}}{2E_0}  
\left[1 - \frac{E_{n,l}^{(0)}}{mc^2} \right]
- 
\frac{1}{m c^2} \, (l+1)^2
\Bigg\}
 \, .
\end{align}
\end{subequations}
The weakly-relativistic ZFS follows from the evaluation of Eq.~\eqref{def:ZFS} using the parameters of Eqs.~\eqref{eq:energiesandderivativese} and \eqref{eq:energiesandderivativespa}.

\subsection{Relativistic correction to the zero-field  susceptibility}
\label{sec:rczfs}

The weakly-relativistic ZFS will have contributions from $\rm p$ and $\rm e$ eigenstates, and it can therefore be written as $\chi^{(\rm wr)} = \chi_{\rm p} + \chi_{\rm e}$. Performing the $m_j$ summation in Eq.~\eqref{def:ZFS} for each subset of eigenstates while keeping up to linear terms in $E_{n,l}^{(0)}/mc^2$, we have
\begin{subequations}
\label{eq:chi}
\begin{align}
\label{eq:chip}
\frac{\chi_{\rm p}}{|\chi_\mathrm{L}|} = & -\frac{9\pi E_0}{k_\mathrm{F}a} \sum_{l=0}^{\infty} \sum_{n=1}^\infty
\left\{\left(l+1\right)^2 \left[1 - \frac{2E_{n,l}^{(0)}}{mc^2} \frac{l+1}{l+3/2} \right] f'_{\bar \mu}\!\left(E_{n,l,(-)}^{(0)} \right) 
\right.
\nonumber \\
 & 
+ 
\left.
\left(
\frac{l+1}{l+3/2} \, \frac{\mathcal{R}_{n,l}}{2E_0} \, \left[1 - \frac{E_{n,l}^{(0)}}{mc^2} \right] - \frac{1}{m c^2} \, (l+1)^2 
\right)
f_{\bar \mu}\!\left(E_{n,l,(-)}^{(0)} \right)\right\} \, , \\
\label{eq:chie}
\frac{\chi_{\rm e}}{|\chi_\mathrm{L}|} = & -\frac{3\pi E_0}{k_\mathrm{F}a}  \sum_{l=1}^{\infty}  \sum_{n=1}^{\infty} l \left\{ \frac{l-1/2}{l+1/2} \, \left(l^2 \left[1 - \frac{2 E_{n,l}^{(0)}}{mc^2} \frac{l}{l-1/2} \right] f'_{\bar \mu}\!\left(E_{n,l,(+)}^{(0)} \right)+\left(l+1 \right)^2 \left[1 - \frac{2 E_{n,l}^{(0)}}{mc^2} \frac{l+1}{l+3/2} \right] f'_{\bar \mu}\!\left(E_{n,l,(-)}^{(0)} \right)\right) \right.
\nonumber \\
 & +  \left[ \left(
 \frac{\mathcal{R}_{n,l} }{E_0} \left[1 - \frac{E_{n,l}^{(0)}}{mc^2} \right] - \frac{l^2-l+1}{mc^2} \right) f_{\bar \mu}\!\left(E_{n,l,(+)}^{(0)} \right)+ \left( \frac{l+1}{l+3/2} \, \frac{\mathcal{R}_{n,l} }{E_0}  \left[1 - \frac{E_{n,l}^{(0)}}{mc^2} \right] - \frac{l(l+1)}{mc^2} \right) f_{\bar \mu}\!\left(E_{n,l,(-)}^{(0)} \right)\right] 
\nonumber \\
&  \left. - \frac{mc^2}{2 E_0 E_{n,l}^{(0)}}
\left[1 - \frac{E_{n,l}^{(0)}}{mc^2} \right]
\frac{l+1}{(l+1/2)^2} 
\left[f_{\bar \mu}\!\left(E_{n,l,(+)}^{(0)} \right)-f_{\bar \mu}\!\left(E_{n,l,(-)}^{(0)} \right)\right] 
\right\} 
 \, .
\end{align}
\end{subequations}
\end{widetext}

In the right-hand side of Eq.~\eqref{eq:chi} we identify the first term of the sums, containing  $f'_{\bar \mu}(E_{n,l,(\pm)}^{(0)})$, which generalize the nonrelativistic paramagnetic contributions of Eqs.~\eqref{eq:chi_Zeeman} and \eqref{eq:chi_para} including the corresponding weakly-relativistic corrections. Similarly, the second term of the sums, containing $f_{\bar \mu}(E_{n,l,(\pm)}^{(0)})$, generalizes the nonrelativistic diamagnetic contribution of Eq.~\eqref{eq:chi_dia}. The third term in the sum of Eq.~\eqref{eq:chie}, containing  $f_{\bar \mu}(E_{n,l,(+)}^{(0)} )-f_{\bar \mu}(E_{n,l,(-)}^{(0)})$, yields a paramagnetic contribution of the van Vleck kind, since it follows from the second-order perturbation correction in $\mathcal{H}^{(\mu)}$ appearing in the coupled basis, together with a weakly-relativistic correction. 

Equation \eqref{eq:chi} can be directly evaluated from the knowledge of the zero-field, weakly-relativistic eigenenergies \eqref{eq:ErelB0}. However, since we are interested in the relativistic corrections to the ZFS, we will consider 
\begin{equation}
\Delta \chi = \chi ^{(\rm wr)} -  \chi^{(\rm nr)} \, ,
\end{equation}
where $\chi^{(\rm nr)}$ is defined in Eq.~\eqref{eq:chi_nr}, and thus associated with the mean chemical potential $\bar \mu_{0}$ of the nonrelativistic problem, while the weakly relativistic $\chi ^{(\rm wr)}$ is associated with the renormalized mean chemical potential 
\begin{equation}
\label{eq:renchempot}
{\bar \mu} = {\bar \mu}_0 + \Delta \mu^{(\rm k)} + \Delta \mu^{(\rm D)} \, ,
\end{equation}
where $\Delta \mu^{(\rm k)}$ and $\Delta \mu^{(\rm D)}$ are, respectively, given by Eqs.~\eqref{eq:deltamuk} and \eqref{eq:deltamuD}. 

Furthermore, for calculational purposes, it is convenient to treat separately the corrections from the $\rm p$ and $\rm e$ contributions. Thus, we write
\begin{equation}
\label{eq:deltachidecomp}
\Delta \chi = \Delta\chi_{\rm p} + \Delta \chi_{\rm e} \, ,
\end{equation}
where $\chi^{(\rm nr)}_{\rm p}$ and $\chi^{(\rm nr)}_{\rm e}$ have been worked out in Appendix \ref{app:nrsucb}. In the weakly-relativistic limit, Eqs.~\eqref{eq:chi}, \eqref{eq:chinrp}, and \eqref{eq:chinre}, together with the form \eqref{eq:ErelB0} of the zero-field, weakly-relativistic energy correction and that of the renormalized mean chemical potential \eqref{eq:renchempot}, yield to first-order in $E_0/mc^2$
\begin{widetext}
\begin{subequations}
\label{eq:deltachi0}
\begin{align}
\label{eq:chi0p}
\frac{\Delta \chi_{\rm p}}{|\chi_\mathrm{L}|} = & - \frac{9\pi E_0}{k_\mathrm{F}a} 
\sum_{l=0}^{\infty} (l+1) \sum_{n=1}^\infty \left\{(l+1) \, f_{\bar{\mu}_0}''(E_{n,l}^{(0)} ) + \frac{1}{l+3/2} \, \frac{ \mathcal{R}_{n,l}}{2 E_0}  \, f'_{\bar{\mu}_0}(E_{n,l}^{(0)}) \right\} 
\left[-\frac{\left(E_{n,l}^{(0)}\right)^2}{2mc^2} + E_{n,l}^{(\rm D)} - \Delta \mu
+ \frac{E_0 \, E_{n,l}^{(0)}}{mc^2}  l \right]
\nonumber \\
 & +
 \frac{9\pi E_0}{k_\mathrm{F}a \, mc^2} 
\sum_{l=0}^{\infty} (l+1) \sum_{n=1}^\infty 
\left\{ \frac{2(l+1)^2}{l+3/2} \, E_{n,l}^{(0)} \, f'_{\bar{\mu}_0}\!\left(E_{n,l}^{(0)} \right) 
+ \left( \frac{1}{l+3/2} \, \frac{\mathcal{R}_{n,l}}{2E_0} \, E_{n,l}^{(0)} + l+1 \right) f_{\bar{\mu}_0}(E_{n,l}^{(0)}) \right\}
\, , \\
\label{eq:chi0e}
\frac{\Delta \chi_{\rm e}}{|\chi_\mathrm{L}|} = & -\frac{3\pi E_0}{k_\mathrm{F}a}  \sum_{l=1}^{\infty} l \, \sum_{n=1}^{\infty} \Bigg\{\left[ \left(2l^2+1\right) \, f''_{\bar{\mu}_0}(E_{n,l}^{(0)} ) +  \frac{\mathcal{R}_{n,l}}{E_0} \, \frac{2l+5/2}{l+3/2} \, f'_{\bar{\mu}_0}\!\left(E_{n,l}^{(0)} \right) \right] \, 
\left[-\frac{\left(E_{n,l}^{(0)}\right)^2}{2mc^2} + E_{n,l}^{(\rm D)} - \Delta \mu \right]
\nonumber \\
 & +
 \left[ \left(l^2-1\right) \, f''_{\bar{\mu}_0}(E_{n,l}^{(0)} ) - \frac{3}{2}  \frac{\mathcal{R}_{n,l}}{E_0} \, \frac{l+1}{l+3/2} \, f'_{\bar{\mu}_0}\!\left(E_{n,l}^{(0)} \right) \right]  \frac{E_0 E_{n,l}^{(0)}}{m c^2} \Bigg\}
\nonumber \\
 & 
+ \frac{3\pi E_0}{k_\mathrm{F}a \, mc^2}  \sum_{l=1}^{\infty} l \, \sum_{n=1}^{\infty} \left\{
E_{n,l}^{(0)} \, \frac{4 l^3 + 6 l^2 + l +1 }{l+3/2}  \, f_{\bar{\mu}_0}'(E_{n,l}^{(0)} ) + \left[ \frac{2l+5/2}{l+3/2} \, \frac{ \mathcal{R}_{n,l}}{E_0} \, E_{n,l}^{(0)} +2l^2+1 \right] f_{\bar{\mu}_0}(E_{n,l}^{(0)})  
 \right\}
 \, .
\end{align}
\end{subequations}
Grouping the two components of Eq.~\eqref{eq:deltachidecomp}, we have
\begin{align}
\label{eq:deltachitotep}
\frac{\Delta \chi}{|\chi_\mathrm{L}|} = & -\frac{6\pi E_0}{k_\mathrm{F}a}  \sum_{l=0}^{\infty} \left(l+\frac{1}{2} \right) \sum_{n=1}^{\infty}  \left\{\left(l^2+l + 3 \right) \, f''_{\bar{\mu}_0}(E_{n,l}^{(0)} ) +  \frac{\mathcal{R}_{n,l}}{E_0} \, f'_{\bar{\mu}_0}\!\left(E_{n,l}^{(0)} \right) \right\} \, 
\left[-\frac{\left(E_{n,l}^{(0)}\right)^2}{2mc^2} + E_{n,l}^{(\rm D)} - \Delta \mu \right]
\nonumber \\
 & -
  \frac{3\pi E_0}{k_\mathrm{F}a \, mc^2}  \sum_{l=0}^{\infty}  \sum_{n=1}^{\infty} 
   \left\{
 4 l \left(l+\frac{1}{2} \right) \left(l+1\right) \, E_0 \, E_{n,l}^{(0)} \, f''_{\bar{\mu}_0}(E_{n,l}^{(0)} )
 - 2 \left( 2 l^3 +3 l^2 + 5 l +2 \right) \, E_{n,l}^{(0)} \, f_{\bar{\mu}_0}'(E_{n,l}^{(0)} )
 \right.
\nonumber \\
 & 
 \left.
 - \left[ 2 \left(l+\frac{1}{2} \right) \, \frac{ \mathcal{R}_{n,l}}{E_0} \,  E_{n,l}^{(0)}  +2l^3+3 l^2 + 7 l +3 \right] f_{\bar{\mu}_0}(E_{n,l}^{(0)})  
 \right\}
 \, .
 \end{align}
\end{widetext}

Similarly to the discussion presented after Eq.~\eqref{eq:finalexpressionchik1}, we remark that the first sum in Eq.~\eqref{eq:deltachitotep} corresponds to the zero-field weakly-relativistic correction of the eigenenergies arising from the kinetic and Darwin terms. It could then be directly obtained by implementing, in the nonrelativistic ZFS expressions  \eqref{eq:chi_Zeeman}--\eqref{eq:chi_paraanddia}, the shift from $E^{(0)}_{n,l}$ to $E^{(0)}_{n,l}[1-E^{(0)}_{n,l}/2mc^2]+E^{(\rm D)}_{n,l}$, with the corresponding renormalization $\Delta \mu$ of the chemical potential. The first contribution in the second sum of Eq.~\eqref{eq:deltachitotep} corresponds to the effect of $\mathcal{H}^{(\rm so)}$, while the remaining contributions arise from the magnetic-field dependence of $\mathcal{H}^{(\rm k)}$ and $\mathcal{H}^{(\rm ame)}$.

\subsection{Numerical evaluation of the relativistic corrections}
\label{sec:nasrczfs}

When we evaluate numerically the weakly-relativistic correction $\Delta \chi$, according to Eq.~\eqref{eq:deltachitotep}, as a function of $k_{\rm F}a$, 
we find no noticeable difference with the results for the kinetic correction $\Delta \chi^{(\rm k)}$ on the scale of Fig.~\ref{fig:chideltak}. We therefore conclude that $\Delta \chi$ is dominated by $\Delta \chi^{(\rm k)}$.

\begin{figure*}[tb]
\begin{center}
\includegraphics[width=.9\linewidth]{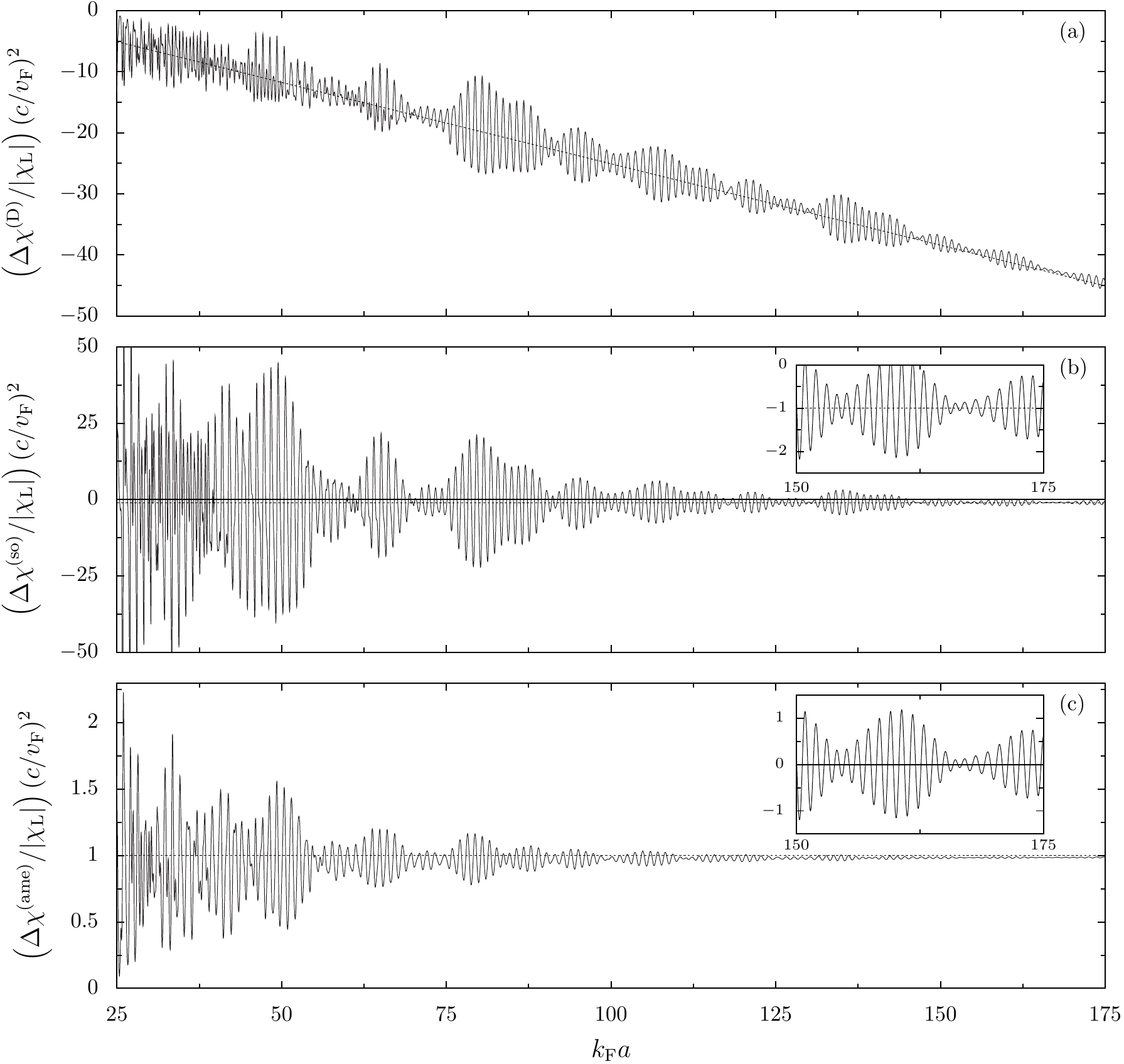}
\caption{\label{fig:deltachiDsoame} 
Weakly-relativistic corrections to the zero-field susceptibility arising from the (a) Darwin, (b) spin-orbit, and (c) angular magneto-electric components, for $k_\mathrm{B}T /{\bar \mu}_0 = 5 \times 10^{-3}$, obtained from Eq.~\eqref{eq:deltachi}, as a function of the nanoparticle radius $a$, using the same scaling and physical parameters as in Fig.~\ref{fig:chideltak}. The dashed lines represent the mean (smooth), leading-order in $k_\mathrm{F}a$, values of the different corrections according to Eq.~\eqref{eq:deltachitot2}. The insets present the corrections $\Delta \chi^{(\rm so)}$
[panel (b)]  and 
$\Delta \chi^{(\rm so-ame)}=\Delta \chi^{(\rm so)}+\Delta \chi^{(\rm ame)}$ [panel (c)] corresponding to large sizes (in the same units as in the main panels), showing the approach to the values predicted  by the one-dimensional semiclassical approach given in Eq.~\eqref{eq:deltachitot2}.}
\end{center}
\end{figure*}

As discussed at the end of the last section, it is straightforward to disentangle in $\Delta \chi$ the different contributions arising from the various Hamiltonian components of Eq.~\eqref{eq:Hamiltonian_Delta}, and thus we write 
\begin{equation}
\Delta \chi = \Delta\chi^{(\rm k)} + \Delta \chi^{(\rm D)} + \Delta \chi^{(\rm so)} + \Delta \chi^{(\rm ame)}  \, ,
\end{equation} 
where the kinetic correction $\Delta\chi^{(\rm k)}$ is given by Eq.~\eqref{eq:finalexpressionchik1}. The corrections arising from the energy shift associated with the Darwin term, the spin-orbit coupling, and the 
angular magneto-electric effect are given, respectively, by  
\begin{widetext}
\begin{subequations}
\label{eq:deltachi}
\begin{align}
\label{eq:deltachiD}
\frac{\Delta \chi^{(\rm D)}}{|\chi_\mathrm{L}|}  & =  -\frac{6\pi \, E_0}{k_\mathrm{F}a} \, \sum_{l=0}^{\infty} \left(l+\frac{1}{2} \right) \sum_{n=1}^{\infty} \left\{ \left(l^2+l+3\right) \, f''_{\bar \mu_0}\!\left(E^{(0)}_{n,l}\right) + \frac{\mathcal{R}_{n,l}}{E_0} \, f'_{\bar \mu_0}\!\left(E^{(0)}_{n,l}\right) \right\} \left[ E^{(\rm D)}_{n,l} - \Delta \mu^{(\rm D)} \right] \, ,
\\
\label{eq:deltachiso}
\frac{\Delta \chi^{(\rm so)}}{|\chi_\mathrm{L}|}  & =  -\frac{12\pi \, E_0^2}{k_\mathrm{F}a \, mc^2} \, \sum_{l=1}^{\infty} l \, \left(l+\frac{1}{2} \right) \left(l+1\right) \sum_{n=1}^{\infty} E^{(0)}_{n,l} \, f''_{\bar \mu_0}\!\left(E^{(0)}_{n,l}\right) \, ,
\\
\label{eq:deltachiame}
\frac{\Delta \chi^{(\rm ame)}}{|\chi_\mathrm{L}|}  & = - \frac{12 \pi \, E_0}{k_\mathrm{F}a \, mc^2} \, \sum_{l=1}^{\infty} \left(l+\frac{1}{2} \right) \sum_{n=1}^{\infty} E^{(0)}_{n,l} \, f'_{\bar \mu_0}\!\left(E^{(0)}_{n,l}\right) \, .
\end{align}
\end{subequations}
\end{widetext}

The behavior of $\Delta\chi^{(\rm k)}$, including its asymptotic dependence in $(k_\mathrm{F}a)^0$, has been discussed in Sec.~\ref{sec:kedc}. The numerical evaluation of the other weakly-relativistic corrections to the ZFS, given by Eq.~\eqref{eq:deltachi}, is presented in the three panels of Fig.~\ref{fig:deltachiDsoame}. In all three cases, the typical values are much smaller than those of $\Delta\chi^{(\rm k)}$, and we observe oscillations as a function of $k_\mathrm{F}a$ around mean values. Using a one-dimensional semiclassical treatment, analogous to that of the nonrelativistic ZFS (see Appendix \ref{app:sc}) and the kinetic correction (see Sec.~\ref{sec:kedc}), we can evaluate the leading-order corrections in $k_\mathrm{F}a$ in the three cases.  For the Darwin, spin-orbit, and angular magneto-electric contributions, in leading-order in $k_\mathrm{F}a$, we have, respectively, 
\begin{subequations}
\label{eq:deltachitot2}
\begin{equation}
\label{eq:deltachid2}
\frac{\overline{\Delta \chi}^{(\rm D)}}{|\chi_\mathrm{L}|}  = -  \frac{1}{5} \, \left(u^{-1}-2\right) \left(\frac{v_\mathrm{F}}{c}\right)^2  \, ,
\end{equation}
\begin{equation}
\label{eq:deltachiSOC2}
\frac{\overline{\Delta \chi}^{(\rm so)}}{|\chi_\mathrm{L}|}  =  - \left(\frac{v_\mathrm{F}}{c}\right)^2
 \, ,
\end{equation}
\begin{equation}
\label{eq:deltachiAME2}
\frac{\overline{\Delta \chi}^{(\rm ame)}}{|\chi_\mathrm{L}|}  =  \left(\frac{v_\mathrm{F}}{c}\right)^2
 \, ,
\end{equation}
\end{subequations}
with $u=\sqrt{E_0/V_0}$.

Using the physical parameters of gold, we have $\overline{\Delta \chi}^{(\rm D)}/|\chi_\mathrm{L}|  =  - 0.27 \, \left(v_\mathrm{F}/c\right)^2 \left(k_\mathrm{F}a\right)$, which is in good agreement with the slope associated to the secular behavior of the very small oscillations present in Fig.~\ref{fig:deltachiDsoame}(a). The unbounded behavior obtained for large $a$ is unphysical, since it prevents of achieving the bulk value $\chi^{(\rm wr)}_{(\rm b)}$. As discussed in Appendix \ref{app:mebsp}, our perturbative approach is problematic for the Darwin term when treating the discontinuous electric field resulting from an abrupt electrostatic potential that confines fermions in a reduced (bag) region of space \cite{greiner1990relativistic}.  

According to Eqs.~\eqref{eq:deltachiSOC2} and \eqref{eq:deltachiAME2}, we have $\overline{\Delta \chi}^{(\rm so-ame)}=0$  to leading order in $k_\mathrm{F}a$, in agreement with the numerical results [see inset in Fig.~\ref{fig:deltachiDsoame}(c)] and with the expectation that in the infinite-volume limit the spin-orbit and magneto-electric couplings become irrelevant, and Eq.~\eqref{eq:weakly-rela_sus} accounts for the weakly-relativistic ZFS. The cancelation between the mean values of $\Delta \chi^{(\rm so)}$ and $\Delta \chi^{(\rm ame)}$ occurs despite the fact that
the oscillations of the former are one order of magnitude larger than those of the latter. 
The comparison between Figs.~\ref{fig:chideltak} and \ref{fig:deltachiDsoame}(b) indicates that $\Delta \chi^{(\rm so)}$ is typically
more than an order of magnitude smaller than $\Delta \chi^{(\rm k)}$.

The smallness of the SOC contribution is understandable in view of the high symmetry of the spherical geometry. In Appendix \ref{app:sdsoc} we present an alternative semiclassical evaluation of this correction, which confirms the results obtained in the previous sections and allows us to highlight the dependence of the SOC effects on the different system parameters. In particular, the spin-orbit correction to the ZFS \eqref{eq:chi_SOC} exhibits the expected $(v_\mathrm{F}/c)^2$ dependence, as well as oscillations between diamagnetic and paramagnetic values with typical amplitudes scaling as $(k_\mathrm{F}a)^{1/2}$, which are out-of-phase with respect to the oscillations of the nonrelativistic ZFS. The previous typical amplitudes should be contrasted with the $(k_\mathrm{F}a)^{3/2}$ dependence of the nonrelativistic ZFS of an individual nanoparticle and the $k_\mathrm{F}a$ dependence resulting from an ensemble average \cite{viloria2018orbital}.
The suppression of the oscillations of $\Delta \chi^{(\rm so)}$ for large nanoparticle sizes is faster than that of $\Delta \chi^{(\rm nr)}$
since in the first case, in addition to the thermal damping \eqref{eq:R_T} acting on the contribution of each family of classical periodic orbits, 
we have the fact that the SOC becomes comparatively weaker as $a$ increases. Indeed, from Eqs.~\eqref{eq:E^0} and \eqref{eq:EpmSO} we have $E_{n,l,(\pm)}^{({\rm so})}/E^{(0)}_{n,l}=[\mp(l+1/2)-1/2]E_0/mc^2$, indicating that the relative importance of the SOC decreases with $a$.

\subsection{Average zero-field  susceptibility}
\label{sec:azfs}

In the previous section we established that the SOC contribution to the ZFS of an individual nanoparticle is much smaller than its nonrelativistic counterpart, as well as the kinetic-energy correction. Since the measurements are typically not done on individual objects, but on ensembles with a large number of nanoparticles, we need to consider the average magnetic response, which, as discussed in Sec.~\ref{sec:msffs}, follows from the finite-$N$ correction to the free energy $\Delta F^{(2)}$ [cf.\ Eq.~\eqref{eq:F2}]. The semiclassical description of the spin-orbit coupling, developed in Appendix \ref{app:sdsoc} as an alternative approach to quantum perturbation theory, becomes extremely useful for calculating $\Delta F^{(2)}$, expressed in terms of the oscillating part of the DOS. Since we are interested in the effect of the SOC, we ignore in the remaining of this section the other weakly-relativistic corrections. Performing the energy integration in Eq.~\eqref{eq:F2} we write
\begin{widetext}
\begin{align}
\label{eq:rho_oscsoc5}
\Delta F^{(2)}  =&\; \frac{1}{2{\bar \varrho}({\bar \mu},0)} 
\left( \frac{\hbar v_\mathrm{F}}{E_0} \right)^2 \frac{k_\mathrm{F}a}{\pi} 
\Bigg\{
\sum_{\substack{\nu>0\\\eta>2\nu}} 
\frac{(-1)^{\nu}}{\sqrt{\eta}} \ \frac{1}{L_{\nu\eta}}
\cos{\varphi_{\nu\eta}} \, \sin^{3/2}{\varphi_{\nu\eta}}
 \ R(L_{\nu\eta}/L_T) 
\nonumber \\ 
&\times 
 \left[- I_{\nu\eta, {\rm c}}({\bar \mu},B) \, \sin{\big(\theta_{\nu\eta}(k)\big)} + 
I_{\nu\eta, {\rm s}}({\bar \mu},B) \, \cos{\big(\theta_{\nu\eta}(k)\big)} \right]
\Bigg\}^2
\, ,
\end{align}
\end{widetext}
where the symbols $I_{\nu\eta, {\rm c}}$ and $I_{\nu\eta, {\rm s}}$ are defined in Eq.~\eqref{eq:I_integrals}. The associated ZFS follows from the magnetic-field dependence of these symbols. To leading order in $v_\mathrm{F}/c$ we have 
$I_{\nu\eta, {\rm c}}({\bar \mu},0)=4$, $I_{\nu\eta, {\rm s}}({\bar \mu},0)=0$, and
$(\partial I_{\nu\eta, {\rm c}}/\partial B)|_{B=0}=(\partial I_{\nu\eta, {\rm s}}/\partial B)|_{B=0}=0$, while the second derivatives are given in Eq.~\eqref{eq:BderIcs2}.

The ensemble average ZFS will be dominated by the diagonal terms in the double sum over families of trajectories, and its nonrelativistic component resulting from \eqref{eq:rho_oscsoc5} is given by 
\begin{align}
\label{eq:chi_av-nr}
\frac{\chi^{(2)-({\rm nr})} }{|\chi_\mathrm{L}|}=&\;36 \pi \ k_\mathrm{F}a
\sum_{\substack{\nu>0\\\eta>2\nu}}
\frac{1}{\eta} \, \cos^4{\varphi_{\nu\eta}} \ \sin^{3}{\varphi_{\nu\eta}} 
\nonumber\\
&\times
\left[1+ \frac{3}{ \cos^2{\varphi_{\nu\eta}}} 
\left( \frac{1}{k_\mathrm{F}a} \right)^2 \right]
\nonumber\\
&\times
 R^{2}(L_{\nu\eta}/L_T) \,
\sin^{2}{\big(\theta_{\nu\eta}(k_\mathrm{F})\big)} \, .  
\end{align} 
%
$\chi^{(2)-({\rm nr})}$ exhibits a paramagnetic character and the expected period-halving in $k_\mathrm{F}a$ with respect to the ZFS of an individual nanoparticle. Its leading-order contribution in $k_\mathrm{F}a$ 
[first term in the square brackets of Eq.~\eqref{eq:chi_av-nr}] represents the diagonal approximation to the average orbital response calculated in 
Ref.~\cite{viloria2018orbital} and reproduced by the red line of Fig.~\ref{fig:chinr} (once the nonrelativistic bulk  ZFS $\chi_\mathrm{b}^{(\rm nr)}$ is included). The other component [second term in the square brackets of Eq.~\eqref{eq:chi_av-nr}] stands from the contribution of the Pauli susceptibility. However, it is not significant within our calculation scheme, as it is of an order in $k_\mathrm{F}a$ which is not taken into account in the semiclassical approximation of the orbital effects. The remaining contribution to the ZFS resulting from $\Delta F^{(2)}$ in Eq.~\eqref{eq:rho_oscsoc5} is the correction $\Delta \chi^{(2)-({\rm so})}$ associated to the SOC, that can be expressed as 
\begin{align}
\label{eq:chi_av-soc}
\frac{\Delta \chi^{(2)-({\rm so})}}{|\chi_\mathrm{L}|}=&\; - \frac{9\pi}{16}  
\left(\frac{v_\mathrm{F}}{c} \right)^2
\sum_{\substack{\nu>0\\\eta>2\nu}} \sin^{4}{(2\varphi_{\nu\eta})} 
\nonumber\\
&\times R^{2}(L_{\nu\eta}/L_T) \,
\sin{\big(2\theta_{\nu\eta}(k_\mathrm{F})\big)} \, . 
\end{align} 
$\Delta \chi^{(2)-({\rm so})}$ also presents the period-halving in $k_\mathrm{F}a$ characteristic of an average response, but it does not have a definite sign. Like in the case of the SOC contribution of an individual nanoparticle discussed in Appendix \ref{app:sdsoc}, $\Delta \chi^{(2)-({\rm so})}$ is much smaller than its nonrelativistic counterpart \eqref{eq:chi_av-nr}, due to the prefactor $(v_\mathrm{F}/c)^2$ and to one less power of  $k_\mathrm{F}a$. In a nanoparticle ensemble with an important size dispersion, the oscillating term $\sin^{2}{\big(\theta_{\nu\eta}(k_\mathrm{F})\big)}$ in Eq.~\eqref{eq:chi_av-nr} averages to $1/2$ and we obtain the result $\chi_\mathrm{ens}^{\rm d}$ plotted as a blue line in Fig.~\ref{fig:chinr} (once the nonrelativistic bulk  ZFS $\chi_\mathrm{b}^{(\rm nr)}$ is included), while the contribution $\Delta \chi^{(2)-({\rm so})}$ vanishes, and thus does not provide an SOC correction to $\chi_\mathrm{ens}^{\rm d}$.

\section{Half-spherical nanoparticles}
\label{sec:hsn}

As discussed at the end of Sec.~\ref{sec:nasrczfs}, the high symmetry of the spherical potential translates into the smallness of the SOC contribution to the ZFS. Thus, a reduction of these symmetries appears as a way to boost the relative importance of the SOC. A first step in the process of progressive destruction of symmetries is to consider a half-spherical confining potential
with a magnetic field directed along the axial axis. Such a geometry has the advantage that its nonrelativistic eigenstates at zero magnetic field can be identified as a subset of those of the sphere \cite{chang03_PRE}, and many of the analytical developments performed for the case of the sphere can be readily adapted for the case in which the magnetic field is applied along the symmetry axis. Moreover, it is in metallic nanoparticles with the approximate shape of a half-sphere (HS) that the smallest $g$-factors (as low as 0.3) have been reported \cite{davidovic1999}. 

The kinetic and Darwin corrections are expected to change minimally when trading the spherical geometry by the HS, while the angular magneto-electric effect was shown, for the sphere, to be typically one order of magnitude smaller than the one arising from the SOC. Therefore, in this section we concentrate on the effect of the SOC for the ZFS  for the reduced symmetry case of half-spherical nanoparticles. Since the total angular momentum is no longer a conserved quantity, the use of the coupled basis is of no aid to treat the SOC, and thus we present our calculations in the product basis, appealing to a numerical diagonalization once the SOC is included in the Hamiltonian.

Similar considerations to those discussed in Sec.~\ref{sec:model}, concerning the passage from the ideal model of electrons confined in a sphere to the case of realistic nanoparticles, also apply for the HS geometries studied in this section.

\subsection{Nonrelativistic susceptibility}
\label{sec:nrshs}

\begin{figure}[t]
\begin{center}
\includegraphics[width=.9\linewidth]{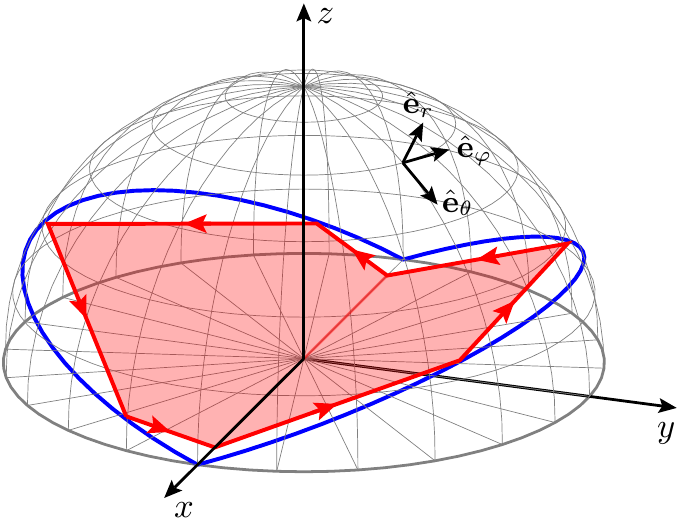}
\caption{\label{fig:half-sphere}
Half-spherical geometry with the coordinate axes used in the text. In red: Classical periodic orbit with the same topological indices $(\nu,\eta)= (1,5)$ than in the case of Fig.~\ref{fig:sphere} for the sphere, but now contained in two planes (whose intersections with the HS are indicated by the two blue semi-circles).}
\end{center}
\end{figure}

We consider noninteracting electrons, described by a Hamiltonian $\mathcal{H}^{(0)}$ of the form \eqref{eq:H_00}, confined by a potential with the shape of a HS, which, with the choice of coordinates of Fig.~\ref{fig:half-sphere}, writes
\begin{equation}
\label{eq:confining_potential_hs}
V(r,\theta) = V_0 \left[ \Theta(r-a) + \Theta(\theta-\pi/2) \right] \, .
\end{equation}
For finite $V_0$ such a potential does not lead to a separable Schrödinger equation, and it has the unphysical feature of taking two different values ($V_0$ and $2V_0$) in the classically forbidden region. However, these shortcomings are no longer found in the limit $V_0 \rightarrow \infty$ of hard walls, where the corresponding eigenstates of $\mathcal{H}^{(0)}$ are characterized by the set of quantum numbers $\{\lambda\}=\{n,l,m_z,m_s\}$ with the condition $l+m_z$ odd, and the associated spinors $\Psi_{n,l,m_z,m_s}^{(\rm HS)}(\mathbf{r})$ have the form \eqref{eq:nlmzmsud} with the orbital wave function given by 
\begin{equation}
\label{eq:orbitalhs}
\psi_{n,l,m_z}^{(\rm HS)}(\mathbf{r})=\sqrt{2} \, R_{n,l}(r) \, Y^{m_z}_l(\vartheta)\; .
\end{equation}
The radial wave function $R_{n,l}(r)$ is given in Eq.~\eqref{eq:rwfs}. The condition $l+m_z$ being odd arises since only the eigenstates of the sphere which are odd with respect to the reflection off the $z=0$ plane can be eigenstates of the HS, together with the property $Y_l^{m_z}(\pi-\theta,\varphi)=(-1)^{l+m_z}Y_l^{m_z}(\theta,\varphi)$ fulfilled by the spherical harmonics. We remark that the previous restriction excludes the isotropic $l=0$ states. The corresponding eigenenergies $E^{(0)}_{n,l}$ are given by Eq.~\eqref{eq:E^0}.

In the nonrelativistic case, the application of a magnetic field $\mathbf{B} =  B\, {\bf {\hat e}}_z$ leads to the same energy corrections of Eqs.~\eqref{eq:Enr}--\eqref{eq:Enrcomponents}. The ZFS of the half-sphere follows from Eq.~\eqref{def:ZFS} by summing over $m_z$ and $m_s$ with the restriction of $l+m_z$ being odd. Including the spin-dependent component, we have 
\begin{align}
\frac{\chi^{(\rm nr)-(HS)}}{|\chi_\mathrm{L}|} =& - \frac{6\pi}{k_\mathrm{F}a}\sum_{l=1}^\infty \sum_{n=1}^\infty \, 
	 l\bigg\{(l^2+2) E_0 f'_{\bar \mu}(E^{(0)}_{n,l})
	 \nonumber\\
	 &+\frac{(l^2-3l/4-1)}{(l-1/2)(l+3/2)} \, \mathcal{R}_{n,l} \, f_{\bar \mu}(E^{(0)}_{n,l}) \bigg\} \, .
\end{align}

The numerical evaluation of the above expression for the nonrelativistic ZFS of the HS leads to a result (not shown) that is indistinguishable from that of the sphere (see the black solid line in Fig.~\ref{fig:chinr}) for the  presented $k_\mathrm{F}a$-interval, while very small differences appear for smaller values of $k_\mathrm{F}a$. For energies corresponding to the first eigenvalues there 
is a phase shift of half-period in the $k_\mathrm{F}a$ oscillations between the ZFS of the sphere
and the half-sphere \cite{fraue98_PRB}, since the lowest-energy state response is always diamagnetic and the HS spectrum starts at the first excited state of the sphere. But this phase shift does not prevail for larger values of $k_\mathrm{F}a$.

The concordance of the ZFS of the sphere and the HS for not too small values of $k_\mathrm{F}a$ is understandable from a semi-classical viewpoint. Indeed, in  Fig.~\ref{fig:half-sphere} we see that the periodic orbits of the HS can be put in correspondence with those of the sphere by simply symmetrizing one of the two containing planes with respect to the $x\!-\!y$ plane. Since the two containing planes are associated with the same angle $\beta$ that defines their orientation with respect to the $z$-axis, Eq.~\eqref{eq:rho_oscsoc2} gives account of the oscillating part of the density of states of the HS (up to a factor of $1/2$ arising from the double counting of the periodic orbits of the sphere). In the absence of SOC the integrations over the angles $\alpha$ and $\gamma$ follow like in the case of the sphere, leading to Eqs.~\eqref{eq:rho_oscsoc4} and \eqref{eq:BderIcs2a}, from where the expression \eqref{eq:chi_1} for the oscillating part of the ZFS $\chi^{(\rm orb)-osc}$ follows (once a factor of $2$ arising from the reduced volume of the HS is included).

\subsection{Spin-orbit coupling for a HS confining potential}
\label{sec:sochs}

In the hard-wall limit of the potential \eqref{eq:confining_potential_hs} defining a half-spherical box, the spin-orbit coupling \eqref{eq:H_so3} is given by the Hamiltonian
\begin{equation}
\mathcal{H}^{(\rm so)-(HS)}=\mathcal{H}^{(\rm dome)}+ \mathcal{H}^{(\rm floor)} \, , 
\end{equation}
where
\begin{subequations}
\begin{align} 
\mathcal{H}^{(\rm dome)}& = \frac{1}{2m^2c^2} \, \frac{V_0}{r} \, \delta(r-a) \, \mathbf{S} \cdot \mathbf{L}  \, ,
\label{eq:H_sosh}
\\ 
\mathcal{H}^{(\rm floor)}&= \frac{1}{2m^2c^2} \, \frac{\hbar V_0}{{\rm i} r} \, \delta\left(\theta-\frac{\pi}{2}\right) 
\nonumber\\
&\hspace{.4cm}\times\mathbf{S} \cdot \left( - {\bf {\hat e}}_{\varphi} \partial_{r} + {\bf {\hat e}}_{r} \frac{1}{r \sin{\theta}} \partial_{\varphi} \right)  \, ,
\label{eq:H_sofl}
\end{align}
\end{subequations}
represent, respectively, the effect of the dome and the floor of the confining potential. 

Writing $\mathbf{S} \cdot \mathbf{L}=(S_{+}L_{-}+S_{-}L_{+})/2 + S_{z}L_{z}$, we see that the contribution $S_{z}L_{z}$ leads to spin-conserving matrix elements of $\mathcal{H}^{(\rm dome)}$ which are diagonal in the indices $m_z$ and $m_s$. That is, 
\begin{align}
\label{eq:H_soshd}
\langle \Psi_{n',l',m_z,m_s}^{(\rm HS)} & |\mathcal{H}^{(\rm dome)}| \Psi_{n,l,m_z,m_s}^{(\rm HS)}  \rangle   = \nonumber\\ 
& (-1)^{(n-n')} \, m_z \, m_s \, \delta_{l',l} \, \zeta_{n',l} \, \zeta_{n,l} \,
\frac{2E_0^2}{mc^2} \, ,
\end{align}
where we have used that  $\mathrm{sgn}\left[j_{l}^{\prime}(\zeta_{n,l})\right] = (-1)^n$, independently of the value of $l$. 
The remaining terms of $\mathbf{S} \cdot \mathbf{L}$ result in spin-flip matrix elements of $\mathcal{H}^{(\rm dome)}$ which are nondiagonal  in the indices $m_z$ and $m_s$, i.e.,
\begin{widetext}
\begin{align}
\langle \Psi_{n',l',m_z^{\prime},m_s^{\prime}}^{(\rm HS)} & |\mathcal{H}^{(\rm dome)}| \Psi_{n,l,m_z,m_s=\pm1/2}^{(\rm HS)}  \rangle = \nonumber\\ 
& (-1)^{(n-n')} \, \delta_{m_z^{\prime},m_z \pm 1} \, \delta_{m_s^{\prime},-m_s} \, \zeta_{n',l'} \, \zeta_{n,l} \, \sqrt{(l\mp m_z^{\prime} +1)(l \pm m_z^{\prime})} \, 
\mathcal{I}^{(\rm dome)}_{l',l,m'_z} \,
\frac{2E_0^2}{mc^2} 
\, ,
\label{eq:H_soshnd}
\end{align}
where we have defined
\begin{equation}
\mathcal{I}^{(\rm dome)}_{l',l,m'_z} = \int_{\rm dome}\mathrm{d}\vartheta  \left(Y_{l'}^{m_z'}(\vartheta)\right)^{*} Y_l^{m_z^{\prime}}(\vartheta) \, .
\label{eq:angint3}
\end{equation}
Since $l+m_z$ and $l'+m_z^{\prime}$ are both odd, the two spherical harmonics in  the previous equation have different parities. Therefore, the integral over the dome does not trivially vanish in the case that interests us, where $m_z^{\prime} = m_z \pm 1$, and then $l$ and $l'$ have a different parity. Equation \eqref{eq:angint2}, in Appendix \ref{app:mehs}, provides the result of the integral \eqref{eq:angint3}.

The Hamiltonian component \eqref{eq:H_sofl} writes
\begin{equation}
\mathcal{H}^{(\rm floor)}= \frac{1}{4m^2c^2} \, \frac{\hbar^2 V_0}{{\rm i} r} \, \delta\left(\theta-\frac{\pi}{2}\right) \left( \left(\sigma_{x} \sin\varphi - \sigma_{y} \cos\varphi  \right) \partial_{r} +  \frac{1}{r} \left(\sigma_{x} \cos\varphi + \sigma_{y} \sin\varphi + \sigma_{z} \cot\theta \right) \partial_{\varphi} \right)  \, ,
\label{eq:H_sofl2}
\end{equation}
and thus, the condition $\delta(\theta-\pi/2)$ results in vanishing spin-conserving matrix elements, while the nondiagonal, spin-flip matrix elements are given, in the limit of large $V_0$, by
\begin{align}
\langle \Psi_{n',l',m_z^{\prime},m_s^{\prime}}^{(\rm HS)} |\mathcal{H}^{(\rm floor)}|  \Psi_{n,l,m_z,m_s=\pm1/2}^{(\rm HS)}  \rangle =&\;  
\frac{\hbar^2 V_0}{4m^2c^2} \, \delta_{m_s^{\prime},-m_s} 
\int_{0}^{a} {\rm d}r \, r \int_{0}^{2\pi} {\rm d}\varphi \, \exp{(\pm {\rm i} \varphi)} 
\, \psi_{m_z^{\prime}}^{(V_0,n',l')}(r,\theta\!=\!\pi/2,\varphi)
\nonumber\\ 
& 
\times
\left\{ \mp \, \partial_{r}  + \frac{1}{{\rm i} r} \partial_{\varphi} \right\} 
\, \psi_{m_z}^{(V_0,n,l)}(r,\theta\!=\!\pi/2,\varphi)
\, .
\label{eq:H_sofloornd}
\end{align}
where $\psi_{m_z}^{(V_0,n,l)}(r,\theta,\varphi)$ and $\psi_{m_z^{\prime}}^{(V_0,n',l')}(r,\theta,\varphi)$ converge, respectively, to the orbital wave functions $\psi_{n,l,m_z}(r,\theta,\varphi)$ and $\psi_{n',l',m_z^{\prime}}(r,\theta,\varphi)$ when $V_0 \rightarrow \infty$. Thus, the limiting condition \eqref{eq:wfv2db} and the form \eqref{eq:rwfs} of the radial wave function for the hard-wall case allow us to write
\begin{align}
\langle \Psi_{n',l',m_z^{\prime},m_s^{\prime}}^{(\rm HS)} |\mathcal{H}^{(\rm floor)}|  \Psi_{n,l,m_z,m_s=\pm1/2}^{(\rm HS)}  \rangle
=&\; \pm \frac{(-1)^{(l+l'+m_z+m_z^{\prime})/2}}{2^{l+l'}} \, \delta_{m_z^{\prime},m_z \pm 1} \, \delta_{m_s^{\prime},-m_s} \, \frac{4E_0^2}{mc^2} \, \mathcal{J}_{n',l',n,l,m_z} 
\nonumber\\ 
& \times
\,  \frac{ \sqrt{(2l+1)(2l'+1)}}{|j_{l'+1}(\zeta_{n',l'}) \, j_{l+1}(\zeta_{n,l})|} \,
\frac{\sqrt{(l+m_z)! \, (l-m_z)!}}{(\frac{l+m_z-1}{2})! \, (\frac{l-m_z-1}{2})! }  \,
\frac{\sqrt{(l^{\prime}+m_z^{\prime})! \, (l^{\prime}-m_z^{\prime})!}}{(\frac{l^{\prime}+m_z^{\prime}-1}{2})! \, (\frac{l^{\prime}-m_z^{\prime}-1}{2})! } 
\, ,
\label{eq:H_sofloornd2}
\end{align}
where $l$ and $l'$ have different parities and we have expressed the integral over the radial coordinate through
\begin{equation}
\mathcal{J}_{n',l',n,l,m_z}   = 
\int_{0}^{1} \frac{{\rm d}\zeta}{\zeta} \, j_{l'}(\zeta_{n',l'} \zeta) 
\left(\frac{l  \mp m_z-1}{\zeta} \, j_{l}(\zeta_{n,l} \zeta) - 
\zeta_{n,l} \, j_{l+1}(\zeta_{n,l} \zeta) \right)
\, .
\label{eq:Ifloor}
\end{equation}
Recalling that $j_{l}(\zeta) \sim \zeta^{l}/(2l+1)!!$ for small values of $\zeta$, and that $l=0$ is not allowed for the HS, we verify that the integral in Eq.~\eqref{eq:Ifloor} is divergence-free.
\end{widetext}

\begin{figure}[tb]
\begin{center}
\includegraphics[width=\linewidth]{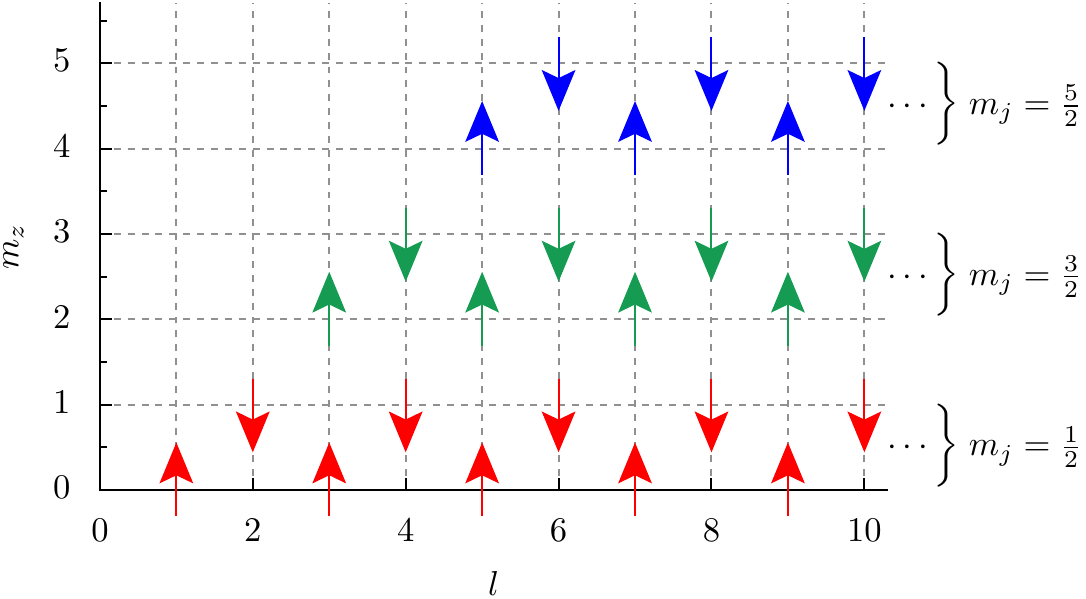}
\caption{\label{fig:colorv2_spinimage} 
The $l$-$m_z$ plane for positive $m_j=m_z+m_s$ for the half-spherical nanoparticle. Each arrow pointing up (down) represents the orientation of the spin $m_s=1/2$ ($m_s=-1/2$) for the allowed states $l+m_z$ odd and $l>m_z$. At zero-field, the Hamiltonian of the half-spherical nanoparticle only couples states with the same $m_j$.}
\end{center}
\end{figure}

The axial symmetry of the HS translates in the conservation of the $z$-component of the total angular momentum. The subspaces with different $m_j=m_z+m_s$ are not coupled by $\mathcal{H}^{(\rm so)-(HS)}$, and within each of these subspaces the coupling between two eigenstates $\Psi_{n,l,m_z,m_s}^{(\rm HS)}$ and $\Psi_{n',l',m'_z,m'_s}^{(\rm HS)}$ of $\mathcal{H}^{(0)}$ only occurs in the following cases: (i) $\{l',m'_z,m'_s\}=\{l,m_z,m_s\}$; (ii) $m_s=1/2$ and $\{m'_z,m'_s\}=\{m_z+1,-1/2\}$; (iii) $m_s=-1/2$ and $\{m'_z,m'_s\}=\{m_z-1,1/2\}$ (with $l$ and $l'$ having different parity in the last two cases). These restrictions can be graphically represented in the plane $l$-$m_z$ (see Fig.~\ref{fig:colorv2_spinimage}), where only the subspaces with positive $m_j$ are considered. The diagonalization within each of these subspaces (those with negative $m_j$ follow from Kramer's degeneracy) leads to the eigenstates of $\mathcal{H}^{(0)} + \mathcal{H}^{(\rm so)-(HS)}$
\begin{equation}
\label{eq:PsiHSdiagon}
\Psi_{m_j,p}(\mathbf{r}) = \sum_{\substack{n,l,m_s=\pm 1/2 \\ l >  |m_j-m_s| \\ l+m_j-m_s \neq \dot{2}}} C_{n,l,m_s}^{m_j,p} \, \Psi_{n,l,m_j-m_s,m_s}^{(\rm HS)}(\mathbf{r}) \, ,
\end{equation}
labeled by the half-integer $m_j$ and the positive integer index $p$, with the associated eigenenergies $E_{m_j,p}$.

\subsection{Perturbative treatment of the magnetic field}
\label{sec:ptmfHS}

Under the application of a weak magnetic field $B$, the energy $E_{m_j,p}$ picks up  a correction $\delta E_{m_j,p}$. According to Eq.~\eqref{def:ZFS}, the latter determines the ZFS of the HS through the parameters 
\begin{subequations}
\label{eq:derivativesenHSsoc}
\begin{align}
\label{eq:derivativesenHSsoca}
\left. \frac{\partial \, \delta E_{m_j,p}}{\partial B}\right|_{B=0}  =&\;  \mu_{\rm B}
\sum_{n,l,m_s} \left| C_{n,l,m_s}^{m_j,p} \right|^2 \,  \left(m_j+ m_s \right) \, , \\
\label{eq:derivativesenHSsocb}
 \left.\frac{\partial^2 \, \delta E_{m_j,p}}{\partial B^2}\right|_{B=0}    =&\;  \frac{\mu_{\rm B}^2}{2E_0}  \sum_{n,l,m_s}  \, \sum_{n'} 
 \sum_{i=-1}^{+1}
 \left( C_{n',l+2i,m_s}^{m_j,p} \right)^{*}
 \nonumber\\
 &\times C_{n,l,m_s}^{m_j,p}
 \mathcal{R}_{n',l+2i,n,l} \,  \mathcal{Y}_{l,i}^{m_j-m_s} 
 \nonumber\\
&\times \Theta\left(l+2i - |m_j-m_s|\right)
  \, ,
\end{align}
\end{subequations}
where the sums over $\{n,l,m_s\}$ have the same restrictions as in Eq.~\eqref{eq:PsiHSdiagon}, and the Heaviside function prevents the consideration of unphysical terms having $l'=l-2$. The angular matrix elements are calculated over the whole sphere, where the diagonal ones $\mathcal{Y}_{l,0}^{m_z} = \mathcal{Y}_{l}^{m_z}$ are given by Eq.~\eqref{eq:Y}, while the nonvanishing off-diagonal  ($l'=l \pm 2$) ones can be expressed, for $i=\pm 1$, as
\begin{align}
\mathcal{Y}_{l,i}^{m_z} =&\; - \frac{1}{4(l+1/2+i)} 
\nonumber\\
&\times
\sqrt{\frac{\left[(l+1+i)^2-m_z^2\right] \left[(l+i)^2-m_z^2\right]}{(l+3/2+i)(l-1/2+i)}}
\, .
\end{align}
The radial matrix elements are
\begin{equation}
\label{eq:Rnd}
\mathcal{R}_{n',l',n,l} = \frac{1}{a^2} \int_{0}^{a} {\rm d}r \, r^4 \, R_{n',l'}(r) \, R_{n,l}(r)
\, .
\end{equation}
The diagonal ones $\{n',l'\}=\{n,l\}$ are given by Eq.~\eqref{eq:R}, while the off-diagonal ones can be obtained by numerical integration or by recurrence formulas, as shown in Appendix \ref{app:mes}.

\subsection{Spin-orbit correction for the half-sphere}
\label{sec:nespchs}

\begin{figure*}[tb]
\begin{center}
\includegraphics[width=.9\linewidth]{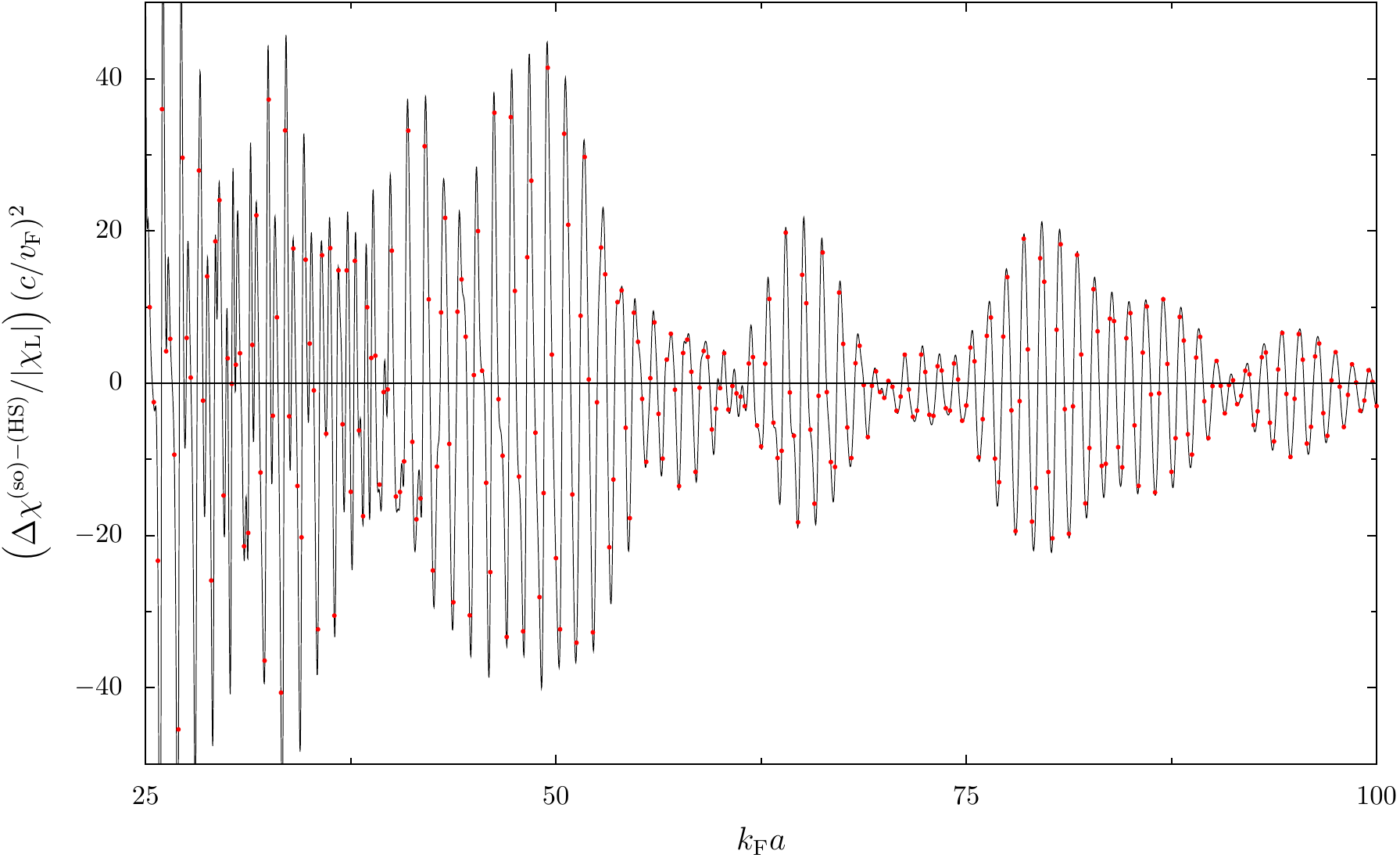}
\caption{\label{fig:SOC-half-sphere}
Red dots: SOC correction $\Delta \chi^{\rm (so)-(HS)}$ to the nonrelativistic ZFS $\chi^{\rm (nr)-(HS)}$ in a half-sphere, from Eqs.~\eqref{def:ZFS} and \eqref{eq:derivativesenHSsoc}, as a function of $k_{\rm F}a$.
Black line: reproduction of the result of Fig.~\ref{fig:deltachiDsoame}(b) obtained for the sphere using Eq.~\eqref{eq:deltachiso}.}
\end{center}
\end{figure*}

In Fig.~\ref{fig:SOC-half-sphere} we present the numerical evaluation of the SOC correction $\Delta \chi^{\rm (so)-(HS)}$ to the nonrelativistic ZFS $\chi^{\rm (nr)-(HS)}$, according to Eqs.~\eqref{def:ZFS} and \eqref{eq:derivativesenHSsoc}, as a function of $k_{\rm F}a$ (red dots). 
These results are indistinguishable from those corresponding to the SOC correction for the sphere presented in Fig.~\ref{fig:deltachiDsoame}(b) 
and reproduced by the black line of Fig.~\ref{fig:SOC-half-sphere}. The symmetry reduction in going from the sphere to the HS is thus not 
enough to yield an enhanced SOC correction in the last case. Upon increasing $a$, the SOC mixing of energy levels of the HS with different 
$(n,l)$ is favored. But such a tendency is countered by the relative weakening of the SOC matrix elements, as discussed at the end of Sec.~\ref{sec:nasrczfs}.

The ZFS correction $\Delta \chi^{\rm (so)-(HS)}$, which is very close to the corresponding correction in the sphere, can be very well approximated by only considering the energy shifts arising from only the diagonal (in $m_z$, $m_s$, and $n$) matrix elements \eqref{eq:H_soshd}, as well as the corresponding diagonal SOC matrix elements for the sphere in the product basis (not shown). These accordances indicate that for the sphere, 
as well as for the HS, for the considered $k_\mathrm{F}a$ values, the SOC can be accounted for in first-order perturbation theory.

\section{Conclusion}
\label{sec:ccl}

Motivated by measurements of the magnetization in ensembles of noble-metal nanoparticles \cite{orrit2013, nealo12_Nanoscale} and the $g$-factor of an individual nanoparticle \cite{salinas1999, davidovic1999}, together with the theoretical proposals pointing to the key role played by the spin-orbit coupling in these experimental results
\cite{brouwer2000, matveev2000, titov2019}, we considered in this work the relevance of such a relativistic effect on the magnetic response of confined electrons. In particular, we attempted to quantify the spin-orbit effect by going beyond the previous phenomenological approaches that assigned arbitrary values to its strength. We focused on the extrinsic SOC originating from the confining potential, treating model systems, and then discussing their applicability to the case of metallic nanoparticles of different shapes. The relevance of the SOC in the magnetic response was gauged against other weakly-relativistic corrections and finite-size effects inherent to the mesoscopic regime. 

For spherical geometries, the inclusion of the SOC and other relativistic effects could be readily done, at the quantum level, by working in the coupled basis of the total angular momentum, within a perturbative treatment of the magnetic field. An equivalent formulation of the SOC correction was derived from a semiclassical approach including spin and orbital effects. 

Analogously to the nonrelativistic zero-field susceptibility characterizing electrons confined in a spherical geometry \cite{viloria2018orbital}, the weakly-relativistic corrections present oscillations as a function of $k_\mathrm{F} a$ between paramagnetic and diamagnetic values, which are typically much larger than the background that sets the bulk response. The oscillations corresponding to the SOC contribution are out-of-phase
with respect to those of the nonrelativistic ZFS.
The typical values of the SOC contribution are 
more than an order of magnitude larger than the ones due to the magneto-electric coupling, and
more than an order of magnitude smaller than those arising from the kinetic energy correction. The latter, being the dominant weakly-relativistic correction to the ZFS, remains much smaller than the nonrelativistic ZFS. These small values of the weakly-relativistic corrections stem from their $(v_\mathrm{F}/c)^2$ dependence, while the reduced effect of the SOC correction is associated with the high spatial symmetry of the spherical geometry.

In order to study the impact of a symmetry reduction on the magnetic response, the formalism developed to treat the SOC in the sphere was adapted to the case of a half-sphere. The nonrelativistic ZFS was shown to be the same for the sphere and the half-sphere in the semiclassical limit of large $k_\mathrm{F} a$, as can be readily understood from semiclassical arguments. The inclusion of the SOC in the HS leads to corrections of the ZFS which are very close to those
of the sphere, and therefore
 much smaller than the correction arising from the kinetic energy shift and the typical values of the nonrelativistic ZFS. The symmetry reduction when going from the spherical geometry to that of the HS is not enough to render relevant the SOC effects due to the electron confinement. 
Even if the HS confinement favors the SOC mixing of the unperturbed states in comparison with the case of the sphere, 
such an effect is offset by the generic suppression of the SOC matrix elements with the parameter $(v_\mathrm{F}/c)^2$ and the size $a$. 
The SOC is thus weak enough to remain perturbative, and therefore it does not induce the transition in the statistical properties of the 
spectra necessary to reverse the sign of the magnetic response of a nanoparticle ensemble. 
 
Further symmetry reduction would lead to more quasi-degeneracies in the $B=0$ spectrum, enhancing the effect of the SOC. However, 
the cases of the quarter- and eighth-sphere, which can be worked out similarly to that of the HS, do not show a significant increase
of the SOC contribution to the ZFS \cite{gomez}. The case of cylindrical nanoparticles is important, as it is the geometry 
considered in Ref.~\cite{orrit2013}. At least in the most symmetric case where the magnetic field is oriented along the cylinder axis, the SOC
due to the confinement seems to have minor importance \cite{gomez_unpublished}. Moving forward 
from integrable to chaotic geometries will increase the importance of the SOC mixing, but at the same time will be accompanied by 
a reduction of the typical values attained by
 the nonrelativistic ZFS \cite{richt96_PhysRep, richt98_EPL}.
 
 Our theoretical results show that the spin-orbit coupling generated by the potential discontinuity
 at the border of a metallic nanoparticle remains far away from the strong-coupling regime 
 \cite{titov2019} where the change of the statistical properties of the spectrum may lead to a 
 diamagnetic response for the case of disordered or classically chaotic systems. The nanorods
 of Ref.~\cite{orrit2013} are considerably larger than the nanoparticles of 
 Refs.~\cite{cresp04_PRL, dutta07_APL, guerr08_Nanotechnology, hori04_PRB} 
 which might result in a stronger extrinsic SOC effect arising from the impurity potentials. However, the high-quality single-crystalline gold employed is not expected to result in important extrinsic
 effects, nor in an underlying chaotic classical dynamics of the conduction electrons. 
 While the SOC contribution to the zero-field susceptibility of an individual nanoparticle is suppressed by a factor $(v_\mathrm{F}/c)^2$ with respect to its nonrelativistic counterpart, the 
 SOC contribution to the average response for ensembles of nanoparticles with an important size dispersion vanishes. As discussed in the introduction, the other mechanisms of SOC (host ionic lattice and impurities) are likewise expected to be ineffective towards driving the transition 
 between statistical ensembles. We thus conclude that spin-orbit effects are by themselves
 not responsible for the large (with respect to the bulk) diamagnetic susceptibility observed in Refs.~\cite{cresp04_PRL, dutta07_APL, guerr08_Nanotechnology, orrit2013, hori04_PRB}.
 
The magnetic response of micro- and nanostructures is characterized by the magnetic susceptibility
in the singly-connected case and by the persistent current in multiply-connected geometries. 
Our work in the first problem contributes to the open discussion about the possible origin 
of the diamagnetic response of a large ensemble of metallic rings \cite{deblo02_PRL}. 
While the typical values of the persistent current measured in the second-generation 
experiments \cite{bluhl09_PRL, Shanks, Harris2009} agree with the theoretical predictions 
\cite{schmi91, felix91, altsh88_PRL, ambeg90_PRL}, the observed average diamagnetic persistent
current forces us to quantify the role of potentially relevant physical effects, such as the spin-orbit
coupling in constrained geometries or the attractive superconducting-like electron-electron interaction, the first of which constituted the object of the present study.

\begin{acknowledgments}
We thank H\'el\`ene Bouchiat, Ga\"etan Percebois, and Pablo Tamborenea for stimulating discussions. 
We acknowledge financial support from the French National Research Agency (ANR) through Grant No.\ ANR-14-CE26-0005 Q-MetaMat. 
This work of the Interdisciplinary Thematic Institute QMat, as part of the ITI 2021--2028 program of the University of Strasbourg, CNRS, and Inserm, was supported by IdEx Unistra (ANR 10 IDEX 0002), and by SFRI STRAT’US project (ANR 20 SFRI 0012) and ANR-17-EURE-0024 under the framework of the French Investments for the Future Program.
\end{acknowledgments}

\appendix
\section{Semiclassical evaluation of the zero-field orbital susceptibility}
\label{app:sc}

In this appendix, we use the semiclassical approximation for the $l$-fixed DOS $\varrho_l(E)$ in order to calculate the contributions of Eq.~\eqref{eq:chi_paraanddia} to the nonrelativistic ZFS. We also show the equivalence between this procedure with the use of the 3D semiclassical approximation employed in Ref.~\cite{viloria2018orbital}.

\subsection{One-dimensional semiclassical approach}

Similarly to the decomposition of the DOS into a smooth (Weyl) and an oscillatory (in energy) part, presented in Eq.~\eqref{trace3D} for the field-dependent case, the semiclassical approximation to the radial problem allows to write the field-independent $l$-fixed spinless DOS as  \cite{weick05_PRB} 
\begin{equation}
\label{trace}
\varrho_l(E) = {\bar \varrho}_l(E) + \varrho_l^{\rm osc}(E) \, ,
\end{equation}
with
\begin{subequations}
\label{eq:rhosemi}
\begin{align}
\label{eq:rhosemismooth}
{\bar \varrho}_l(E) &=  \frac{\tau_l(E)}{2 \pi \hbar}  \, ,
\\
\label{eq:rhosemiosc}
\varrho_l^{\rm osc}(E) &= \frac{\tau_l(E)}{\pi \hbar} \sum_{\eta=1}^{\infty} 
 \cos{\left( \eta \left[ \frac{S_l(E)}{\hbar}
 -\frac{3\pi}{2} \right] \right)}  \, .
\end{align}
\end{subequations}
The sum is over the repetitions $\eta$ of the periodic orbit in the effective ($l$-dependent) 1D radial potential
\begin{equation}
\label{eq:effective_potential}
V_l^{\rm eff}(r) = V(r) + \frac{\hbar^2}{2m} \, \frac{(l+1/2)^2}{r^2} \, ,
\end{equation}
having an energy $E$, a period
\begin{equation}
\label{eq:tau}
\tau_l(E) = \frac{\hbar}{E} \, \sqrt{\frac{E}{E_0} - \left(l+\frac{1}{2}\right)^2} \, ,
\end{equation}
and a classical action
\begin{align}
S_l(E) =&\; 2 \hbar \left[ \sqrt{\frac{E}{E_0} - \left(l+\frac{1}{2}\right)^2} \right.
\nonumber\\
& \left.-\left(l+\frac{1}{2}\right)  \arccos{\left( \frac{ l +1/2}{\sqrt{E/E_0}}\right)} \right] \, .
\end{align}
For a given $E$, the maximum allowed $l$ is given by
\begin{equation}
\label{eq:lmax}
l_{\rm max}=\left \lfloor{\sqrt{\frac{E}{E_0}}} -\frac{1}{2} \right \rfloor \, ,
\end{equation}
where the floor function $\left \lfloor\zeta \right \rfloor$ yields the integer part of $\zeta$. 

The radial problem, being a one-dimensional configuration, results in a ${\bar \varrho}_l(E)$ with a $1/\sqrt{E}$ behavior for $E/E_0 \gg (l+1/2)^2$ and the same $\hbar$ dependence than $\varrho_l^{\rm osc}(E)$. The decomposition \eqref{trace}, when used in Eq.~\eqref{eq:reldosldos}, reproduces \cite{weick05_PRB, weick2006} the well-known zero-field semiclassical DOS of the sphere \citep{balian92_AnnPhys,tanaka96_PRB}, 
with a smooth part $g^{(\mathrm{3D})}(E) {\mathcal{V}}$ and an oscillating one, that  like Eq.~\eqref{trace3Dp}, has the form
\begin{equation}
\label{eq:rho_osc}
\varrho^\mathrm{osc}(E,0)= \sum_{\nu=1}^\infty\sum_{\eta=2\nu+1}^\infty \varrho_{\nu\eta}(E) \, ,
\end{equation}
where the spin degeneracy factor is included, and 
\begin{equation}
\label{eq:rho_osc2}
\varrho_{\nu\eta}(E)=\frac{4}{E_0}\sqrt{\frac{ka}{\pi}} \, \frac{(-1)^{\nu}}{\sqrt{\eta}} \cos{\varphi_{\nu\eta}} \, \sin^{3/2}{\varphi_{\nu\eta}} 
\, \cos{\big(\theta_{\nu\eta}(k)\big)}
\, .
\end{equation}
The sum is over families of classical periodic orbits lying on the equatorial plane of the sphere, labeled by the topological indexes $(\nu, \eta)$, with $\nu$ the number of turns around the center (i.e., the winding number) and $\eta$ the number of specular reflections at the boundary (i.e., the number of bounces).\footnote{We have not included the contribution of the diametral (self-retracing)
orbits with $\eta=2\nu$ having only a twofold degeneracy, and thus associated with a contribution to the DOS which is of higher order in $\hbar$ than the ones of Eq.~\eqref{eq:rho_osc}. Moreover, diametral trajectories not enclosing a magnetic flux do not contribute to the ZFS, and as we will see in the sequel, the SOC is ineffective in their cases. \label{footnote:diameter}} We note $k=\sqrt{2mE}/\hbar$ and define 
\begin{equation}
\label{eq:anglephi}
\varphi_{\nu\eta}=\frac{\pi\nu}{\eta} \, ,
\end{equation}
corresponding to half the angle spanned between two consecutive bounces (see Fig.~\ref{fig:scattering_circle}). Furthermore, we have defined the $k$-dependent phase
\begin{equation}
\label{eq:thetaphase}
\theta_{\nu\eta}(k)=kL_{\nu\eta}+\frac{\pi}{4}-\frac{3\pi}{2}\eta \, ,
\end{equation}
with $L_{\nu\eta}=2\eta a\sin{\varphi_{\nu\eta}}$ the trajectory length.  

In Ref.~\cite{tanaka96_PRB} it is shown that the effect of a magnetic field weak enough to produce a cyclotron radius much larger than $a$ is to affect each term of the sum \eqref{eq:rho_osc} representing the oscillating part of the density of states by a field-dependent orbital modulation factor for the $(\nu, \eta)$ family, given by
\begin{equation}
\label{eq:mf}
{\cal M}_{\nu\eta}^{\rm (orb)}(B)=j_0\left(2\pi\phi_{\nu\eta}/\phi_0\right).
\end{equation}
We note $j_0(\zeta)=\sin{\zeta}/\zeta$ the zeroth-order spherical Bessel function of the first kind, while
\begin{equation}
\label{eq:magflux}
\phi_{\nu\eta}=B\mathcal{A}_{\nu\eta}
\end{equation}
is the magnetic flux enclosed by the orbits $(\nu,\eta)$ covering the area 
$\mathcal{A}_{\nu\eta}=\frac 12\eta a^2\sin{(2\varphi_{\nu\eta})}$, and 
$\phi_0=hc/e$ is the flux quantum.

\subsection{Equivalence between 1D and 3D semiclassics}

The semiclassical evaluation of $\chi^{(\rm orb)}$ can be addressed by trading in Eqs.~\eqref{eq:chi_paraanddia} the sums over the principal quantum number $n$ by energy-integrals and the use of the Poisson summation formula for the sum over $l$, resulting in   
\begin{subequations}
\label{eq:chi_para_diap}
\begin{align}
\label{eq:chi_paraap}
\frac{\chi^{(\rm para)}}{|\chi_\mathrm{L}|} =&\;  - \frac{6\pi E_0}{k_\mathrm{F}a}   \int_{0}^{\infty} {\rm d}E  f'_{\bar \mu_0}(E) \sum_{\nu=-\infty}^{+\infty}  
\int_{-1/2}^{\sqrt{E/E_0}-1/2} {\rm d}l 
\nonumber \\
& \; 
\times \exp{\left(2\pi \mathrm{i} \nu l\right)} \,
l \left(l+\frac{1}{2} \right) \left(l+1\right) \varrho_l(E)  \, ,
\\
\label{eq:chi_diaap}
\frac{\chi^{(\rm dia)}}{|\chi_\mathrm{L}|} =&\; - \frac{6\pi}{k_\mathrm{F}a}
\int_{0}^{\infty} {\rm d}E \, f_{\bar \mu_0}(E) \sum_{\nu=-\infty}^{+\infty}
\int_{-1/2}^{\sqrt{E/E_0}-1/2} {\rm d}l 
\nonumber \\
&\;
\times \exp{\left(2\pi \mathrm{i} \nu l\right)} 
\left(l+\frac{1}{2} \right)
 \, \mathcal{R}_{l}(E) \, \varrho_l(E)  \, ,
\end{align}
\end{subequations}
where $l$ is now understood as a continuous variable. Following Eq.~\eqref{eq:R}, we have defined $\mathcal{R}_{l}(E) = (1/3)\left[1+(2E_0/E)(l+3/2)(l-1/2) \right]$.

In the leading order in $k_\mathrm{F}a \gg 1$, the smooth part of $\chi^{(\rm orb)}$ is obtained by using ${\bar \varrho}_l(E)$ in Eqs.~\eqref{eq:chi_para_diap} and only keeping the $\nu=0$ term of the sum,
\begin{subequations}
\label{eq:chi_para_diaps}
\begin{align}
\label{eq:chi_paraaps}
\frac{\bar{\chi}^{(\rm para)}}{|\chi_\mathrm{L}|} =&\;  - \frac{3}{k_\mathrm{F}a}  \int_{0}^{\infty} {\rm d}E \, f'_{\bar \mu_0}(E)  \left(\frac{E}{E_0} \right)^{1/2} 
\nonumber \\
& \; 
\times \int_{0}^{1} {\rm d}\zeta \, \zeta
\left(\frac{E}{E_0} \, \zeta^2 -\frac 14 \right) \sqrt{1-\zeta^2}\, ,
\\
\label{eq:chi_diaaps}
\frac{\bar{\chi}^{(\rm dia)}}{|\chi_\mathrm{L}|} =&\;  - \frac{1}{k_\mathrm{F}a\,E_0} 
\int_{0}^{\infty} {\rm d}E \, f_{\bar \mu_0}(E) \left(\frac{E}{E_0} \right)^{1/2} 
\nonumber \\
& \; 
\times \int_{0}^{1} {\rm d}\zeta \, \zeta
\left(1+2\zeta^2-\frac{2}{E/E_0} \right) \sqrt{1-\zeta^2}\,  .
\end{align}
\end{subequations}
Performing the integration over the variable $\zeta=(l+1/2)\sqrt{E_0/E}$ we have
\begin{subequations}
\label{eq:chi_para_diapss}
\begin{align}
\label{eq:chi_paraapss}
\frac{\bar{\chi}^{(\rm para)}}{|\chi_\mathrm{L}|} &=  - \frac{1}{k_\mathrm{F}a}  \int_{0}^{\infty} {\rm d}E \, f'_{\bar \mu_0}(E)  \left(\frac{E}{E_0} \right)^{1/2} 
\left(\frac{2E}{5E_0} - \frac{1}{4} \right) \, ,
\\
\label{eq:chi_diaapss}
\frac{\bar{\chi}^{(\rm dia)}}{|\chi_\mathrm{L}|} &=  - \frac{1}{k_\mathrm{F}a E_0}  \int_{0}^{\infty} {\rm d}E \, f_{\bar \mu_0}(E) 
\left(\frac{E}{E_0} \right)^{1/2} \left(\frac{3}{5} -\frac{2E_0}{3E} \right) \,  .
\end{align}
\end{subequations}

An integration by parts in Eq.~\eqref{eq:chi_diaapss} leads to
\begin{equation}
\label{eq:chi_diapss}
\frac{\bar{\chi}^{(\rm dia)}}{|\chi_\mathrm{L}|} = \frac{1}{k_\mathrm{F}a}  
\int_{0}^{\infty} {\rm d}E \, f'_{\bar \mu_0}(E) \left(\frac{E}{E_0} \right)^{1/2} 
\left(\frac{2E}{5E_0} - \frac{4}{3} \right)  \,  .
\end{equation}
Thus, the leading-order term in $k_\mathrm{F}a$ of $\bar{\chi}^{(\rm para)}$ and $\bar{\chi}^{(\rm dia)}$ cancel each other, and $\bar{\chi}^{(\rm orb)}$ is then given by next-order contributions.\footnote{For the degenerate case, in the leading order in $k_\mathrm{F}a$, we have  that $\bar{\chi}^{(\rm para)}=|\bar{\chi}^{(\rm dia)}|= (2/5) \left(k_\mathrm{F}a\right)^2$. Notwithstanding, we stress that the separation between $\chi^{(\rm para)}$ and $\chi^{(\rm dia)}$ is only for computational purposes, and lacks physical reality.}
However, such terms are not captured by the expressions of Eqs.~\eqref{eq:chi_para_diapss}, which are only valid in the leading order in $k_\mathrm{F}a$, since they result from $E$ and $l$ integrations in which the form \eqref{eq:tau} was used beyond its regime of validity of $E/E_0 \gg (l+1/2)^2$. In Ref.~\cite{richt96_PhysRep} it is shown that the magnetic field dependence of  
${\bar \varrho}(E,B)$, given by the so-called zero-length trajectories, results in $\bar{\chi}^{(\rm orb)}=\chi_\mathrm{L}$, while the numerical evaluation of Eqs.~\eqref{eq:chi_Zeeman}--\eqref{eq:chi_paraanddia} presented in Fig.~\ref{fig:chinr} approaches, in the limit of large radius $a$ where the role of the confinement potential becomes irrelevant, the bulk result given by Eqs.~\eqref{eq:Landau_sus}--\eqref{eq:Pauli_sus} (see the inset in Fig.~\ref{fig:chinr}). 

The fact that $\bar{\chi}^{(\rm orb)}/|\chi_\mathrm{L}|$ is of order $(k_\mathrm{F}a)^0$, points to the importance of $\chi^{\rm osc}$, which is obtained by using $\varrho_l^{\rm osc}(E)$ in Eq.~\eqref{eq:chi_para_diap}. The rapidly oscillating (in $E$) phases $2 \pi \nu \pm \eta S_l(E)/\hbar$ allow us to perform a stationary-phase (sp) integration over $l$ with the condition $l_{\rm sp}= \sqrt{E/E_0} \cos{(\pi\nu/\eta)} - 1/2$, yielding
\begin{subequations}
\label{eq:chi_para_diaposc}
\begin{align}
\label{eq:chi_paraaposc}
\frac{{\chi}^{(\rm para)-osc}}{|\chi_\mathrm{L}|} =&  - \frac{6 \sqrt{\pi}}{k_\mathrm{F}a}  \int_{0}^{\infty} {\rm d}E \, f'_{\bar \mu_0}(E) \left(\frac{E}{E_0} \right)^{1/4} 
\nonumber \\
&  \times 
\sum_{\substack{\nu>0\\\eta>2\nu}} \frac{(-1)^\nu
}{\sqrt{\eta}}
\sin^{3/2}{\varphi_{\nu\eta}}  \, \cos{\varphi_{\nu\eta}}
\nonumber\\
&\times
\left[ \frac{E}{E_0} \, \cos^2{\varphi_{\nu\eta}} -\frac 14 \right] \cos{\big(\theta_{\nu\eta}(k)\big)}
 \, ,
\end{align}
\begin{align}
\label{eq:chi_diaaposc}
\frac{{\chi}^{(\rm dia)-osc}}{|\chi_\mathrm{L}|} =& - \frac{2 \sqrt{\pi}}{k_\mathrm{F}aE_0}  \int_{0}^{\infty} {\rm d}E \, f_{\bar \mu_0}(E)  \left(\frac{E}{E_0} \right)^{1/4} 
\nonumber\\
&\times
\sum_{\substack{\nu>0\\\eta>2\nu}} \frac{(-1)^\nu
}{\sqrt{\eta}}
\sin^{3/2}{\varphi_{\nu\eta}} \,  \cos{\varphi_{\nu\eta}}
\nonumber\\
&\times
\left[1+ 2 \cos^2{\varphi_{\nu\eta}} -\frac{2E_0}{E} \right] \cos{\big(\theta_{\nu\eta}(k)\big)}
  \,  .
\end{align}
\end{subequations}

The couple of integers $(\nu, \eta)$ of the one-dimensional problem, introduced respectively in Eqs.~\eqref{eq:chi_para_diap} and \eqref{eq:rhosemiosc}, take now the role of the topological indexes labeling the families of classical periodic orbits in the sphere. The restriction of $\nu \neq 0$ appears since these contributions, considered in Eqs.~\eqref{eq:chi_para_diaps}, lead to $\bar{\chi}^{(\rm orb)}$. Only positive values of $\nu$ are kept, as the sign of $\nu$ is associated with the orientation in which the periodic orbit is traveled. The condition $\eta>2\nu$ appears as a restriction for the stationary-phase value $l_{\rm sp}$ to be within the integration interval. The special case of $\eta=2\nu$, where $l_{\rm sp}$ is at the lower limit of the integration interval, corresponds to the diametral orbits (see footnote \ref{footnote:diameter}). The angle $\varphi_{\nu\eta}$ and the $k$-dependent phase $\theta_{\nu\eta}$ have been defined in Eqs.~\eqref{eq:anglephi} and \eqref{eq:thetaphase}, respectively. The stationary-phase procedure yielding \eqref{eq:chi_para_diaposc} is analogous to that allowing to link $\varrho_l^{\rm osc}(E)$ and $\varrho^{\rm osc}(E,0)$ \cite{weick2006}. Since we work in leading-order in $k_\mathrm{F}a$, we can neglect the last terms in the square brackets of Eqs.~\eqref{eq:chi_para_diaposc}. 

In the low-temperature limit, we use $f_{\bar \mu_0}(E)=\Theta({\bar \mu}_0-E)$ and, with the help of Fresnel integrals, to leading order in $k_\mathrm{F}a$, we find
\begin{align}
\frac{{\chi}^{(\rm dia)-osc}}{|\chi_\mathrm{L}|} =& - 2 \sqrt{\pi}(k_\mathrm{F}a)^{1/2}
\sum_{\substack{\nu>0\\\eta>2\nu}}
\frac{(-1)^\nu
}{\eta^{3/2}}
\sin^{1/2}{\varphi_{\nu\eta}}  \cos{\varphi_{\nu\eta}} 
\nonumber\\
&\;\times(1+\cos^2{\varphi_{\nu\eta}} )\,\sin{\big(\theta_{\nu\eta}(k_\mathrm{F})\big)} \, .
\end{align}
Thus, ${\chi}^{(\rm dia)-osc}\sim (k_\mathrm{F}a)^{1/2}$ is of lower order in $k_\mathrm{F}a$ than 
${\chi}^{(\rm para)-osc}\sim (k_\mathrm{F}a)^{3/2}$, and in the semiclassical limit we have ${\chi}^{(\rm orb)-osc}={\chi}^{(\rm para)-osc}$.  

The integration of $f_{\bar{\mu}_0}'(E)$ multiplied by a rapidly oscillating function of $E$ results in a general expression \cite{richt96_PhysRep}, that when applied to Eq.~\eqref{eq:chi_paraaposc} yields
\begin{align}
\label{eq:chi_1}
\frac{\chi^{(\rm orb)-osc}}{|\chi_\mathrm{L}|}=&\;6\sqrt{\pi}(k_\mathrm{F}a)^{3/2}
\sum_{\substack{\nu>0\\\eta>2\nu}}
\frac{(-1)^\nu
}{\sqrt{\eta}}\, \sin^{3/2}{\varphi_{\nu\eta}}  \cos^3{\varphi_{\nu\eta}}
\nonumber\\
&\times
R(L_{\nu\eta}/L_T) \,
\cos{\big(\theta_{\nu\eta}(k_\mathrm{F})\big)} \, ,
\end{align}
which is the result obtained in Ref.~\cite{viloria2018orbital} starting from the three-dimensional DOS \eqref{eq:rho_osc} with the modulation factor \eqref{eq:mf}.  The thermal factor $R$, given by Eq.~\eqref{eq:R_T}, depends on the ratio $L/L_T$, where $L$ is the trajectory length and $L_T=\hbar v_\mathrm{F}/\pi k_\mathrm{B}T$ the thermal length.

When $\chi^{(\rm nr)}_{\rm b}$ is added to the semiclassical evaluation of Eq.~\eqref{eq:chi_1}, the result is almost indistinguishable from the nonrelativisitc ZFS shown in Fig.~\ref{fig:chinr} (in black), obtained from the quantum perturbative Eqs.~\eqref{eq:chi_Zeeman} and \eqref{eq:chi_paraanddia}.

\section{Nonrelativistic susceptibility using the coupled basis}
\label{app:nrsucb}

As discussed in Secs.~\ref{sec:socsn} and \ref{sec:ptmf}, the treatment of the SOC is greatly simplified by working in the coupled basis of total angular momentum within the subspace decomposition \eqref{eq:decomposition_spaces}. It is therefore useful to recast the nonrelativistic ZFS as the sum of two components arising from the $\rm p$ (product) and $\rm e$ (entangled) states associated, respectively, with the subspaces ${\cal S}_{n,l,({\rm d/u})}^{\rm p}$ and ${\cal S}_{n,l,m_j}^{\rm e}$, writing 
\begin{equation}
\chi^{(\rm nr)}=\chi^{(\rm nr)}_{\rm p} + \chi^{(\rm nr)}_{\rm e} \, .
\end{equation}

The $\rm p$-states ($\rm d$/$\rm u$) are characterized, respectively, by the quantum numbers $\{n,l,m_z= - l,m_s= - 1/2 \}$ and $\{n,l,m_z= l,m_s= +1/2 \}$ of the product basis, that can be expressed in a compact way as $\{n,l,m_z=\mp l,m_s=(1/2) (m_z/l) \}$. According to Eq.~\eqref{eq:Enr}, at finite magnetic field, these states are associated with the energies
\begin{equation}
E_{n,l}^{\rm{(nr) \, ( d/u)}}=E_{n,l}^{(0)} \mp \mu_{\rm B} B  (l+1) +E^{(\rm dia)}_{n,l,\mp l} \, . 
\label{eq:Enrp}
\end{equation}
Applying Eq.~\eqref{def:ZFS} and summing over $m_z=\mp l$ we have 
\begin{align}
\label{eq:chinrp}
\frac{\chi^{(\rm nr)}_{\rm p}}{|\chi_\mathrm{L}|} =& 
-\frac{9\pi E_0}{k_\mathrm{F}a} \sum_{l=0}^\infty\sum_{n=1}^{\infty} 
\bigg\{
\left(l+1\right)^2 f'_{\bar{\mu}_0}\!\left(E^{(0)}_{n,l}\right) 
\nonumber\\
&+
\frac{\mathcal{R}_{n,l}}{2E_0} \, \frac{l+1}{l+3/2} \, f_{\bar{\mu}_0}\!\left(E^{(0)}_{n,l}\right)\bigg\} \, .
\end{align}

In the nonrelativistic case, the subspace ${\cal S}_{n,l,m_j}^{\rm e}$ can be characterized by the product-basis states with quantum numbers $\{n,l,m_z=\mp l,m_s=- (1/2) (m_z/l) \}$ and $\{n,l,m_z\in \left[-l+1,l-1\right],m_s= \pm 1/2 \}$. For each couple $(n,l)$, there are two states of the first kind and $4l-2$ of the second one, whose energies at finite magnetic field follow from Eq.~\eqref{eq:Enr}. Once the sum over $m_z$ is performed in Eq.~\eqref{def:ZFS}, they yield
\begin{align}
\label{eq:chinre}
\frac{\chi_{\rm e}^{\rm (nr)}}{|\chi_\mathrm{L}|} =&
 -\frac{3\pi E_0}{k_\mathrm{F}a} \sum_{l=1}^\infty\sum_{n=1}^{\infty} l \,
 \bigg\{
 \left(2l^2+1\right) f'_{\bar{\mu}_0}\!\left(E^{(0)}_{n,l}\right) 
 \nonumber\\
& 
+\frac{\mathcal{R}_{n,l}}{E_0} \, \frac{2l+5/2}{l+3/2} \, f_{\bar{\mu}_0}\!\left(E^{(0)}_{n,l}\right) 
\bigg\} \, .
\end{align}
Obviously, the addition of Eqs.~\eqref{eq:chinrp} and \eqref{eq:chinre} results in the expression of $\chi^{\rm (nr)}$ presented in Sec.~\ref{sec:nrssg} [cf.\ Eq.~\eqref{eq:chi_nr}].

\section{Darwin correction}
\label{app:mebsp}

In this appendix, we consider the Darwin correction for a spherically symmetric confinement, given by the Hamiltonian contribution $\mathcal{H}^{(\rm D)}$ of Eq.~\eqref{eq:H_Darwin2}. In order to address the corresponding matrix elements, we first tackle a technical issue concerning the eigenvalues and eigenvectors of a spherical box confined by a finite-height potential.

\subsection{Finite-box spherical potential}

The spherical symmetry of the potential \eqref{eq:confining_potential} allows to write the eigenstates as in Eqs.~\eqref{eq:nlmzmsud}--\eqref{eq:orbital}, where the radial wave function of the bound states is given by
\begin{equation}
R_{n,l}(r) = \sqrt{\frac{2}{a^3}} \, \frac{1}{|c_{n,l}|^{1/2}}
\begin{cases}
j_{l}\left(k_{n,l} \, r\right) \, , & r \leqslant a\, , \\[.2truecm]
\displaystyle \frac{j_{l}(k_{n,l} \, a)}{\mathtt{k}_{l}(\kappa_{n,l} \, a)} \, {\mathtt{k}}_{l}
\left(\kappa_{n,l} \, r\right) \, , &r > a\, . 
\end{cases}
\label{eq:rwf}
\end{equation}
We follow the standard convention of using $j_l(\zeta)$ for the spherical Bessel function of the first kind and order $l$. Similarly, we note $\mathtt{k}_{l}(\zeta)$ the modified spherical Bessel function of the second kind and order $l$, obtained from the imaginary-argument spherical Hankel function $h_l^{(1)}(\zeta)$, i.e., $\mathtt{k}_{l}(\zeta) = -\mathrm{i}^l h_l^{(1)}(\mathrm{i} \zeta)$. The condition of a bound state implies that its energy $E^{(0)}_{n,l}=E_0 (k_{n,l} \, a)^2 $ is smaller than $V_0$. We have defined $E_0=\hbar^2/2m a^2$ and $\kappa_{n,l}=\sqrt{2m(V_0-E^{(0)}_{n,l})}/\hbar$. The normalization is settled by the constant
\begin{align} 
c_{n,l}  =&\;  
\left( 
\frac{j_{l}(k_{n,l} \, a)}{\mathtt{k}_{l}(\kappa_{n,l} \, a)} 
\right)^{2}
\, \mathtt{k}_{l-1}\left(\kappa_{n,l} \, a\right) \, \mathtt{k}_{l+1}\left(\kappa_{n,l} \, a\right) 
\nonumber \\
& -  j_{l-1}\left(k_{n,l} \, a\right) \, j_{l+1}\left(k_{n,l} \, a\right)  \, .
\end{align} 

The allowed $E^{(0)}_{n,l}$ and $k_{n,l}$ result from the solutions of the quantization condition
\begin{equation}
\label{eq/quantcond}
\kappa_{n,l} \, a \, \frac{\mathtt{k}^{\prime}_{l}(\kappa_{n,l} \, a)}{\mathtt{k}_{l}(\kappa_{n,l} \, a)} = k_{n,l} \, a \, \frac{j^{\prime}_{l}(k_{n,l} \, a)}{j_{l}(k_{n,l} \, a)}   \, ,
\end{equation}
which can be recast as
\begin{equation}
\label{eq/quantcond2}
\kappa_{n,l} \, a \, \frac{\mathtt{k}_{l+1}(\kappa_{n,l} \, a)}{\mathtt{k}_{l}(\kappa_{n,l} \, a)} = k_{n,l} \, a \, \frac{j_{l+1}(k_{n,l} \, a)}{j_{l}(k_{n,l} \, a)}   \, ,
\end{equation}
once we employ the useful recurrence relationship  
\begin{equation}
\label{eq:recurrence}
q^{\prime}_l(\zeta)=\frac{l}{\zeta} q_l(\zeta) - q_{l+1}(\zeta) \, ,
\end{equation}
valid for $q_l(\zeta)=j_{l}(\zeta)$, as well as for $q_l(\zeta)=\mathtt{k}_{l}(\zeta)$.

In the limiting case of a hard-wall potential ($V_0 \rightarrow \infty$), the support of the wave function is $r<a$, and the previous expressions result in
\begin{subequations}
\label{eq:knlandnorma}
\begin{align} 
k_{n,l} \, a & \rightarrow  \zeta_{n,l} \, ,
\label{eq:knl}
\\ 
c_{n,l} & \rightarrow  \left[ j_{l+1}(\zeta_{n,l}) \right]^2 \, ,
\label{eq:norma}
\end{align} 
\end{subequations}
where $\zeta_{n,l}$ stands for the $n$th root of $j_l(\zeta)$, and thus the radial wave function \eqref{eq:rwf} takes the simpler form of Eq.~\eqref{eq:rwfs}.

In the case where $V_0$ is large but remains finite, the second-order expansion of 
\eqref{eq/quantcond2} around $\zeta_{n,l}$ in the small parameter $u=\sqrt{E_0/V_0}$ allows to write the corrections to Eq.~\eqref{eq:knlandnorma} and to other important parameters as
\begin{subequations}
\label{eq:knlandnorma2}
\begin{align} 
k_{n,l} \, a & \simeq  \zeta_{n,l} \left(1 -  u +  u^2 \right) \, ,
\label{eq:knl2}
\\ 
c_{n,l} & \simeq  \left[ j_{l+1}(\zeta_{n,l}) \right]^2  \left(1 + 3  u  + 3 u^2 \right) \, ,
\label{eq:norma2}
\\ 
j_{l}(k_{n,l} \, a) & \simeq  j_{l+1}(\zeta_{n,l}) \, \zeta_{n,l} \, u \, ,
\label{eq:jl}
\\ 
j_{l}(k_{n,l} \, a) \, j_{l}^{\prime}(k_{n,l} \, a) & \simeq  - \left[ j_{l+1}(\zeta_{n,l}) \right]^2 \zeta_{n,l}  \left(u  + 2 u^2 \right)  \, ,
\label{eq:jljpl}
\\ 
\kappa_{n,l} \, a & \simeq  u^{-1}  - \frac{1}{2} \, \zeta_{n,l}^2 \, u +  \zeta_{n,l}^2 \, u^2 \, .
\label{eq:kappaa}
\end{align} 
\end{subequations}

It is important to remark that Eq.~\eqref{eq:jl} is valid up to quadratic order in $u$, and this limiting condition leads to the expression 
\eqref{eq:I_SO} of the SOC radial matrix element, which does not depend on $V_0$.

\subsection{Matrix elements for the Darwin correction}

According to Eq.~\eqref{eq:H_Darwin2}, the diagonal Darwin energy correction is
\begin{equation}
E_{n,l}^{({\rm D})} =\frac{E_0}{4mc^2} \, I^{(\rm D)}_{n,l} \, ,
\end{equation}
with the radial matrix element
\begin{equation}
\label{eq:rmed}
I^{(\rm D)}_{n,l} = a^2 \int_{0}^{\infty} \mathrm{d}r  \left[ R_{n,l}(r) \right]^2 \,  \frac{{\rm d}}{{\rm d}r}  \left(r^2 \, \frac{{\rm d}V}{{\rm d}r}  \right) \, .
\end{equation}
For the potential \eqref{eq:confining_potential} we have $V^{\prime}(r)=V_0 \, \delta(r-a)$ and therefore the integral \eqref{eq:rmed} can be trivially performed, yielding 
\begin{equation}
\label{eq:rmed1p}
I^{(\rm D)}_{n,l} = -\left. V_0 \, a^4 \, \frac{\rm d}{{\rm d} r} \left[ R_{n,l}(r) \right]^2\right|_{r=a} \, ,
\end{equation}
that, using the form \eqref{eq:rwf} of the radial wave function, results in
\begin{equation}
\label{eq:rmed2}
I^{(\rm D)}_{n,l} = - \,  4 V_0 \, \frac{k_{n,l} \, a}{|c_{n,l}|} \, j_l\left(k_{n,l} \, a\right) \, j_l^{\prime}\left(k_{n,l} \, a\right) \, .
\end{equation}

According to Eq.~\eqref{eq:jl}, the radial matrix element $I^{(\rm D)}_{n,l}$ diverges as $\sqrt{V_0}$ in the large $V_0$-limit. Such a divergence is unphysical since an infinite $V_0$ would imply an infinite electric field, for which the weakly-relativistic approach leading to Eq.~\eqref{eq:Hamiltonian} would not be valid. The subtleties related with a relativistic particle hitting a steep wall are extensively discussed in the literature \cite{greiner1990relativistic, Alberto_1998}. The appearance of a divergence in our weakly-relativistic approach calls for a systematic expansion in the small parameter $u=\sqrt{E_0/V_0}$, as presented in the first part of this appendix. In addition, we need to verify that the physical constants of the problem are such that we work in the regime of validity of the perturbative approach. For the case of gold nanoparticles, the values of $E_\mathrm{F}=\unit[5.5]{eV}$ and $W=\unit[4.3]{eV}$ \cite{AM} result in $u \simeq 0.75 \, (k_\mathrm{F} a)^{-1}$. Therefore, in the semiclassical limit of $k_\mathrm{F} a \gg 1$, to which our study is restricted, $u$ is indeed a small parameter. 
 
Using the second-order expressions of Eq.~\eqref{eq:knlandnorma2}, the radial matrix element \eqref{eq:rmed2} can be approximated by
\begin{equation}
\label{eq:rmed3}
I^{({\rm D})}_{n, l} = 4V_0 \, \zeta^2_{n,l} \, u\left(1-2u\right) \, ,
\end{equation}
and thus, the Darwin energy correction is
\begin{equation}
\label{eq:ED}
E_{n,l}^{({\rm D})} 	=  \frac{E_0^2\zeta^2_{n,l}}{mc^2} \left(u^{-1}-2\right) \, . 
\end{equation}
Notice that the forms \eqref{eq:ED} and \eqref{eq:esoj}, respectively, of the Darwin and spin-orbit energy corrections, are both valid up to terms of order $(V_0)^0$. We remark that, while $\mathcal{H}^{(\rm D)}$ couples states with different $n$, the resulting second-order corrections in $v_\mathrm{F}/c$ can be neglected. 

Similarly to the case of the kinetic correction treated in Sec.~\ref{sec:kedc}, the Darwin energy \eqref{eq:ED} induces, at $B=0$, a renormalization of the chemical potential  
\begin{equation}
\label{eq:deltamuD}
\Delta \mu^{(\rm D)} = \left(u^{-1}-2\right) \, \frac{\bar \mu_{0} \, E_0}{m c^2}  \, .
\end{equation}
%

\section{Matrix elements for the angular magneto-electric coupling}
\label{app:memec}

In this appendix we calculate the matrix elements of the angular magneto-electric Hamiltonian $\mathcal{H}^{(\rm ame)}$ restricted to the subspace ${\cal S}_{n,l,m_j}^{\rm e}$, as discussed in Sec.~\ref{sec:ptmf}. According to Eq.~\eqref{eq:H_ame2} the diagonal matrix element is

\begin{align}
\label{eq:dmeamec}
{\cal E}_{n,j,m_j,(\pm)}^{({\rm ame})} &= \langle \Phi_{n,j,m_j}^{(\pm)} |\mathcal{H}^{(\rm ame)}|  \Phi_{n,j,m_j}^{(\pm)} \rangle \nonumber\\
&= \frac{\mu_{\rm B}  B}{4mc^2} \, I^{(\rm so)}_{n, j \pm 1/2} \, {\cal I}^{(\rm d-ame)(\pm)}_{j,m_j}
\, .
\end{align}
The radial matrix element coincides with that arising from the spin-orbit coupling \eqref{eq:I_SO} since, in the hard-wall limit, 
\begin{equation}
\int_0^a \mathrm{d}r  \, r^3 \left[ R_{n,l}(r) \right]^2  V^{\prime}(r)=I^{(\rm so)}_{n, l}. 
\end{equation}
Using the standard notation $\sigma_{\pm}=\sigma_{x} \pm {\rm i} \sigma_{y}$, the angular matrix element is
\begin{widetext}
\begin{equation}
\label{eq:dmeamecai}
{\cal I}^{(\rm d-ame)(\pm)}_{j,m_j}
= \langle {\Upsilon}^{(\pm)}_{j,m_j} | \, \sin^2{\theta} \, \sigma_{z} -\frac{\sin{\theta} \, \cos{\theta} }{2} \left(\mathrm{e}^{-{\rm i}\varphi}\sigma_{+} + \mathrm{e}^{{\rm i}\varphi}\sigma_{-} \right) 
| {\Upsilon}^{(\pm)}_{j,m_j} \rangle 
\, .
\end{equation}
The spin-conserving component of ${\cal I}^{(\rm d-ame)(+)}_{j,m_j}$ is given by
\begin{equation}
\label{eq:dmeamecsc}
\langle {\Upsilon}_{j,m_j}^{(+)} \left| \, \sin^2{\theta} \, \sigma_{z} \right|  {\Upsilon}_{j,m_j}^{(+)} \rangle 
= - \frac{m_j}{2} \, \left( \frac{ j(j+2)+j+1+ m_j^2}{j(j+1)(j+2)} \right)
\, .
\end{equation}
Since the two spin-flip components coincide, we only need 
\begin{align}
\label{eq:aimeame}
 \langle {\Upsilon}^{(+)}_{j,m_j} | & \sin{\theta} \, \cos{\theta} \, \mathrm{e}^{-{\rm i}\varphi}\sigma_{+} | {\Upsilon}^{(+)}_{j,m_j} \rangle = (-1)^{(m_j+1/2)} \, 
 \sqrt{\frac{2 \pi}{15}} \, \frac{\sqrt{(j+1)^2-m_j^2}}{j+1} \, \int {\rm d}\vartheta \, Y_{j+1/2}^{-(m_j+1/2)}(\vartheta) \, Y_{2}^{1}(\vartheta) \, Y_{j+1/2}^{m_j-1/2}(\vartheta)
 \nonumber \\
 & = (-1)^{(m_j+1/2)} \, \sqrt{\frac{2}{3}} \, \sqrt{(j+1)^2-m_j^2} \,
 \begin{pmatrix}
j+1/2 & \, j+1/2 & \, 2 \\
0 & \, 0 & \, 0
\end{pmatrix} \,
 \begin{pmatrix}
j+1/2 & \, j-1/2 & \, 2 \\
- (m_j+1/2) & \, m_j-1/2 & \, 1
\end{pmatrix} \, ,
\end{align}
where we have used the integration formula of three spherical harmonics in terms of Wigner-$3j$ symbols. The first of the $3j$ symbols can be trivially calculated, while the use of Regge and permutation symmetries for the second one leads to 
\begin{equation}
 \langle {\Upsilon}^{(+)}_{j,m_j} | \sin{\theta} \, \cos{\theta} \, \mathrm{e}^{-{\rm i}\varphi}\sigma_{+} | {\Upsilon}^{(+)}_{j,m_j} \rangle = \frac{m_j}{2} \, \frac{(j+1)^2-m_j^2}{j(j+1)(j+2)} \, .
\end{equation}
Proceeding analogously with the other basis vector ${\Upsilon}^{(-)}_{j,m_j}$, while combining the spin-conserving and spin-flipping components, we have
\begin{equation}
\label{eq:dmeamecai2}
{\cal I}^{(\rm d-ame)(\pm)}_{j,m_j} = \mp \, \frac{m_j(j+1/2)}{j(j+1)}
\, .
\end{equation}

The off-diagonal matrix element of $\mathcal{H}^{(\rm ame)}$ restricted to the subspace ${\cal S}_{n,l,m_j}^{\rm e}$ is
\begin{equation}
\label{eq:odmeame}
\langle \Phi_{n,j,m_j}^{(-)} |\mathcal{H}^{(\rm ame)}|  \Phi_{n,j,m_j}^{(+)} \rangle 
= \frac{\mu_{\rm B}  B}{4mc^2} \, I^{(\rm so)}_{n, j \pm 1/2} \, {\cal I}^{(\rm od-ame)(\pm)}_{j,m_j}
\, ,
\end{equation}
with
\begin{equation}
\label{eq:odmeamecai0}
{\cal I}^{(\rm od-ame)}_{j,m_j} 
= \langle {\Upsilon}^{(-)}_{j+1,m_j} |  \sin^2{\theta} \, \sigma_{z} -\frac{\sin{\theta} \, \cos{\theta}}{2} \left(\mathrm{e}^{-{\rm i}\varphi}\sigma_{+} + \mathrm{e}^{{\rm i}\varphi}\sigma_{-} \right) 
| {\Upsilon}^{(+)}_{j,m_j} \rangle 
\, .
\end{equation}
The spin-conserving component of ${\cal I}^{(\rm od-ame)(+)}_{j,m_j}$ is given by
\begin{equation}
\label{eq:odmeamecsc}
\langle {\Upsilon}_{j+1,m_j}^{(-)} |  \sin^2{\theta} \, \sigma_{z} |  {\Upsilon}_{j,m_j}^{(+)} \rangle 
= - \frac{1}{2} \, \frac{\sqrt{(j+1)^2 - m_j^2}}{(j+1)} \, 
\left(1+\frac{m_j^2}{j(j+2)} \right)
\, .
\end{equation}
Since here too the two spin-flip components coincide, we only need 
\begin{align}
\label{eq:odmeamecai}
\langle {\Upsilon}^{(-)}_{j+1,m_j} | \sin{\theta} \, \cos{\theta} \, \mathrm{e}^{-{\rm i}\varphi}\sigma_{+} | {\Upsilon}^{(+)}_{j,m_j} \rangle &= 
 - (-1)^{(m_j+1/2)} \, 
 \sqrt{\frac{8 \pi}{15}} \, \frac{m_j}{j+1} \, \int {\rm d}\vartheta \, Y_{j+1/2}^{-(m_j+1/2)}(\vartheta) \, Y_{2}^{1}(\vartheta) \, Y_{j+1/2}^{m_j-1/2}(\vartheta)
 \nonumber \\
 &  =  - \, \frac{m_j^2}{2} \, \frac{\sqrt{(j+1)^2 - m_j^2}}{j(j+1)(j+2)}
\, ,
\end{align}
\end{widetext}
and thus 
\begin{equation}
\label{eq:odmeamecai2}
{\cal I}^{(\rm od-ame)}_{j,m_j} = - \, \frac{1}{2} \, \frac{\sqrt{(j+1)^2-m_j^2}}{(j+1)}
\, .
\end{equation}
which leads to the off-diagonal matrix element \eqref{eq:odmeamel}. The simple structure of the angular matrix elements \eqref{eq:dmeamecai2} and \eqref{eq:odmeamecai2} is a consequence of the Wigner--Eckart theorem applied to the diagonal and off-diagonal matrix elements of $\mathcal{H}^{(\rm ame)}$, given by Eqs.~\eqref{eq:dmeamec}  and \eqref{eq:odmeame}, respectively.

\section{Semiclassical description of the spin-orbit coupling}
\label{app:sdsoc}
In this appendix we apply the semiclassical description of the spin dynamics in order to calculate the effect of the spin-orbit coupling on the ZFS, showing that the corresponding correction constitutes the semiclassical limit of the quantum perturbative results obtained in Sec.~\ref{sec:rczfs}.

\subsection{Semiclassics without orbital effects}
\label{app:swoe}

A trace formula describing the semiclassical DOS for a system with SOC has been developed from the Dirac equation \cite{BolteKeppeler87_PRL}, or alternatively, from a coherent-state path-integral description of the spin variables \cite{Grabert99_JPhysA,Pletyukhov03_JPhysA,zaitsev2005,ribeiro07_JMathPhys}. In the so-called weak-coupling limit, in which the SOC is weak enough to provide a negligible perturbation of the orbital motion, and assuming that the applied magnetic field only acts on the spin variables without affecting the orbital part of the wave functions, the oscillating part of the DOS can be written as \cite{Zaitsev02_JPhysA}
\begin{equation}
\label{eq:rho_oscsoc}
\varrho^\mathrm{osc}(E,B)= \sum_{\xi} {\cal M}_{\xi}^{\rm (s)}(E,B)  
{\cal A}_{\xi}(E)  \cos{\left( \frac{S_{\xi}(E)}{\hbar}-\frac{\pi}{2} \mu_{\xi} \right)}
\, .
\end{equation}
The sum is over the periodic orbits $\xi$ of the uncoupled system, while ${\cal A}_{\xi}(E)$, $S_{\xi}(E)$, and $\mu_{\xi}$ are, respectively, the corresponding stability determinant, classical action, and Maslov index. No assumption is made on whether the periodic orbits are isolated or not. The only difference of Eq.~\eqref{eq:rho_oscsoc} with the standard Berry--Tabor or Gutzwiller DOS is the appearance of the spin modulation factor, that for a spin $1/2$ particle takes the form 
\begin{equation}
\label{eq:smf}
{\cal M}_{\xi}^{\rm (s)}(E,B) = 2 \, \cos{\left( \frac{1}{2} \, \Zhe_{\xi}(E) \right)}
\, ,
\end{equation}
where $\Zhe_{\xi}(E)$ is the phase accumulated by the precession of the spin under the effect of the Zeeman \eqref{eq:H_Zeeman} and the SOC \eqref{eq:H_so3} terms of the Hamiltonian when the particle travels along the periodic orbit $\xi$. Writing 
\begin{equation}
\label{eq:H_ZeemanplusHsoc}
\mathcal{H}^{(\rm Z)} + \mathcal{H}^{(\rm so)} = \frac{\hbar}{2} \, \boldsymbol{\sigma} \cdot \mathbf{C}(\mathbf{r},\mathbf{p})
\end{equation}
with
\begin{equation}
\label{def:Crp}
\mathbf{C}(\mathbf{r},\mathbf{p}) =
\frac{g_0 \mu_\mathrm{B}}{\hbar} \, \mathbf{B} +
\frac{1}{2m^2c^2} \, \boldsymbol{\nabla}V \times \mathbf{p} \, , 
\end{equation}
we have, for the case in which $\mathbf{C}$ remains oriented along the $z$-axis, i.e., 
$\mathbf{C}\big(\mathbf{r}(t),\mathbf{p}(t)\big)=C\big(\mathbf{r}(t),\mathbf{p}(t)\big) \, {\hat {\mathbf{e}}}_z$, a simple expression for the accumulated phase \cite{Zaitsev02_JPhysA}
\begin{equation}
\label{eq:phi}
\phi_{\xi}(E) = \int_{0}^{\tau_{\xi}} {\rm d}t \, C\big(\mathbf{r}(t),\mathbf{p}(t)\big)
\, ,
\end{equation}
where $\tau_{\xi}$ is the period of the classical orbit.

In the absence of magnetic field and SOC, ${\cal M}_{\xi}^{\rm (s)}=2$ for all $\xi$, and therefore the spin modulation factor merely represents spin degeneracy. Without SOC and under a magnetic field $\mathbf{B} =  B\, {\bf {\hat e}}_z$, we have
\begin{equation}
\label{eq:phiwithoutSOC}
\Zhe_{\xi}(E) = \frac{g_0 \mu_\mathrm{B}}{\hbar} \, B \, \tau_{\xi}
\, .
\end{equation}
Using Eqs.~\eqref{eq:rho_oscsoc} and \eqref{eq:chi1}, the resulting ZFS in this particular case can be expressed as 
\begin{align}
\label{eq:chi_Zosc}
{\chi}^{(\rm Z)-osc} &= - \frac{1}{\mathcal{V}} \, \frac{\partial^2}{\partial B^2} 
\Bigg[ \sum_{\xi} R(\tau_{\xi}/\tau_{T}) \left(\frac{\hbar}{\tau_{\xi}} \right)^2 
\nonumber\\
&\hspace{.5cm}\times
\cos{\left(\frac{g_0 \mu_\mathrm{B}B}{\hbar} \tau_{\xi}\right)}
\varrho^\mathrm{osc}_\xi(E_\mathrm{F},0)
    \Bigg]
\Bigg|_{B=0}\ ,
\end{align}
where the thermal factor $R$ has been defined in Eq.~\eqref{eq:R_T} and $\tau_{T}=\hbar/\pi k_{\rm B}T$. In the limit of $T \rightarrow 0$, 
Eq.~\eqref{eq:chi_Zosc} simplifies to
${\chi}^{(\rm Z)-osc}=(\mu_\mathrm{B}^2/\mathcal{V}) \varrho^\mathrm{osc}(E_\mathrm{F},0) $,
yielding the small component of the Pauli susceptibility associated with the oscillatory part of the DOS [see the discussion after Eq.~\eqref{eq:chi_Zeeman2}]. 

\begin{figure}[tb]
\begin{center}
\includegraphics[width=.7\linewidth]{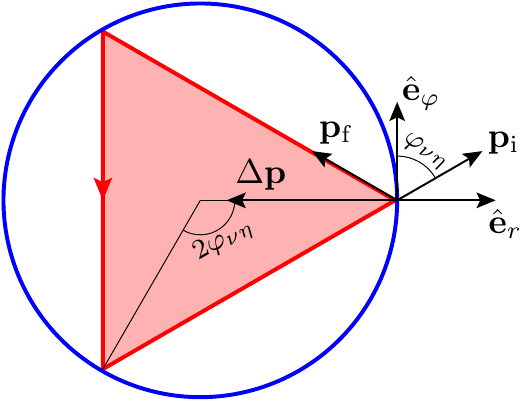}
\caption{\label{fig:scattering_circle} 
Classical periodic orbit $(1,3)$ (in red) with $\varphi_{\nu\eta}$ the half angle spanned between two collisions and the 
centripetal vector $\Delta\mathbf{p}=\mathbf{p}_\mathrm{f}-\mathbf{p}_\mathrm{i}$ being the impulse acquired in the collision.}
\end{center}
\end{figure}

We address now the accumulated phase, considering the effect of the SOC in the absence of a magnetic field, and focusing ourselves on the case of the sphere. Since the impulse at each bounce is given by the change in momentum $\Delta \mathbf{p}$, for a collision occurring at a time $t_{\rm c}$ we have 
\begin{equation}
\label{eq:impulse}
\int_{t_{\rm c}^{-}}^{t_{\rm c}^{+}} {\rm d}t \, \boldsymbol{\nabla}V\big(\mathbf{r}(t)\big) \times \mathbf{p}(t) = - \Delta \mathbf{p} \times \mathbf{p}
\, ,
\end{equation}  
where $\mathbf{p}$ can be taken indistinctly as the momentum before or after the collision, as the angular momentum $\mathbf{L}$ is conserved over the collision (see Fig.~\ref{fig:scattering_circle}). Specializing ourselves in the case of a periodic orbit in the equatorial plane defined by the topological indices $(\nu, \eta)$, we have
\begin{equation}
\label{eq:impulse2}
 - \Delta \mathbf{p} \times \mathbf{p} = p^2 \sin{(2\varphi_{\nu\eta})} \, {\bf {\hat e}}_z
\, ,
\end{equation}  
where the angle $\varphi_{\nu\eta}$, defined in Eq.~\eqref{eq:anglephi}, corresponds to half the angle spanned between two consecutive bounces (see Fig.~\ref{fig:scattering_circle}). Taking into account the $\eta$ bounces of the periodic orbit,
\begin{equation}
\label{eq:phiwithoutB}
\Zhe_{\nu\eta}(E) = \pm \, \frac{1}{2} \left(\frac{v}{c} \right)^2 \eta \, \sin{(2\varphi_{\nu\eta})}
\, ,
\end{equation}
where the $\pm$ corresponds to the counterclockwise (clockwise) orientation of the trajectory with respect to ${\bf {\hat e}}_z$ and $v=p/m$. 
Since the spin modulation factor of Eq.~\eqref{eq:smf} is obtained as a trace over the spin Hilbert space \cite{Zaitsev02_JPhysA}, a change of the spin basis can be implemented without altering the results, and thus the expression \eqref{eq:phiwithoutB} is valid for any periodic orbit of the family defined by the indices $(\nu, \eta)$, not necessarily on the equatorial $z=0$ plane. We notice that the SOC is irrelevant for the highly symmetric diametral orbits, having $\varphi_{\nu\eta}=0$. Furthermore, we remark that, in analogy with the quantum SOC matrix element \eqref{eq:esoj}, the accumulated phase $\Zhe_{\nu\eta}(E)$ is independent of $V_0$, and that it scales as $(v/c)^2$.

\begin{figure}[b]
\begin{center}
\includegraphics[width=.9\linewidth]{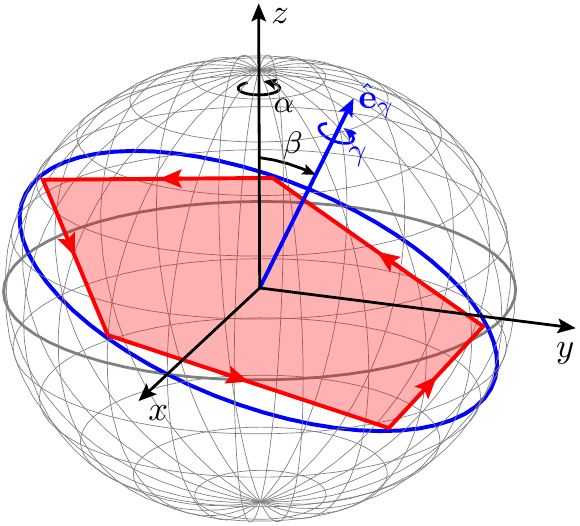}
\caption{\label{fig:sphere} 
In red: Classical periodic orbit $(\nu, \eta, \alpha, \beta, \gamma, (+))$ belonging to the degenerate family with topological indices $(\nu,\eta)= (1,5)$, labeled by the Euler angles $(\alpha, \beta, \gamma)$, with a counterclockwise orientation with respect to ${\bf {\hat e}}_{\gamma}$. This unit vector, associated with the polar angles $(\alpha, \beta)$, is perpendicular to the plane of motion (whose intersection with the sphere is indicated by the blue circle). We note $\gamma$ the rotation angle setting the trajectory in its plane.}
\end{center}
\end{figure}

The individual trajectories of the family $(\nu, \eta)$ can be labeled by these two topological indices, together with the three Euler angles $(\alpha, \beta, \gamma)$, where $(\alpha, \beta)$ are the polar angles defining the unit vector ${\bf {\hat e}}_{\gamma}$ perpendicular to the plane of the motion, and $\gamma$ is the rotation angle in that plane. We choose the above defined angles in such a way that $\beta$ is the angle formed by ${\bf {\hat e}}_{\gamma}$ 
and ${\bf {\hat e}}_z$ (see Fig.~\ref{fig:sphere}).

Comparison between Eqs.~\eqref{eq:phiwithoutSOC} and \eqref{eq:phiwithoutB} shows that the kicked precessions occurring at the $\eta$ bounces of the periodic orbit labeled by $(\nu, \eta,\alpha, \beta, \gamma,(\pm))$ would have on the accumulated phase $\Zhe_{\nu\eta\alpha \beta \gamma(\pm)}(E)$ the same effect than that of a fictitious magnetic field
\begin{equation}
\label{eq:Bfict}
\mathbf{B}_{\nu \eta \alpha \beta \gamma}^{\rm f(\pm)} = \pm \,
\frac{\hbar}{\mu_\mathrm{B}}\, \frac{v^3}{4 a c^2} \, \cos{\varphi_{\nu\eta}} \, {\bf {\hat e}}_{\gamma}
\, ,
\end{equation}
where the $\pm$ corresponds to the orientation of the trajectory with respect to ${\bf {\hat e}}_{\gamma}$.

\begin{widetext}
\subsection{Semiclassics with orbital effects}
\label{app:swoe2}

We now include both terms of the vector $\mathbf{C}$ of Eq.~\eqref{def:Crp}, comprising the effects of a magnetic field  and the SOC. Towards simplifying the calculation of the corresponding modulation factor, we consider an effective field
\begin{equation}
\label{eq:Beff}
\mathbf{B}_{\nu \eta \alpha \beta \gamma}^{\rm eff(\pm)} =  B\, {\bf {\hat e}}_z + \mathbf{B}_{\nu \eta \alpha \beta \gamma}^{\rm f(\pm)} 
\, ,
\end{equation}
which allows to obtain the spin modulation factor with the help of Eqs.~\eqref{eq:smf} and \eqref{eq:phiwithoutSOC} when trading in the last equation $B$ by the magnitude of $\mathbf{B}_{\nu \eta \alpha \beta \gamma}^{\rm eff(\pm)}$, 
\begin{equation}
B_{\nu \eta \beta}^{\rm eff(\pm)} = \left[B^2 
\pm \frac{\hbar}{\mu_\mathrm{B}}\, \frac{v^3 B}{2 a c^2} \, \cos{\varphi_{\nu\eta}} \, \cos{\beta} 
+
\left(\frac{\hbar}{\mu_\mathrm{B}}\,\frac{v^3}{4 a c^2} \, \cos{\varphi_{\nu\eta}} \right)^2
\right]^{1/2} \, ,
\end{equation}
which is independent on the angles $\alpha$ and $\gamma$.

As indicated above, Eq.~\eqref{eq:rho_oscsoc} only considers the effect of the magnetic field on the spin variables. In order to lift up such a restriction and include the perturbation of the classical action of the periodic orbits by a weak magnetic field, we must go one step backwards from Eq.~\eqref{eq:rho_oscsoc} and write the oscillating part of the DOS as a sum over individual periodic orbits $(\nu, \eta, \alpha, \beta, \gamma(\pm))$, i.e.,
\begin{equation}
\label{eq:rho_oscsoc2}
\varrho^\mathrm{osc}(E,B)= - \frac{1}{\pi} {\rm Im} \Bigg\{ 
\sum_{\substack{\nu>0\\\eta>2\nu}} \sum_{(\pm)} 
\frac{1}{4 \pi^2} \int_{0}^{2\pi} {\rm d}\alpha \int_{0}^{2\pi} {\rm d}\gamma
\int_{0}^{\pi/2} {\rm d}\beta \sin{\beta} \,
{\cal M}_{\nu \eta \alpha \beta \gamma}^{(\pm)}(E,B) \, 
{\cal B}_{\nu\eta}(E) \, \exp{\left(\mathrm{i} \theta_{\nu\eta}(k)  \right)} \Bigg\}
\, ,
\end{equation}
where \cite{keller1960asymptotic,tanaka96_PRB}
\begin{equation}
\label{eq:Bnueta}
{\cal B}_{\nu\eta}(E)= - \mathrm{i}\, \frac{\sqrt{\pi ka}}{E_0} \, \frac{(-1)^{\nu}}{\sqrt{\eta}} \cos{\varphi_{\nu\eta}} \, \sin^{3/2}{\varphi_{\nu\eta}} 
\, ,
\end{equation}
and the $k$-dependent phase $\theta_{\nu\eta}(k)$ has been defined in Eq.~\eqref{eq:thetaphase}. The integral over $\beta$ stops at $\pi/2$ since the angles between  $\pi/2$ and  $\pi$ correspond to trajectories within the integration interval, but traveled in the clockwise direction. It is easy to recover from  Eq.~\eqref{eq:rho_oscsoc2} the particular cases of Eqs.~\eqref{eq:rho_oscsoc} and \eqref{eq:rho_osc2}, for which the orbit-dependent modulation factor ${\cal M}_{\nu \eta \alpha \beta \gamma}^{(\pm)}(E,B)$ leads, upon summing over the individual orbits of the family $(\nu, \eta)$, to the form \eqref{eq:smf} and \eqref{eq:mf}, respectively. In the general case \eqref{def:Crp}, having a magnetic field and SOC, ${\cal M}_{\nu \eta \alpha \beta \gamma}^{(\pm)}(E,B)$ factors out in a spin and an orbital part for the individual trajectories, even if such separation is not possible at the level of families of periodic orbits. 

Considering the contribution from the counterclockwise and the clockwise orientations, performing the trivial integrals over the angles $\alpha$ and $\gamma$, and taking into account the perturbation on the classical action of the periodic orbits by the effect of a weak magnetic field, we have
\begin{align} 
\label{eq:rho_oscsoc3}
\varrho^\mathrm{osc}(E,B)   =&  - \frac{1}{\pi} {\rm Im} 
\Bigg\{ 
\sum_{\substack{\nu>0\\\eta>2\nu}} 
\int_{0}^{\pi/2} {\rm d}\beta \, \sin{\beta} \, {\cal B}_{\nu\eta}(E) 
\left[
{\cal M}_{\nu \eta}^{\rm (s)}(E,B_{\nu \eta \beta}^{\rm eff(+)}) \, 
\exp{\left(\mathrm{i} \left[ \theta_{\nu\eta}(k) - \frac{2\pi\phi_{\nu\eta}}{\phi_0} \, \cos{\beta} \right] \right)} 
\right. 
\nonumber \\
 & + 
\left. 
{\cal M}_{\nu \eta}^{\rm (s)}(E,B_{\nu \eta \beta}^{\rm eff(-)}) \, 
\exp{\left(\mathrm{i} \left[ \theta_{\nu\eta}(k) + \frac{2\pi\phi_{\nu\eta}}{\phi_0} \, \cos{\beta}\right] \right)} 
\right]
\Bigg\}
\, .
\end{align}
The $k$-dependent phase $\theta_{\nu\eta}$ has been defined in Eq.~\eqref{eq:thetaphase}, while $\pm\, \phi_{\nu\eta} \cos \beta$ is the flux of the magnetic field $B \,{\bf {\hat e}}_z$ piercing the trajectories 
$(\nu, \eta, \alpha, \beta, \gamma, (\pm))$, and $\phi_{\nu\eta}$ is given by Eq.~\eqref{eq:magflux}. Towards the calculation of the ZFS, it is convenient to recast Eq.~\eqref{eq:rho_oscsoc3} as
\begin{equation}
\label{eq:rho_oscsoc4}
\varrho^\mathrm{osc}(E,B) = \frac{1}{E_0}\sqrt{\frac{ka}{\pi}} 
\sum_{\substack{\nu>0\\\eta>2\nu}} \frac{(-1)^{\nu}}{\sqrt{\eta}} \cos{\varphi_{\nu\eta}} \, \sin^{3/2}{\varphi_{\nu\eta}} 
\left[I_{\nu\eta, {\rm c}}(E, B) \, \cos{\big(\theta_{\nu\eta}(k)\big)} + 
I_{\nu\eta, {\rm s}}(E, B) \, \sin{\big(\theta_{\nu\eta}(k)\big)} \right]
\end{equation}
with
\begin{subequations}
\label{eq:I_integrals}
\begin{align} 
I_{\nu\eta, {\rm c}}(E, B)   &=  
\int_{0}^{\pi/2} {\rm d}\beta \, \sin{\beta} \,
\cos{\left( \frac{2\pi\phi_{\nu\eta}}{\phi_0} \, \cos \beta \right)} 
\left[ 
{\cal M}_{\nu \eta}^{\rm (s)}(E,B_{\nu \eta \beta}^{\rm eff(+)}) +
{\cal M}_{\nu \eta}^{\rm (s)}(E,B_{\nu \eta \beta}^{\rm eff(-)}) 
\right] \,  ,
\\
I_{\nu\eta, {\rm s}}(E, B)   &=  
\int_{0}^{\pi/2} {\rm d}\beta\, \sin{\beta} \,
\sin{\left( \frac{2\pi\phi_{\nu\eta}}{\phi_0} \, \cos \beta \right)} 
\left[ 
{\cal M}_{\nu \eta}^{\rm (s)}(E,B_{\nu \eta \beta}^{\rm eff(+)}) -
{\cal M}_{\nu \eta}^{\rm (s)}(E,B_{\nu \eta \beta}^{\rm eff(-)}) 
\right] 
\, .
\end{align}
\end{subequations}
In order to obtain $\chi^\mathrm{osc}$ from Eq.~\eqref{eq:chi1}, we need the second derivative of $\varrho^\mathrm{osc}$ with respect to $B$, which will be determined by
\begin{subequations}
\begin{align} 
\label{eq:BderIcs}
\frac{\partial^2 I_{\nu\eta, {\rm c}}}{\partial B^2}\bigg|_{B=0}  & =  
\int_{0}^{\pi/2} {\rm d}\beta\, \sin{\beta} \,
\left[ 
- \left( \frac{e}{\hbar c} \, \mathcal{A}_{\nu\eta} \right)^2 \cos^2{\beta} 
\left( {\cal M}_{\nu \eta}^{\rm (s)}(E,B_{\nu \eta \beta}^{\rm eff(+)}) +
{\cal M}_{\nu \eta}^{\rm (s)}(E,B_{\nu \eta \beta}^{\rm eff(-)})
\right) 
\bigg|_{B=0}
\right. 
\nonumber
\\
& \left.\hspace{.5cm}+
\frac{\partial^2}{\partial B^2}
\left( {\cal M}_{\nu \eta}^{\rm (s)}(E,B_{\nu \eta \beta}^{\rm eff(+)}) +
{\cal M}_{\nu \eta}^{\rm (s)}(E,B_{\nu \eta \beta}^{\rm eff(-)})
\right)
\bigg|_{B=0} 
\right] 
\,  ,
\\
\frac{\partial^2 I_{\nu\eta, {\rm s}}}{\partial B^2}\bigg|_{B=0}   & =  
\frac{2 e}{\hbar c} \, \mathcal{A}_{\nu\eta} 
\int_{0}^{\pi/2} {\rm d}\beta\, \sin{\beta} \, \cos{\beta} \,
 \frac{\partial}{\partial B}
\left( {\cal M}_{\nu \eta}^{\rm (s)}(E,B_{\nu \eta \beta}^{\rm eff(+)}) -
{\cal M}_{\nu \eta}^{\rm (s)}(E,B_{\nu \eta \beta}^{\rm eff(-)})
\right)
\bigg|_{B=0} 
\, .
\end{align}
\end{subequations}
\end{widetext}

Performing the $\beta$ integrals and neglecting the terms of order $(v/c)^4$ we have
\begin{subequations}
\label{eq:BderIcs2}
\begin{align} 
\frac{\partial^2 I_{\nu\eta, {\rm c}}}{\partial B^2}\bigg|_{B=0}  & =  
- \frac{4}{3} \left( \frac{e}{\hbar c} \, \mathcal{A}_{\nu\eta} \right)^2 
- 4 \left( \frac{\mu_\mathrm{B}}{\hbar} \, \frac{L_{\nu\eta}}{v} \right)^2
\,  ,
\label{eq:BderIcs2a}
\\
\frac{\partial^2 I_{\nu\eta, {\rm s}}}{\partial B^2}\bigg|_{B=0}   & =  - \frac{1}{3} \,
\frac{e^2}{\hbar c^4} \, \frac{v}{m a} \, \mathcal{A}_{\nu\eta} \, L_{\nu\eta}^2 \, \cos{\varphi_{\nu\eta}} 
\, .
\label{eq:BderIcs2b}
\end{align}
\end{subequations}
Using Eqs.~\eqref{eq:rho_oscsoc4} and \eqref{eq:chi1}, we verify that the first term on the right-hand side of Eq.~\eqref{eq:BderIcs2a} leads to the expression \eqref{eq:chi_1} of $\chi^{(\rm orb)-osc}$, the oscillating part for the orbital component of the ZFS, obtained in Ref.~\cite{viloria2018orbital} and rederived in Appendix \ref{app:sc}. Similarly, the second term on the right-hand side of Eq.~\eqref{eq:BderIcs2a} leads to the expression \eqref{eq:chi_Zosc} of the oscillating part of the Pauli spin susceptibility. While Eq.~\eqref{eq:BderIcs2a} simply yields the oscillating part of the nonrelativistic ZFS, its relativistic counterpart follows from Eq.~\eqref{eq:BderIcs2b} and can be written as
\begin{align}
\label{eq:chi_SOC}
\frac{\chi^{(\rm so)-osc}}{|\chi_\mathrm{L}|}=&\;6 \left(\frac{v_\mathrm{F}}{c} \right)^2
(\pi k_\mathrm{F}a)^{1/2}
\sum_{\substack{\nu>0\\\eta>2\nu}}
(-1)^\nu \sqrt{\eta} \, \sin^{5/2}{\varphi_{\nu\eta}} 
\nonumber\\
&\times
\cos^3{\varphi_{\nu\eta}} R(L_{\nu\eta}/L_T) \,
\sin{\big(\theta_{\nu\eta}(k_\mathrm{F})\big)} \, .
\end{align}

Comparing $\chi^{(\rm so)-osc}$ with $\chi^{(\rm orb)-osc}$ [cf.\ Eq.~\eqref{eq:chi_1}], we see that the former is smaller than the latter by the factor $(v_\mathrm{F}/c)^2$ and due to one less power of  $k_\mathrm{F}a$. In addition, these two terms are ``out of phase", concerning the $k_\mathrm{F}a$ oscillations.

\subsection{Equivalence between the semiclassical and quantum perturbative approaches}
\label{app:ebsqpa}

In the remaining part of this appendix, we show that the previously obtained form \eqref{eq:chi_SOC} of the spin-orbit correction to the ZFS can also be derived by applying the one-dimensional semiclassical approximation to the quantum perturbative result \eqref{eq:deltachiso}. For that purpose, we follow the same steps as in Appendix \ref{app:sc}, addressing the quantum result with the help of the Poisson summation rule and only keeping the highest-order terms in $k_\mathrm{F}a$ [see Eqs.~\eqref{eq:chi_para_diap}]. The smooth part of $\chi^{(\rm so)}$ resulting from the $\nu = 0$ term has been evaluated in Sec.~\ref{sec:nasrczfs} [see Eq.~\eqref{eq:deltachiSOC2}]. Using for the oscillating part the same stationary-phase condition than in Appendix \ref{app:sc}, we have, to leading order in $k_\mathrm{F}a$,
\begin{align}
\label{eq:chi_soc_osc}
\frac{\chi^{(\rm so)-osc}}{|\chi_\mathrm{L}|} =&  - \frac{12 \sqrt{\pi}}{k_\mathrm{F}a} \, \frac{E_0^2}{mc^2} \, 
\int_{0}^{\infty} {\rm d}E \, f_{\bar \mu_0}^{\prime\prime}(E) \left(\frac{E}{E_0} \right)^{9/4} 
\nonumber\\
&\times\sum_{\substack{\nu>0\\\eta>2\nu}} \frac{(-1)^\nu
}{\sqrt{\eta}}
\sin^{3/2}{\varphi_{\nu\eta}} \cos^3{\varphi_{\nu\eta}}
 \cos{\big(\theta_{\nu\eta}(k)\big)}
.
\end{align}
An integration by parts leaves us with the integral of $f_{\bar \mu_0}'(E)$ multiplied by a rapidly oscillating function of $E$, that leads to \cite{richt96_PhysRep}
\begin{align}
\label{eq:chi_soc_osc2}
\frac{\chi^{(\rm so)-osc}}{|\chi_\mathrm{L}|} =&\;
\frac{12 \sqrt{\pi}}{k_\mathrm{F}a} \, \frac{E_0^2}{mc^2}    
\sum_{\substack{\nu>0\\\eta>2\nu}} \frac{(-1)^\nu}{\sqrt{\eta}}
\, \sin^{3/2}{\varphi_{\nu\eta}} 
\nonumber\\
&\times \cos^3{\varphi_{\nu\eta}} \, R(\tau_{\nu\eta}/\tau_T)
\nonumber\\
&\times
\left.\frac{{\rm d}}{{\rm d} E}
\left(
\left(\frac{E}{E_0} \right)^{9/4} 
 \cos{\big(\theta_{\nu\eta}(k)\big)}
 \right)\right|_{E=E_\mathrm{F}} 
  \,  ,
\end{align}
which to leading order in $k_\mathrm{F}a$ coincides with the semiclassical result \eqref{eq:chi_SOC}. Such an agreement demonstrates that the semiclassical treatment of the spin-orbit interaction \cite{Zaitsev02_JPhysA} is valid not only in the large-spin limit, but also for a finite spin, including spin $1/2$.

\section{Matrix elements for the case of the half-sphere}
\label{app:mehs}
In this appendix we prove a general result concerning the limiting values of the wave function and its derivative
at a potential discontinuity, in one- and three-dimensional geometries, which is crucial in order to obtain the 
spin-orbit matrix elements \eqref{eq:H_soshd} and \eqref{eq:H_sofloornd}. We also use nontrivial recurrence relations
allowing to calculate the angular matrix element \eqref{eq:H_soshnd} and the radial one, Eq.~\eqref{eq:Rnd}.

\subsection{Wave-function values close to a potential barrier}

\begin{figure}[t]
\begin{center}
\includegraphics[width=.75\linewidth]{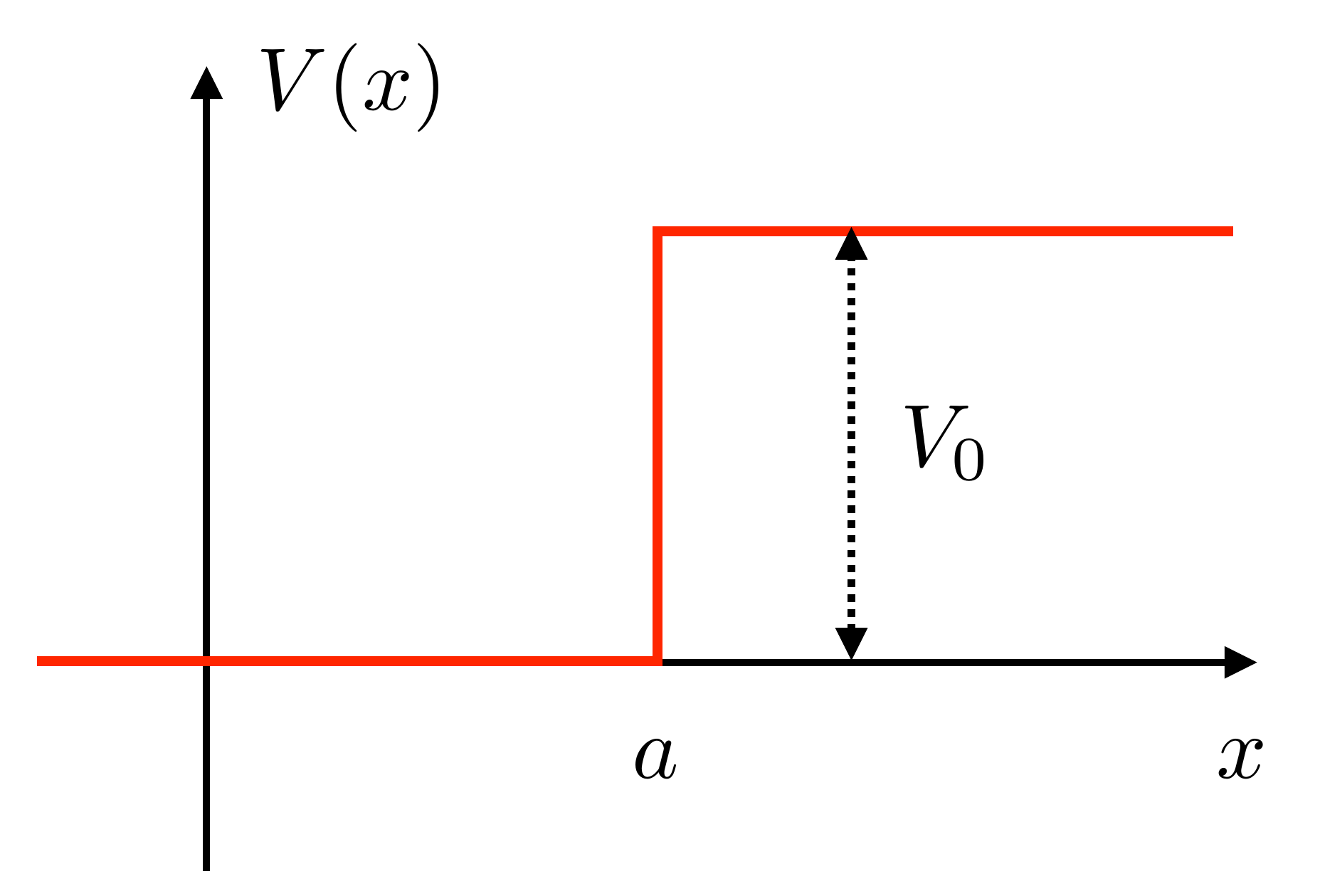}
\caption{\label{fig:theorem} 
Sketch of a one-dimensional sharp potential barrier of height $V_0$ at $x=a$.
}
\end{center}
\end{figure}

We consider a one-dimensional Schrödinger equation with a potential barrier of height $V_0$ at $x=a$ (see Fig.~\ref{fig:theorem}). We will prove that its solutions $\psi(x)$ verify
\begin{equation}
\label{eq:theorem1d}
\lim_{V_0\rightarrow \infty} \frac{\sqrt{2mV_0}}{\hbar} \, \psi(a) = - \psi^{\prime}(a^{-}) \, .
\end{equation}

Without loss of generality, we take $V(x)=0$ for $x \in (a-\epsilon,a)$ and $V(x)=V_0$ for $x \in (a, a+\epsilon)$, where $\epsilon$ is a small, but finite length. Integrating the Schrödinger equation in the small interval $(a-\epsilon, a+\epsilon)$ yields 
\begin{align}
\label{eq:schr1d}
- \frac{\hbar^2}{2m}\left[ \psi^{\prime}(a+\epsilon) - \psi^{\prime}(a-\epsilon)  \right] & +  V_0 \int_{a}^{a+\epsilon} {\rm d}x \, \psi(x) = \nonumber \\
& E  \int_{a-\epsilon}^{a+\epsilon} {\rm d}x \, \psi(x)  \, .
\end{align}
Assuming that the barrier extends to $x=+ \infty$, we have $\psi(x)=\psi(a) \exp{(-\kappa x)}$  in the interval $(a, a+\epsilon)$, with $\kappa=\sqrt{2m(V_0-E)}/\hbar$. Taking the limit of large $V_0$ while keeping a finite value of $\epsilon$ we obtain
\begin{subequations}
\begin{align} 
\psi^{\prime}(a+\epsilon) & = - \kappa \, \psi(a) \, \exp{(-\kappa \epsilon)} 
\nonumber\\
&\simeq 0
\,  ,
\\
V_0 \int_{a}^{a+\epsilon} {\rm d}x \, \psi(x) & \simeq \frac{V_0}{\kappa} \, \psi(a) 
\, .
\end{align}
\end{subequations}
Thus, in the leading order in $V_0$,  Eq.~\eqref{eq:schr1d} can be recast as 
\begin{equation}
\label{eq:schr1dp}
\frac{\hbar^2}{2m} \, \psi^{\prime}(a-\epsilon)  +  \sqrt{\frac{V_0\hbar^2}{2m}}   \, \psi(a)=0  \, .
\end{equation}
Since $\psi^{\prime}(x)$ is a continuous function in the open interval $(a-\epsilon,a)$, we promptly obtain the announced result \eqref{eq:theorem1d}. The particular case of the radial wave function is contained in Eq.~\eqref{eq:knlandnorma2} and has been used in obtaining Eq.~\eqref{eq:I_SO}.

The calculation of the spin-orbit matrix elements in the case of the HS necessitates the generalization of \eqref{eq:theorem1d} to a two-dimensional potential barrier. Since the potential \eqref{eq:confining_potential_hs} defining the confinement in the HS with a finite height $V_0$ has axial symmetry, $m_z$ is a good quantum number, and the eigenfunctions have the form 
$\psi_{m_z}(r,\theta,\varphi) = \mathtt{f}_{m_z}(r,\theta) \exp{({\rm i} m_z \varphi)}$, with $\mathtt{f}_{m_z}$ verifying
\begin{widetext}
\begin{equation}
\label{eq:schr2d}
- \frac{\hbar^2}{2m} \, \left( \frac{1}{r^2} \, \partial_{r} \, r^2 \, \partial_{r} + \frac{1}{r^2 \sin{\theta}} \, \partial_{\theta} \, \sin{\theta} \, \partial_{\theta} - \frac{m_z^2}{r^2 \sin^2{\theta}} \right) \mathtt{f}_{m_z}(r,\theta) 
+ V(r,\theta) \, \mathtt{f}_{m_z}(r,\theta)  = E \, \mathtt{f}_{m_z}(r,\theta)  \, .
\end{equation}

Marching on the footprints of the derivation for the one-dimensional case, we integrate the previous equation in the small angular interval $(\pi/2-\epsilon, \pi/2+\epsilon)$, obtaining
\begin{align}
\label{eq:schr2d2}
- \frac{\hbar^2}{2m} \,  \left( \frac{2 \, \epsilon}{r^2} \, \partial_{r} \, r^2 \, \partial_{r} 
\mathtt{f}_{m_z}(r,\theta) \bigg|_{\theta=\pi/2} + 
\frac{1}{r^2} \, \partial_{\theta} \, \mathtt{f}_{m_z}(r,\theta) \bigg|_{\theta=\pi/2-\epsilon}^{\theta=\pi/2+\epsilon} -  \frac{2 \, \epsilon \, m_z^2}{r^2} \, \mathtt{f}_{m_z}(r,\theta)\bigg|_{\theta=\pi/2} \right) + & V_0 \int_{\pi/2}^{\pi/2+\epsilon} {\rm d}\theta \, \mathtt{f}_{m_z}(r,\theta) 
\nonumber \\ 
& = 2 \, \epsilon \, E \, \mathtt{f}_{m_z}(r,\theta)\bigg|_{\theta=\pi/2} \, .
\end{align}

In the leading order in $V_0$ we have $\mathtt{f}_{m_z}(r,\theta)=\mathtt{f}_{m_z}(r,\theta\!=\!\pi/2)  \exp{(-\sqrt{2mV_0r^2}  [\theta-\pi/2]/\hbar)}$  in the interval $(\pi/2, \pi/2+\epsilon)$. Thus, taking the limit of large $V_0$ and then that of $\epsilon \rightarrow 0$ we obtain
\begin{equation}
\frac{r\sqrt{2mV_0}}{\hbar} \, \mathtt{f}_{m_z}(r,\theta\!=\!\pi/2) \simeq - 
\, \partial_{\theta} \, \mathtt{f}_{m_z}(r,\theta) \bigg|_{\theta=\pi/2^{-}} \, ,
\end{equation}
which generalizes Eq.~\eqref{eq:theorem1d} to the case of a two-dimensional potential barrier with axial symmetry.

Taking the hard-wall limit of $V_0 \rightarrow \infty$ for the potential \eqref{eq:confining_potential_hs}, the solutions $\psi_{m_z}^{(V_0,n,l)}(r,\theta,\varphi)$ converge towards the orbital wave functions \eqref{eq:orbitalhs}, with the condition of $l+m_z$ being odd. Thus, we can write 
\begin{equation}
\label{eq:wfv2db}
\frac{\sqrt{mV_0}}{\hbar} \, \psi_{m_z}^{(V_0,n,l)}(r,\theta\!=\!\pi/2,\varphi) \simeq  
-  \sqrt{l(l+1)-m_z(m_z+1)} \, \frac{R_{nl}(r)}{r} \,  \mathrm{e}^{-\mathrm{i}\varphi} \, Y_l^{m_z+1}\left(\theta\!=\!\pi/2,\varphi\right)
\, ,
\end{equation}
where we have used
\begin{equation}
	\frac{\partial Y_l^{m_z}(\theta,\varphi)}{\partial \theta} = m_z \cot(\theta) \, Y_l^{m_z}(\theta,\varphi)+\sqrt{(l-m_z)(l+m_z+1)} \, \mathrm{e}^{-\mathrm{i}\varphi} \, Y_l^{m_z+1}(\theta,\varphi) \, ,
\end{equation}
taking $\theta=\pi/2$.
\end{widetext}

\subsection{Angular integral of the matrix element \eqref{eq:H_soshnd} }

The angular integral \eqref{eq:angint3} appearing in the nondiagonal matrix element \eqref{eq:H_soshnd} of $\mathcal{H}^{\rm (so)-(HS)}$ can be expressed as
\begin{equation}
\mathcal{I}^{(\rm dome)}_{l',l,m'_z} = \frac{1}{2} \sqrt{(2l+1)(2l'+1)} \sqrt{\frac{(l-m)! (l'-m'_z)!}{(l+m'_z)!  (l'+m'_z)}} \mathcal{Y}^{m'_z}_{l',l} \, ,
\label{eq:angint}
\end{equation}
where we have used the definition of the spherical harmonic $Y_l^{m_z}(\vartheta)$ in terms of the associated Legendre function $P_l^{m_z}(\cos\theta)$ [with the standard convention of assigning the Condon-Shortley phase $(-1)^{m_z}$ to the latter], and introduced  
\begin{equation}
\label{eq:legendreint}
\mathcal{Y}^{m'_z}_{l',l}= \int_0^1 \mathrm{d}x\; P_{l'}^{m'_z}(x) \, P_l^{m'_z}(x) \, .
\end{equation}

In the case that interests us, with $l'+m'_z$ odd and $l+m'_z$ even, we can prove, directly from the differential equations fulfilled by the associated Legendre functions $P_l^{m'_z}$ and $P_{l'}^{m'_z}$ with $l \ne l'$, the following useful identity:
\begin{equation}
\mathcal{Y}^{m'_z}_{l',l}=- \frac{l'+1-m'_z}{l'(l'+1)-l(l+1)} \, P_l^{m'_z}(0)\, P_{l'+1}^{m'_z}(0)	\, ,
\end{equation} 
which, together with the relationship
\begin{equation}
\label{eq:plmaz}
P_{l}^{m_z^{\prime}}(0)= \frac{(-1)^{(l+m_z^{\prime})/2}}{2^{l}} \, \frac{(l+m_z^{\prime})!}{(\frac{l+m_z^{\prime}}{2})! \, (\frac{l-m_z^{\prime}}{2})!} \, , 
\end{equation}
valid for even $l+m_z^{\prime}$, lead to
\begin{align}
\mathcal{I}^{(\rm dome)}_{l',l,m} =&\; \frac{(-1)^{m+(l+l'+1)/2}}{2^{l+l'}} \, 
 \frac{ \sqrt{(2l+1)(2l'+1)}}{l(l+1)-l'(l'+1)} 
 \nonumber\\
 &\times
\frac{\sqrt{(l+m)! \, (l-m)!}}{(\frac{l+m}{2})! \, (\frac{l-m}{2})! }  \,
\frac{\sqrt{(l'+m)! \, (l'-m)!}}{(\frac{l'+m-1}{2})! \, (\frac{l'-m-1}{2})! }
\, ,
\label{eq:angint2}
\end{align}
allowing to give a closed expression to the nondiagonal matrix element \eqref{eq:H_soshnd} of $\mathcal{H}^{(\rm dome)}$.

\subsection{Definite integrals of two spherical Bessel functions}
\label{app:mes}

The radial matrix elements \eqref{eq:Rnd}, appearing in the perturbative treatment of the magnetic field for the HS of Sec.~\ref{sec:ptmfHS}, as well as in the diagonalization to obtain the finite-field spectrum of the sphere \cite{viloria2018orbital}, can be expressed as
\begin{equation}
\label{eq:Rndapp}
\mathcal{R}_{n',l',n,l} = \frac{2 \, L_{l',l}^{(4)}\left(\zeta_{n',l'},\zeta_{n,l}\right)}
{|j_{l'+1}(\zeta_{n',l'}) \, j_{l+1}(\zeta_{n,l}) |} 
\, ,
\end{equation}
with
\begin{equation}
L_{l',l}^{(q)}\left(\alpha,\beta \right) = \int_{0}^{1} {\rm d}\zeta \, \zeta^q \, j_{l'}(\alpha\zeta) \, j_{l}(\beta\zeta) \, ,
\end{equation}
and where $\alpha$ and $\beta$ are such that $j_{l-2}(\alpha)=0$ and $j_l(\beta)=0$, respectively. 

The particular case of $\mathcal{R}_{n',l,n,l}$ appearing in the first term of Eq.~\eqref{eq:derivativesenHSsocb} requires a simpler term
\begin{equation}
K_{l}^{(q)}\left(\alpha,\beta \right) = L_{l,l}^{(q)}\left(\alpha,\beta \right) \, ,
\end{equation}
invoking the definite integral of two spherical Bessel functions with the same order. Using repetitively the recursive formulas developed for the indefinite integrals of spherical Bessel functions of the same order \cite{Bloomfield_2017}, we have
\begin{widetext}
\begin{align}
\label{eq:Kl4}
K_{l}^{(4)}\left(\alpha,\beta \right)  = &
\left(\frac{\alpha^2+\beta^2}{2\alpha \beta} \right)^{l} 
K_{0}^{(4)}\left(\alpha,\beta \right) 
- \sum_{d=1}^{l} \frac{\left(\alpha^2+\beta^2\right)^{d-1}}{(2\alpha \beta)^d} \, 
\Bigg\{ 2 j_{l-d}(\alpha) \, j_{l-d}(\beta) 
\nonumber
\\
&-  \alpha\left( \frac{2\left[2(l-d)+3 \right]}{\alpha^2-\beta^2 } - 1 \right)
j_{l+1-d}(\alpha) \, j_{l-d}(\beta) 
 + \beta\left(\frac{2\left[2(l-d)+3 \right]}{\alpha^2-\beta^2 } + 1  \right)
  j_{l-d}(\alpha) \, j_{l+1-d}(\beta) 
 \Bigg\}
\end{align}
with
\begin{align}
K_{0}^{(4)}\left(\alpha,\beta \right)  = &\;
\frac{1}{2\alpha \beta} 
\left\{
	\frac{1}{(\alpha-\beta)^3} 
		\left[
			2(\alpha-\beta) \, \cos{(\alpha-\beta)} + 
				\left( (\alpha-\beta)^2-2 \right) \sin{(\alpha-\beta)} \right] 
\right.
\nonumber
\\
& \left. -
\frac{1}{(\alpha+\beta)^3} 
		\left[
			2(\alpha+\beta) \, \cos{(\alpha+\beta)} + 
				\left( (\alpha+\beta)^2-2 \right) \sin{(\alpha+\beta)} \right] 
 \right\} \, .
\end{align}

The recursive formulas developed for the indefinite integrals of spherical Bessel functions of different orders \cite{Bloomfield_2017} allow us to write
\begin{equation}
L_{l-2,l}^{(4)}(\alpha,\beta) = \frac{2l-1}{2} \, \frac{\alpha}{\beta} \, 
K_{l-1}^{(4)}\left(\alpha,\beta \right)-
\frac{2l+1}{2} \, K_{l-2}^{(4)}\left(\alpha,\beta \right)
\, ,
\end{equation}
which can be directly evaluated from Eq.~\eqref{eq:Kl4}, and also expressed as
\begin{align}
\label{eq:Ll4}
L_{l-2,l}^{(4)}\left(\alpha,\beta \right)  =&\, 
\frac 14 \left[(2l-1) \left( \frac{\alpha}{\beta} \right)^2 - 2l-3 \right]
 \Bigg[
\left(\frac{\alpha^2+\beta^2}{2\alpha \beta} \right)^{l-2} 
K_{0}^{(4)}\left(\alpha,\beta \right) 
- \sum_{d=1}^{l-2} 
\frac{\left(\alpha^2+\beta^2\right)^{d-1}}{(2\alpha \beta)^d} \, 
\Bigg\{ 2 j_{l-2-d}(\alpha) \, j_{l-2-d}(\beta)
\nonumber
\\
 &-  \alpha\left(\frac{2\left[2(l-d)-1 \right]}{\alpha^2-\beta^2 } -1  \right)
j_{l-1-d}(\alpha) \, j_{l-2-d}(\beta) 
 + \beta \left( \frac{2\left[2(l-d)-1 \right]}{\alpha^2-\beta^2 } +1  \right)
  j_{l-2-d}(\alpha) \, j_{l-1-d}(\beta) 
 \Bigg\} \Bigg]
 \nonumber
 \\
 &+\frac{2l-1}{4}\frac{\alpha}{\beta^2}
\left(\frac{2\left[2l-1 \right]}{\alpha^2-\beta^2 } -1  \right)j_{l-1}(\alpha)\, j_{l-2}(\beta)
 \, . 
\end{align}
The integral $L_{l+2,l}^{(4)}\left(\alpha,\beta \right)$ can be obtained from the expression of $L_{l-2,l}^{(4)}\left(\alpha,\beta \right)$ above by implementing the shift of $l$ to $l+2$ while exchanging the role of $\alpha$ and $\beta$.
\end{widetext}

\bibliography{references}
\end{document}